\documentclass[onecolumn]{pasj01_arxiv}
\usepackage{comment}

\begin{document} 
\Received{2019/07/26}
\Accepted{2020/01/29}

\title{FOREST Unbiased Galactic plane Imaging survey with the Nobeyama 45 m telescope (FUGIN). VI. {Dense gas and mini-starbursts in {the W43 giant} molecular cloud complex} }

\author{Mikito \textsc{Kohno}\altaffilmark{1}$^{*}$
\thanks{Present Address is Astronomy Section, Nagoya City Science Museum, 2-17-1 Sakae, Naka-ku, Nagoya, Aichi 460-0008, Japan.} }
\email{kohno@nagoya-p.jp}
\email{k.tachihara@a.phys.nagoya-u.ac.jp}
\author{Kengo \textsc{Tachihara}\altaffilmark{1}$^{*}$}
\author{Kazufumi \textsc{Torii}\altaffilmark{2}$^{*}$}
\author{Shinji \textsc{Fujita}\altaffilmark{1,3}$^{*}$}
\author{Atsushi \textsc{Nishimura}\altaffilmark{1,3}}
\author{Nario \textsc{Kuno}\altaffilmark{4,5,12}}
\author{Tomofumi \textsc{Umemoto}\altaffilmark{2,6}}
\author{Tetsuhiro \textsc{Minamidani}\altaffilmark{2,6,7}}
\author{Mitsuhiro \textsc{Matsuo}\altaffilmark{2}}
\author{Ryosuke \textsc{Kiridoshi}\altaffilmark{3}}
\author{Kazuki \textsc{Tokuda}\altaffilmark{3,7}}
\author{Misaki \textsc{Hanaoka}\altaffilmark{1}}
\author{Yuya \textsc{Tsuda}\altaffilmark{8}}
\author{Mika \textsc{Kuriki}\altaffilmark{4}}
\author{Akio \textsc{Ohama}\altaffilmark{1}}
\author{Hidetoshi \textsc{Sano}\altaffilmark{1,7,9}}
\author{Tetsuo \textsc{Hasegawa}\altaffilmark{7}}
\author{Yoshiaki \textsc{Sofue}\altaffilmark{10}}
\author{{Asao \textsc{Habe}\altaffilmark{11}}}
\author{Toshikazu \textsc{Onishi}\altaffilmark{3}}
\author{Yasuo \textsc{Fukui}\altaffilmark{1,9}}

\altaffiltext{1}{Department of Physics, Graduate School of Science, Nagoya University, Furo-cho, Chikusa-ku, Nagoya, Aichi 464-8602, Japan}
\altaffiltext{2}{Nobeyama Radio Observatory, National Astronomical Observatory of Japan (NAOJ), National Institutes of Natural Sciences (NINS), 462-2, Nobeyama, Minamimaki, Minamisaku, Nagano 384-1305, Japan}
\altaffiltext{3}{Department of Physical Science, Graduate School of Science, Osaka Prefecture University, 1-1 Gakuen-cho, Naka-ku, Sakai, Osaka 599-8531, Japan}
\altaffiltext{4}{Department of Physics, Graduate School of Pure and Applied Sciences, University of Tsukuba, 1-1-1 Ten-nodai, Tsukuba, Ibaraki 305-8577, Japan}
\altaffiltext{5}{Tomonaga Center for the History of the Universe, University of Tsukuba, Ten-nodai 1-1-1, Tsukuba, Ibaraki 305-8571, Japan}
\altaffiltext{6}{Department of Astronomical Science, School of Physical Science, SOKENDAI (The Graduate University for Advanced Studies), 2-21-1, Osawa, Mitaka, Tokyo 181-8588, Japan}
\altaffiltext{7}{National Astronomical Observatory of Japan (NAOJ), National Institutes of Natural Sciences (NINS), 2-21-1 Osawa, Mitaka, Tokyo 181-8588, Japan}
\altaffiltext{8}{Department of Physics, School of Science and Engineering, Meisei University, 2-1-1 Hodokubo, Hino-shi, Tokyo 191-8506, Japan}
\altaffiltext{9}{Institute for Advanced Research (IAR), Nagoya University, Furo-cho, Chikusa-ku, Nagoya 464-8601, Japan}
\altaffiltext{10}{Institute of Astronomy (IoA), The University of Tokyo, 2-21-1, Osawa, Mitaka, Tokyo 181-0015, Japan}
\altaffiltext{11}{{Department of Physics, Faculty of Science, Hokkaido University, Kita 10 Nishi 8, Kita-ku, Sapporo, Hokkaido 060-0810, Japan}}
\altaffiltext{12}{Department of Physics, School of Science and Technology, Kwansei Gakuin University, 2-1 Gakuen,Sanda, Hyogo 669-1337, Japan}



\maketitle

\begin{abstract}
{We performed new large-scale $^{12}$CO, $^{13}$CO, and C$^{18}$O $J=$1--0  observations of the W43 giant molecular cloud complex {in} the tangential direction of the Scutum arm ($l\sim \timeform{30D}$) as a part of the FUGIN project.
The low-density gas traced by $^{12}$CO {is distributed over} 150 pc $\times$ 100 pc ($l \times b$), and has {a} large velocity dispersion ({20-30 $\>$km s$^{-1}$}). 
However, the dense gas traced by C$^{18}$O {is localized in the} W43 Main, G30.5, and W43 South (G29.96-0.02) high-mass star-forming regions in the W43 GMC complex, which have clumpy structures.
We found at least two clouds with a velocity difference of $\sim$ 10-20 $\>$km s$^{-1}$, {both of} which are likely to be physically associated with these high-mass star-forming regions based on the results of high $^{13}$CO $J=$ 3--2 to $J =$ 1-0 intensity ratio and morphological correspondence with the infrared dust emission. 
{The velocity separation of these clouds in W43 Main, G30.5, and W43 South is too large {for each cloud} to be gravitationally bound.}
{We also revealed that the dense gas in the W43 GMC has a high local column density, while “the current SFE" of entire the GMC is low ($\sim 4\%$) compared with the W51 and M17 GMC.}
{We argue that the supersonic cloud-cloud collision hypothesis can explain the origin of the local mini-starbursts and dense gas formation in the W43 GMC complex.}}
\end{abstract}

\section{Introduction}
\subsection{Giant molecular clouds and mini-starbursts {in} the Milky Way}
Giant molecular clouds (GMCs), whose masses {are $\gtrsim10^4 M_{\odot}$} (e.g., Blitz 1993), have {been} studied by CO surveys of the Milky Way using single-dish radio telescopes since the 1970s (e.g., Dame et al. 1986, 2001; Solomon et al. 1987; Scoville et al. 1987; Mizuno \& Fukui 2006; see also reviews of Combes 1991; Heyer \& Dame 2015). 
Recently, high-angular-resolution CO surveys have revealed the internal structures {of} GMCs using {the} Nobeyama 45-m (FUGIN: Umemoto et al. 2017), JCMT 15 m, (COHRS: Dempsey et al. 2013, CHIMPS: Rigby et al. 2017), PMO 13.7 m (MWISP: Su et al. 2019), and Mopra 22 m (Burton et al. 2013; Braiding et al. 2018) {telescopes}.
GMCs {have} ideal environments, {with massive and dense gas, for} {the formation of} high-mass stars and cluster formation (e.g., Lada \& Lada 2003; McKee \& Ostriker 2009), and their evolution and formation mechanism has {been} widely studied {in} the Milky Way and Local Group Galaxies (e.g., Blitz et al. 2007; Fukui \& Kawamura 2010).
In particular, massive GMCs often {form aggregations} called “GMC complexes", {which} {are} {regarded as} {sites for {mini-starbursts including O-type stars} (e.g., W43, W49, W51, Sagittarius B2, NGC 6334-6357: see a review of Motte et al. 2018a).
On the other hand, it is not yet clear {how high-mass stars and {natal massive dense cores} are formed in GMCs} (e.g., Gao and Solomon 2004; Lada et al. 2012; Torii et al. 2019).

\subsection{{The initial condition of O-type star formation produced by a {supersonic} cloud-cloud collision}}
{{ To form the O-type stars ($> 20\ M_{\odot}$), it is necessary to achieve high mass accretion {at the rate} of $\sim 10^{-3} \ M_{\odot}\ {\rm yr}^{-1}$ (e.g., Wolfire \& Cassinelli 1987) to overcome their strong radiation pressure. 
Zinnecker \& Yorke ({2007}) argued that “rapid external shock compression" such as the supersonic gas motions might be an important initial condition for O-type star formation. }}

{Cloud-cloud collisions have been discussed as one of the external triggering mechanisms that produce gravitational instability and high-mass star formation (e.g., Scoville et al. 1986; Elmegreen 1998). Recently, observational evidence has been presented of many super star clusters and O-type star-forming regions in the Milky Way.
{For example,}
\begin{itemize}
\item Super star clusters (young massive clusters): Compact distributions of O-type stars in {1 pc} \\
Westerlund 2 (Furukawa et al. 2009; Ohama et al. 2010), NGC 3603 (Fukui et al. 2014), RCW 38 (Fukui et al. 2016), etc.
\item Multiple O-type star formation in the GMCs\\
Sagittarius B2 (Hasegawa et al. 1994; Sato et al. 2001), W49A (Mufson \& Liszt 1977; Miyawaki et al. 1986; 2009), W51A (Okumura et al. 2001; Kang et al. 2010; Fujita et al. 2019b), 
NGC 6334-6357 (Fukui et al. 2018a), M17 (Nishimura et al. 2018), M42 (Fukui et al. 2018d), RCW 79 (Ohama et al. 2018a), W33 (Kohno et al. 2018a), DR21 (Dobashi et al. 2019), etc.
\item Single O-type star formation in mid-infrared bubbles and H\,\emissiontype{II} regions \\
M20 (Torii et al. 2011,2017), RCW 120 (Torii et al. 2015), N37 (Baug et al. 2016), Sh2-235 (Dewangan \& Ojha 2017), N49 (Dewangan et al. 2017), RCW 34 (Hayashi et al. 2018), RCW 36 (Sano et al. 2018), RCW 32 (Enokiya et al. 2018), GM 24 (Fukui et al. 2018b), S116-118 (Fukui et al. 2018c), N35 (Torii et al. 2018a), {N36 (Dewangan et al. 2018)}, Sh 2-48 (Torii et al. 2018b), RCW 166 (Ohama et al. 2018b), S44 (Kohno et al. 2018b), G8.14+0.23 (Dewangan et al. 2019), N4 (Fujita et al. 2019a), etc
\end{itemize}

{Numerical simulations show that gravitationally{-}unstable, dense cores are formed in {the} compressed layer of two colliding clouds (e.g., Habe \& Ohta 1992; Anathpindika 2010; Takahira et al. 2014, 2018; Balfour et al. 2015, 2017; Matsumoto et al. 2015; Shima et al. 2018; Wu et al. 2015, 2017a, 2017b, 2018). 
{Inoue \& Fukui (2013) demonstrated from the results of {a} magneto-hydrodynamical (MHD) simulation that the shock compression from a collision can produce massive molecular cores by amplifying the turbulence and magnetic field. These massive cores can achieve a high enough accretion rate ($> 10^{-4} M_{\odot} {\rm yr}^{-1}$) to form O-type stars (Inoue et al. 2018). 
Hence, a supersonic cloud-cloud collision is a prominent scenario in which the initial conditions for O-type stars and dense gas formation of galaxies can be met.}

{In order to clarify the origin of dense gas and O-type star formation in GMC complexes, we have performed {a} new high-resolution and complete survey {in the $^{12}$CO, $^{13}$CO, and C$^{18}$O $J=$ 1--0 lines} of the entire {W43 GMC complex, where active star formation is ongoing}}.
{This paper is {constructed} as follows: section \ref{sec:W43} introduces the current knowledge on W43; section \ref{sec:data} presents the observational properties; section \ref{sec:result} gives the FUGIN results and comparisons with infrared wavelengths; in section \ref{sec:discussion} we discuss the formation mechanism of the W43 GMC complex; {and} {in} section \ref{sec:conclusion} {we provide the conclusions of this paper.}}

\section{{W43 as a massive GMC complex in the Milky Way}}
\label{sec:W43}

W43 is {a} Galactic mini-starburst region {located in Aquila}, which {was} first cataloged by the 1390 MHz thermal radio continuum survey (Westerhout 1958).
It is one of the most massive GMC complexes in the Milky Way, {corresponding} to giant molecular {associations} (GMA) of external galaxies (Nguyen-Luong et al. 2011).
Figure \ref{W43_spitzer}(a) shows a three-color composite image obtained by the Spitzer space telescope, where blue, green, and red correspond to the 3.6 $\>\mu$m, 8 $\>\mu$m, and 24 $\>\mu$m emissions, respectively (GLIMPSE: Benjamin et al. 2003; Churchwell et al. 2009; MIPSGAL: Carey et al. 2009). The X-marks indicate W43 Main (Blum et al. 1999) and W43 South (Wood \& Churchwell 1989), which {are active star forming regions.} {G30.5 exists between W43 Main and W43 South, which contains five star-forming regions. {A} 100 pc bow-like structure, vertical to the Galactic plane, was found by the Nobeyama 10 GHz radio continuum survey (Sofue 1985; Handa et al. 1987). }
The 3.6 $\>\mu$m, 8 $\>\mu$m, and 24 $\>\mu$m emission {traces} thermal emission from stars, Polycyclic Aromatic Hydrocarbon (PAH) features (e.g., Draine 2003; Draine \& Li 2007), and hot dust grains ($\sim$ 120 K) excited by the high-mass stars (Carey et al. 2009), respectively. The 8 $\>\mu$m emission {shows} diffuse {distribution covering} the whole GMC complex. The 24 $\>\mu$m emission {is} bright at W43 Main and W43 South, {indicating the existence of H\,\emissiontype{II} regions.} The properties of these objects are described in Table \ref{pro_W43Main_W43South}.

The {distances of W43 South and compact H\,\emissiontype{II} regions around G31.5 have been measured by trigonometric parallax, yielding 5.49 kpc {as} the variance-weighted average of four maser sources {of} G029.86-00.04, G029.95-00.01, G031.28+00.06, and G031.58+00.07 from the Bar and Spiral Structure Legacy Survey (BeSSeL: Zhang et al. 2014; Sato et al. 2014). In this paper, we assume the same distance (5.49 kpc) toward W43 Main{,} because it coincides with a near-side kinematic distance of $v_{\rm LSR}  \sim 95$ $\>$km s$^{-1}$ at $l \sim \timeform{30.8D}$. } 
{Figure \ref{MW_large} shows the position of W43 superposed on the top-view of the Milky Way (R. Hurt: NASA/JPL-Caltech/SSC).}
We point out that W43 exists {close to} the tangential point of the Scutum arm (e.g., Vall\'ee 2014; Hou \& Han 2014; Nakanishi \& Sofue 2016). In addition, {this region} is suggested {to be} the meeting point of the Galactic long-bar and {the} Scutum arm (e.g., Nguyen-Luong et al. 2011; Veneziani et al. 2017). Therefore, W43 is an important object {for} studying the spiral arm and bar interaction.

Large-scale CO observations of W43 have {been} performed by {single-dish radio telescopes (Nguyen Luong et al. 2011; Carlhoff et al. 2013; Motte et al. 2014). They showed that W43 is a molecular cloud complex with {a} size of $\sim$ 100-200 pc, and a broad velocity dispersion of {20-30 $\>$km s$^{-1}$}}
{They proposed that a cloud-cloud collision or {a} colliding flow scenario is likely to explain the origin of the mini-starbursts in the W43 GMC at the tip of the Galactic bar (e.g., Nguyen-Luong et al. 2011, 2013; Motte et al. 2014; Louvet et al. 2016; see also review section 4.3 of Motte et al. 2018a).}
Furthermore, a number of papers have {been} published {on} the W43 GMC complex (e.g., Pipher et al. 1974; Lester et al. 1985; Liszt et al. 1993; Liszt 1995; {Mooney et al. 1995}; Subrahmanyan \& Goss 1996; Balser et al. 2001; Lemoine-Goumard et al. 2011; Beuther et al. 2012; Nishitani et al. 2012; Eden et al. 2012; Moore et al. 2015; Bihr et al. 2015; Walsh et al. 2016; G\'enova-Santos et al. 2017; Langer et al. 2017; Bialy et al. 2017). 
{On the other hand, these papers do not deal with detail analysis toward individual objects in the W43 GMC complex.}
In the next subsection, we introduce the individual objects of W43 Main, G30.5, and W43 South{, on which we base our analysis.}  

\subsection{{W43 Main (G030.8-00.0) as {a} mini-starburst region in the Milky Way}}
Figure \ref{W43_spitzer}(b) shows {a} close-up image of W43 Main. The X-mark indicates the {center position} of {an OB cluster} (Blum et al. 1999), and the white circles present 51 {protocluster candidates} identified by sub-millimeter dust continuum observations (Motte et al. 2003). W43 Main is the most active high-mass star-forming region in the W43 GMC complex ({e.g., Nguyen-Luong et al. 2013, 2017; Cortes et al. 2019}), {associated with a} giant H\,\emissiontype{II} region {and a} young massive cluster (Blum et al. 1999; Longmore et al. 2014). We find {a} ring-like structure {in} 8~$\>\mu$m emission, which {corresponds} to the infrared bubble N52 (Churchwell et al. 2006).
The total {infrared} luminosity and ultraviolet photon number ($N_{\rm Ly}$) are estimated to be $7-10 \times 10^6 \ L_{\odot}$ (e.g., Hattori et al. 2016; Hanaoka et al. 2019; Lin et al. 2016) and $2.3 \times 10^{50}$ s$^{-1}$ from the flux density at 5 GHz (Smith et al. 1978; Deharveng et al. 2010), respectively. The $N_{\rm Ly}$ is equivalent to $\sim 14$ O5V or $\sim 40$ O7V stars, based on the stellar parameters of Martins et al. (2005). 
{The W43 Main cluster contains one Wolf-Rayet star (W43\#1) and two O-type giant stars (W43 \#2 and W43 \#3), identified by Blum et al. (1999).
W43\#1 is also listed {as} WR 121a (van der Hutch 2001).}
 W43 \#1 and \#3 are likely to be binary O-type stars concentrated in a narrow space of $<$ 1 pc {from} the W43 Main cluster (Luque-Escamilla et al. 2011; Binder \& Povich 2018). 
 {Their stellar {ages are} estimated {to be} 1--6 Myr{,} based on the typical main-sequence lives of the {O-type stars and Wolf-Rayet stars} (Motte et al. 2003; Bally et al. 2010).}
We {summarize} the properties of {the} W43 Main cluster in Table \ref{Ostar_W43Main}.

Recently, {several} papers have {been} published on W43-MM1 (e.g., Cortes \&  Crutcher 2006; Cortes et al. 2010; Cortes 2011; Cortes et al. 2016; Herpin et al. 2012; Sridharan et al. 2014; Louvet et al. 2014; Jacq et al. 2016; Nony et al. 2018; Molet et al. 2019). {These papers revealed that W43-MM1 is {the} most massive protocluster in W43 Main by the interferometer observations.}
In particular, Motte et al. (2018b) identified {131 massive dust clumps} in W43-MM1 from  Atacama Large Millimeter/Submillimeter Array (ALMA) observations. They {showed that} the core mass function (CMF) {is} flatter (more top-heavy) than the stellar initial mass function (IMF: Salpeter 1955), which suggests that massive dense cores are likely to be formed more efficiently in W43-MM1.

\subsection{G30.5}
Figure \ref{W43_spitzer}(c) shows {a} close-up image of G30.5, which exists between W43 Main and W43 South. 
Five star-forming regions exist at G30.5: IRAS 18445-0222, IRAS 18447-0229, G030.489-00.364, G030.213-00.156, and G30.404-00.238 (Beichman et al. 1988; Anderson et al. 2014\footnote{http://astro.phys.wvu.edu/wise/}; Lockman 1989). 
There are 870 $\>\mu$m cold dust clumps {corresponding} to these infrared sources except for G030.489-00.364 (Nguyen-Luong et al. 2011).
{The authors} also {found} multiple velocity components {by using} the $^{13}$CO $J=$ 1--0 emission (see Figure 4 of Nguyen-Luong et al. 2011).  
{Recently, Sofue et al. ({2019}) analyzed the FUGIN data on the W43 GMC complex.  They revealed {a molecular bow structure in G30.5, which is suggested} to be formed by the Galactic shock compression (Fujimoto 1968; Roberts 1969; Tosa 1973). We point out that these star-forming regions correspond to the head of the molecular bow.}

\subsection{W43 South (G29.96-0.02)}
Figure \ref{W43_spitzer}(d) presents {a} close-up image of W43 South (G29.96-0.02). The crosses indicate the radio continuum sources identified by the NRAO/VLA Sky Survey (NVSS: Condon et al. 1998), and correspond to the {late-O or early-B} type stars from the {radio continuum flux} (Beltran et al. 2013). The numbering is the same as in Beltran et al. (2013). The total luminosity is derived to be 2--6 $\times 10^6\ L_{\odot} $ (Beltr\'an et al. 2013, Lin et al. 2016). The stellar age is derived to be $\sim$ 0.1 Myr {from the evolutionary stage of the ultra-compact H\,\emissiontype{II} region} (Watson \& Hanson 1997), which is younger than W43 Main. 
The brightest infrared source (\#1=IRAS 18434-0242) {is regarded} as an {ultra}-compact H\,\emissiontype{II} region or {hot} core, {and has been well studied by many authors} (e.g., Pratap et al. 1994, 1999; Afflerbach et al. 1994;  Cesaroni et al. 1994, 1998, 2017; Fey et al. 1995; Lumsden \& Hoare 1996, 1999; Ball et al. 1996; Watarai et al. 1998; Maxia et al. 2001; De Buizer et al. 2002; Morisset et al. 2002; Mart\'in-Hern\'andez et al. 2003; Olmi et al. 2003; Rizzo et al. 2003; Hoffman et al. 2003; Zhu et al. 2005; Kirk et al. 2010; Beltr\'an et al. 2011; Pillai et al. 2011; Townsley et al. 2014; Roshi et al. 2017). We {summarize} the properties of W43 South in Table \ref{Ostar_W43South}.

{{Several} papers have been published on the} W43 GMC complex and its individual clusters, {so far} {a comprehensive study of diffuse molecular clouds and dense cores {does} not exist. In this paper, we have performed the analysis of $^{12}$CO, $^{13}$CO, and C$^{18}$O $J=$1--0 emissions from the W43 GMC complex, {obtained from} the FUGIN data.}

\section{Data sets}
\label{sec:data}

\subsection{{The} FUGIN project: {the} Nobeyama 45-m telescope $^{12}$CO,  $^{13}$CO, and C$^{18}$O $J =$ 1--0 observations}
We performed the CO  $J=$1--0 survey using the 45-m telescope at the Nobeyama Radio Observatory (NRO). 
We simultaneously observed {the present region} in $^{12}$CO (115.271 GHz),  $^{13}$CO (110.201 GHz), and C$^{18}$O (109.782 GHz) $J=$1--0 transitions  as part of the FUGIN project\footnote{https://nro-fugin.github.io}.
The area of inner Galaxy {covering} $l=$ \timeform{10D}-\timeform{50D}, $b=\timeform{-1D}$-$\timeform{1D}$ {as shown Figure \ref{MW_large}} {has been surveyed} {using} the on-the-fly (OTF) mapping mode (Sawada et al. 2008) from April 2014 to March 2017 (Minamidani et al. 2015; Umemoto et al. 2017). 
The front-end {consists of} four beams, dual polarization, and {sideband-separating} (2SB) superconductor-insulator-superconductor (SIS) receiver FOur-beam REceiver System on the 45 m Telescope (FOREST: Minamidani et al. 2016, {Nakajima et al. 2019}), with a typical system noise temperature ($T_{\rm sys}$) of $\sim 250$ K ($^{12}$CO) and 150 K ($^{13}$CO).
The back-end is an FX-type digital spectrometer named Spectral Analysis Machine for the 45 m telescope (SAM45: Kuno et al. 2011), {which is} {the same as} the digital spectro-correlator system for the  ALMA Atacama Compact Array (Kamazaki et al. 2012).
It has 4096 channels with a bandwidth and frequency resolution of 1 GHz and 244.14 kHz, which correspond to 2600 $\>$km s$^{-1}$ and 0.65 $\>$km s$^{-1}$ at 115 GHz, respectively. 
The effective velocity resolution is 1.3 km s$^{-1}$. 
{The half power beam width (HPBW) of the 45 m telescope is \timeform{14"} and \timeform{15"} at 115 GHz and 110 GHz, respectively.}
The effective beam size {of the final data {cube}, convolved {with} a Bessel $\times$ Gaussian function}, is \timeform{20"} {for} $^{12}$CO and \timeform{21"} {for} $^{13}$CO, {corresponding} to $\sim 0.5 $ pc at a distance of 5.5 kpc.
{The pointing accuracy {is smaller} than \timeform{2"}-\timeform{3"}}{,and measured} by {observing} 43 GHz SiO maser sources every hour using the H40 High Electron Mobility Transistor (HEMT) receiver.
We used the chopper-wheel method (Ulich \& Haas 1976; Kutner \& Ulich 1981) to convert to the antenna temperature ($T_a^*$) scale from the raw data.
The data were calibrated to the {main-beam} temperature ($T_{\rm mb}$) by using the equation $T_{\rm mb}=T_a^* / \eta_{\rm mb}$ with a main beam efficiency ($\eta_{\rm mb}$) of 0.43 for $^{12}$CO, and 0.45 for $^{13}$CO and C$^{18}$O.
The {relative} intensity uncertainty {is} estimated {at} 10--20\% for $^{12}$CO, 10\% for $^{13}$CO, and 10\% for C$^{18}$O by observation of the standard source M17 SW (Umemoto et al. 2017).
The final 3D FITS cube has a voxel size of $(l, b, v) = (\timeform{8.5"}, \timeform{8.5"}, 0.65 $ $\>$km s$^{-1}$).
We smoothed the data by {convolving it with} the two-dimensional Gaussian function (\timeform{35"}) to achieve a spatial resolution of $\sim$ \timeform{40"}.
The root-mean-square (rms) noise levels after smoothing are $\sim$1.0 K, 0.35 K, and 0.35 K for $^{12}$CO, $^{13}$CO, and C$^{18}$O, respectively.
More detailed information on the FUGIN project, {we refer to} described by Umemoto et al. (2017).

\subsection{The CHIMPS project: JCMT $^{13}$CO $J=$3--2 archive data}
We also used the James Clerk Maxwell Telescope (JCMT) archival data of the CO ($J=$3--2) Heterodyne Inner Milky Way Plane Survey (CHIMPS\footnote{https://www.eaobservatory.org/jcmt/science/large-programs/chimps2/}) project (Rigby et al. 2016; {2019}). The cube data were obtained from the JCMT web page\footnote{https://www.canfar.net/storage/list/AstroDataCitationDOI/CISTI.CANFAR/16.0001/data/CUBES/13CO} with an effective resolution of \timeform{15"} obtained by the Heterodyne Array Receiver Program (HARP: Buckle et al. 2009).
We carried out the regridding {in} the spatial and velocity {areas} with reference to the FUGIN data. 
We adopted $\eta_{\rm mb} = 0.72$ as the main beam efficiency  (Buckle et al. 2009), and converted {the intensity from $T_a^*$} to the $T_{\rm mb}$ scale.
The {relative} intensity uncertainty was estimated within 20 \% for $^{13}$CO $J=$3--2 (Rigby et al. 2016).
We convolved the data {with} {a} \timeform{37"} Gaussian to achieve a spatial resolution of $\sim$ \timeform{40"}. 
The rms noise level after smoothing was 0.14 K for $^{13}$CO $J=$3--2.

\subsection{{Spitzer {infrared archive data}}}
We used infrared archival data obtained by the Spitzer space telescope ({Werner et al. 2004}).
The near-infrared 3.6 $\>\mu$m {and} 8.0 $\>\mu$m images were obtained by the Infrared Array Camera (IRAC: Fazio et al. 2004) as a part of the Galactic Legacy Infrared Midplane Survey Extraordinaire (GLIMPSE) project (Benjamin et al. 2003, Churchwell et al. 2009).
The mid-infrared 24 $\>\mu$m {images were} taken from the Multi-band Imaging Photometer for Spitzer (MIPS: Rieke et al. 2004) as a part of the 24 and 70 Micron Survey of the Inner Galactic Disk with MIPS (MIPSGAL) project (Carey et al. 2009).
The {resolutions of these images} are $\sim \timeform{2"}$, $\sim \timeform{2"}$, and $\sim \timeform{6"} $ for 3.6 $\>\mu$m,  8.0 $\>\mu$m, and 24 $\>\mu$m, respectively. We summarize the observational properties and archival information in Table \ref{obs_param}.

\section{Results}
\label{sec:result}

{This} section {is} organized as follows: {subsection} \ref{sec:Galactic-scale} {presents} the CO distribution {in the first galactic quadrant}; {subsection} \ref{sec:W43GMC} gives CO distributions and velocity structures of the W43 GMC complex; and {subsections} \ref{sec:W43Main}-\ref{sec:W43South} focus on the three regions of W43 Main, G30.5, and W43 South, respectively. {These} are active star-forming regions in the W43 GMC complex. 

\subsection{Galactic-scale CO distributions and velocity structures {in} the first {galactic} quadrant with FUGIN}
\label{sec:Galactic-scale}

{The upper panels of Figures \ref{FUGIN_large_12CO}, \ref{FUGIN_large_13CO}, and \ref{FUGIN_large_C18O}} show the integrated intensity maps of the inner Galaxy ($l=\timeform{10D}$--$\timeform{50D}$, $b=\timeform{-1D}$--$\timeform{1D})$ of $^{12}$CO, $^{13}$CO, and C$^{18}$O, respectively. The integrated velocity {ranges} from $-30$ to 160 $\>$km s$^{-1}$. The lower panels show the Galactic longitude-velocity {diagrams}, where {the} dotted lines indicate the spiral arm {suggested} by Reid et al. (2016). We find the Norma, Sagittarius, Scutum, Perseus, {Near 3 kpc, Far 3 kpc}, and Outer (Cygnus) spiral arms {lie in this region of the inner Galactic plane.}
The Aquila Rift and Aquila Spur are also identified as the local components of the solar neighborhood and the inter-arm region, respectively. 
The active high-mass star-forming regions and a supernova remnant exist in each spiral arm (e.g., W33, M17, M16, W43, W49, W51, and W44). 
W43 exists {in the Galactic plane} in the direction of $l=\timeform{30D}$, which contains rich gas{,} {as seen in} the integrated intensity maps and longitude-velocity diagram of the first quadrant ({Figures \ref{FUGIN_large_12CO}-\ref{FUGIN_large_C18O}}).
This region is close to the tangential point of the Scutum Arm and the bar-end of the Milky Way (Nguyen-Luong et al. 2011). 
{The $^{13}$CO also traces the spiral structures more clearly (Figure \ref{FUGIN_large_13CO}).}
We note that dense gas traced by C$^{18}$O (Figure \ref{FUGIN_large_C18O}) has {localized} distribution compared with the low-density gas traced by $^{12}$CO (Figure \ref{FUGIN_large_12CO}).

\subsection{CO distributions of the W43 giant molecular cloud complex}
\label{sec:W43GMC}

Figure \ref{W43_integ} demonstrates the integrated intensity maps of the W43 GMC complex, {in}  (a) $^{12}$CO, (b) $^{13}$CO , and (c) C$^{18}$O. 
The integrated velocity {ranges} from 78 to 120 $\>$km s$^{-1}$, which corresponds to {those defined by} the previous IRAM CO $J=$2-1 observations by Carlhoff et al. (2013).
The X marks show the center positions of W43 Main and W43 South (Blum et al. 1999; Wood \& Churchwell 1989). The entire GMC complex is distributed over $\sim 150$ pc of the Galactic plane. The CO intensity has peaks at W43 Main, G30.5, and W43 South. {Each region corresponds to a} high-mass star-forming region. {We point out that the C$^{18}$O {exists locally} at W43 Main, G30.5, and W43 South.
{The C$^{18}$O is} surrounded by the {diffuse molecular gas} traced by $^{12}$CO. }

Figures \ref{W43_12COch}, \ref{W43_13COch}, and \ref{W43_C18Och} show velocity channel maps of $^{12}$CO,  $^{13}$CO,  and C$^{18}$O, respectively. The velocity {ranges} from 66 to 125 $\>$km s$^{-1}$,{with an interval of} $\sim$ 4 $\>$km s$^{-1}$. {The velocity of the peak positions in W43 Main, G30.5, and W43 South {ranges between} 90-105 $\>$km s$^{-1}$, 100-109 $\>$km s$^{-1}$, and 93-105 $\>$km s$^{-1}$, respectively}.
We find that the dense gas traced by C$^{18}$O has clumpy structures over a wide velocity range (Figure \ref{W43_C18Och}). 

Figure \ref{W43_lv} shows {the} longitude-velocity diagrams integrated {{over the latitudes between} $\timeform{-0.4D}$ {and} $\timeform{0.4D}$}.
 The $^{12}$CO has a {much larger} velocity width {(} {$\sim$ 20-30 $\>$km s$^{-1}$}; {Figure \ref{W43_lv}a, and \ref{W43_spectra}} ) than is {typical} for a GMC, with {a size of} 100 pc ($\Delta v \sim$ 10 $\>$km s$^{-1}$) calculated by  Larson's law ($\Delta v \sim R^{0.5}$: Larson 1981; Heyer \& Brunt 2004). 
 {The velocity width of the $^{13}$CO line is narrower than that of $^{12}$CO (Figure \ref{W43_lv}a and \ref{W43_lv}b).}
However, C$^{18}$O has clumpy structures with velocity widths of $< 10$ $\>$km s$^{-1}$ (Figure \ref{W43_lv}c), {and the large cloud complex in $^{12}$CO breaks into multiple velocity components in C$^{18}$O.}

The total molecular masses {of the whole W43 GMC} are derived { from $^{12}$CO, $^{13}$CO, and C$^{18}$O to be} $\sim 1\times 10^7\ M_{\odot}$, $\sim 1\times 10^7\ M_{\odot}$, and $\sim 2\times 10^6\ M_{\odot}$ {above $5\sigma$ noise levels ($\sim 5$ K of $^{12}$CO and $1.75$ K of $^{13}$CO, C$^{18}$O $J =$1--0)}, respectively. 
Table 5 presents column densities and molecular masses of the W43 GMC complex. {The molecular masses estimated from $^{12}$CO are roughly consistent with {the masses calculated from} $^{13}$CO.}
{On the other hand, The mass estimation from C$^{18}$O is a factor of 4 lower than the others. This difference might be caused by the variation of the CO abundance ratio due to the UV radiation from OB-type stars in the W43 GMC complex. 
{According to Bally and Langer (1982), the selective photodissociation occurs in the surface layer of molecular clouds. Because the self-shielding against UV radiation is more efficient for more abundant molecule, dissociation of C$^{18}$O becomes more effective at low density region, resulting in making the C$^{18}$O molecule an empirical dense gas tracer. 
Indeed, Paron et al. (2018) reported that the $^{13}$CO/C$^{18}$O abundance ratio ($X^{13/18}$) is low in the dense region of W43 South (G29.96-0.02). }
We {present details} of the method calculating the physical properties of the molecular gas in Appendix 1. 

{Figure \ref{W43_spectra} shows the spectra of (a) W43 Main, (b) G30.5, and (c) W43 South. We find a large velocity width (20-30 $\>$km s$^{-1}$) in $^{12}$CO and multiple velocity components in $^{13}$CO from 80 to 120 $\>$km s$^{-1}$ (see red arrows of Figure \ref{W43_spectra}). We observe that the 70 $\>$km s$^{-1}$ component is the foreground cloud{, as} suggested by Carlhoff et al. (2013).}

In the next subsection, we present the {results of} detailed analysis {in} W43 Main, G30.5, and W43 South, aiming {to reveal} the origin of the dense gas and high-mass stars.

\subsection{W43 Main}
\label{sec:W43Main}

\subsubsection{$^{13}$CO and C$^{18}$O spatial distributions and velocity structures}
{We find four clouds at different {velocities} (82 $\>$km s$^{-1}$, 94 $\>$km s$^{-1}$, 103 $\>$km s$^{-1}$, and 115 $\>$km s$^{-1}$) {in the} W43 Main cluster.}
{Figure \ref{W43Main_integ}} presents integrated intensity maps of each cloud {in the} $^{13}$CO ({Figures  \ref{W43Main_integ}} a-d) and C$^{18}$O ({Figures  \ref{W43Main_integ}} e-h) emission.
The X-mark indicates a stellar cluster, and the white circles show 51 protocluster candidates identified by Motte et al. (2003).

The 82 $\>$km s$^{-1}$ cloud ({Figures  \ref{W43Main_integ}a, \ref{W43Main_integ}e}) has an intensity peak at $(l,b)\sim (\timeform{30.66D}, \timeform{0.03D})$, and extends over 20 pc on the western side of the cluster. 
The 94 $\>$km s$^{-1}$ cloud ({Figures  \ref{W43Main_integ}b, \ref{W43Main_integ}f}) is the brightest component, and {is distributed} over 20 pc around the cluster. {Two} peaks {in} $^{13}$CO correspond to the W43-MM1 and W43-MM2 ridges identified {in} the N$_2$H$^+$ and SiO emission (Nguyen-Luong et al. 2013). In particular, the W43-MM2 ridge is brightest {in} the C$^{18}$O emission ({Figure  \ref{W43Main_integ}f}).
The 103 $\>$km s$^{-1}$ cloud ({Figures  \ref{W43Main_integ}c, \ref{W43Main_integ}g}) has intensity peaks at {$(l,b)\sim (\timeform{30.83D}, \timeform{-0.05D})$ and $(l,b)\sim (\timeform{30.86D}, \timeform{-0.08D})$}, $\sim 5$ pc apart from the cluster. These peaks also correspond to the W43-MM1 ridge.
The 115 $\>$km s$^{-1}$ cloud ({Figures  \ref{W43Main_integ}d, \ref{W43Main_integ}h}) shows the diffuse component of $^{13}$CO near the cluster center. The two peaks exist at {$(l,b)\sim (\timeform{30.63D}, \timeform{-0.11D})$ and $(l,b)\sim (\timeform{30.81D}, \timeform{-0.19D})$}, which is {separate} from the cluster.
We also show the velocity channel maps in the Appendix {Figures \ref{W43Main_ch1}-\ref{W43Main_ch3}}

{Figure  \ref{W43Main_lv}} displays the latitude-velocity diagram of {W43 Main in} (a) $^{13}$CO and (b) C$^{18}$O $J=$1--0. 
The black boxes indicate the radio recombination line (RRL) velocity of 91.7 $\>$km s$^{-1}$ at $(l,b)\sim (\timeform{30.780D}, \timeform{-0.020D})$ obtained by Luisi et al. (2017).
The yellow line {indicates} the position of the W43 Main cluster. 
The 94 $\>$km s$^{-1}$ cloud {has} $\sim 10$ $\>$km s$^{-1}$ line width, which is the broadest line width of all clouds and also traced by C$^{18}$O $J=$1--0. {The RRL velocity coincides with the velocity range of the 94 km s$^{-1}$ cloud.}
This cloud {has} the highest peak column density $\sim 10^{23}$ cm$^{-2}$ {among} all clouds, and {is connected} with the 82 $\>$km s$^{-1}$ and 103 $\>$km s$^{-1}$ clouds in the velocity space of $^{13}$CO $J=$1--0.
However, {the} 115 $\>$km s$^{-1}$ cloud is not connected to other clouds. {More detailed analysis of these clouds, including the intermediate velocity components, are described in Section \ref{sec:bridge}.}

\subsubsection{$^{13}$CO $J=$3--2/1--0 intensity ratios and {a} comparison {to} the Spitzer 8 $\>\mu$m images}
{Figures \ref{W43Main_ratio}} (a), (b), (c), and (d) {show} the $^{13}$CO $J=$3--2/1--0 integrated intensity ratio ($R^{13}_{\rm 3-2/1-0}$) maps of the 82 $\>$km s$^{-1}$, 94 $\>$km s$^{-1}$, 103 $\>$km s$^{-1}$, and 115 $\>$km s$^{-1}$ cloud, respectively. {We mask the region with low $^{13}$CO $J=$3--2 intensity {($<3\sigma$).} 
The intensity ratio of different {rotational} transition levels is a useful tool to investigate the physical state (gas density {and} kinematic temperature) of the molecular clouds (e.g., the Large Velocity Gradient model: Goldreich \& Kwan 1974).
We also present a comparison of {$R^{13}_{\rm 3-2/1-0}$} for each cloud with the Spitzer 8 $\>\mu$m images to clarify the physical association with W43 Main. 
{Figures  \ref{W43Main_ratio}} (e), (f), (g), and (h) show the {$^{13}$CO $J=$3--2 intensity of the} 82 $\>$km s$^{-1}$, 94 $\>$km s$^{-1}$, 103 $\>$km s$^{-1}$, and 115 $\>$km s$^{-1}$ {clouds by} contours superposed on the Spitzer 8 $\>\mu$m image, respectively. 
The ring-like structure traced by the 8 $\>\mu$m emission corresponds to the infrared bubble N52 (see Figure 19 of Deharveng et al. 2010).

The 82 $\>$km s$^{-1}$ cloud ({Figures  \ref{W43Main_ratio}a and  \ref{W43Main_ratio}e}) has a high-intensity ratio ($R^{13}_{\rm 3-2/1-0}>0.5$) around the N52 bubble, and {corresponds} to the {brightest} part of the 8 $\>\mu$m image. 
The 94 $\>$km s$^{-1}$ cloud ({Figures  \ref{W43Main_ratio}b and  \ref{W43Main_ratio}f}) is distributed over 10 pc around W43 Main, and the intensity ratio ($R^{13}_{\rm 3-2/1-0}>0.5$) {is enhanced} at the protocluster candidates identified by Motte et al. (2003). 
In particular, the positions of high-intensity ratio correspond to the W43-MM1 and W43-MM2 ridge identified by Nguyen-Luong et al. (2013). 
The 103 $\>$km s$^{-1}$ cloud ({Figures  \ref{W43Main_ratio}c and  \ref{W43Main_ratio}g}) has a high-intensity ratio around the bubble, which is a similar {trend} to the 82 $\>$km s$^{-1}$ cloud. The {central} part of the bubble {has} a cavity-like structure. 
The intensity ratio {at the W43 MM1 ridge} is also enhanced, {where} the dark {cloud is recognized in} the 8 $\>\mu$m image. 
The 115 $\>$km s$^{-1}$ cloud ({Figures  \ref{W43Main_ratio}d and  \ref{W43Main_ratio}h}) shows a locally high-intensity ratio near the W43 Main cluster, which corresponds to the bright part of the 8 $\>\mu$m image.
In addition,  clouds {at} 10 pc away {from W43 Main to the north and south} also {have} {enhanced} intensity ratio ($R^{13}_{\rm 3-2/1-0}>0.8$).
{The average H$_2$ density is estimated to be {around} $6 \times 10^{22}$cm$^{-2}/20\ {\rm pc} \sim 10^3$ cm$^{-3}$ from the $^{13}$CO average column density at W43 Main, if we assume that the line of sight {extends to} 20 pc.  The kinetic temperature is {found to be} $\sim$50 K with $R^{13}_{\rm 3-2/1-0}\sim 0.4$ using the non-LTE code (LADEX\footnote{http://var.sron.nl/radex/radex.php}: Van der Tak et al. 2007) assuming the input parameters as the line width of 5 $\>$km s$^{-1}$, CO column density of $7.8 \times 10^{16}$ cm$^{-2}$ ($=6 \times 10^{22}/X[^{13}{\rm CO}]$), and H$_2$ density of $10^3$ cm$^{-3}$, respectively. This kinetic temperature is higher than {typical} for a molecular cloud without star formation.}
To summarize our results, these four clouds {show} the effect of heating by ultraviolet radiation from high-mass stars {in} W43 Main. 
Therefore, these clouds are likely to be {physically associated with} W43 Main {from} the velocity difference of $\sim 10$-20$\>$km s$^{-1}$.

\subsection{G30.5}
\label{sec:G30.5}

\subsubsection{$^{13}$CO and C$^{18}$O spatial distributions and velocity structures}
We find four different velocity clouds {at} 88 $\>$km s$^{-1}$, 93 $\>$km s$^{-1}$, 103 $\>$km s$^{-1}$, and 113 $\>$km s$^{-1}$ for G30.5. 
{Figures  \ref{G30.5_integ} (a)-(h)} show the integrated intensity maps of $^{13}$CO and C$^{18}$O $J=$1--0 emissions {of these clouds}.
The 88 $\>$km s$^{-1}$ cloud ({Figures  \ref{G30.5_integ}a and  \ref{G30.5_integ}e}) shows a peaked structure at $(l,b)\sim (\timeform{30.39D}, \timeform{-0.10D})$, which corresponds to the infrared source of IRAS 18445-0222. 
The 93 $\>$km s$^{-1}$ cloud ({Figures  \ref{G30.5_integ}b and  \ref{G30.5_integ}f}) has a peak of $(l,b)\sim (\timeform{30.45D}, \timeform{-0.14D})$, and is distributed over 15 pc of the {northern side}. 
The 103 $\>$km s$^{-1}$ cloud ({Figures  \ref{G30.5_integ}c and  \ref{G30.5_integ}g}) {exhibits} three CO peaks, which correspond to the infrared sources of G030.404-00.238, IRAS 18447-0229, and G030.213-00.156.
The 113 $\>$km s$^{-1}$ cloud ({Figures  \ref{G30.5_integ}d and  \ref{G30.5_integ}h}) extends over 20 pc at the direction of the galactic longitude.

We also made a position-velocity diagram to investigate the relationship {between} these four clouds. 
{Figure  \ref{G30.5_lv}} shows the Galactic latitude-velocity diagram of (a) $^{13}$CO and  (b) C$^{18}$O $J=$1--0. The dense gas traced by C$^{18}$O $J=$1--0 {demonstrates the four discrete}  velocity components {more clearly}.
The black boxes present the RRL velocity ($\sim 102.5$ $\>$km s$^{-1}$) at $(l,b)\sim (\timeform{30.404D}, \timeform{-0.238D})$ obtained by Lockman (1989). 
The RRL {{lies} close to the 103 $\>$km s$^{-1}$ clouds}. We also present the velocity channel maps in the Appendix {Figures \ref{G30.5_ch1}-\ref{G30.5_ch3}}.
{We observe that G30.5 contains a high amount of molecular gas, while the present star-formation activity {appears} lower than W43 Main.}

\subsubsection{$^{13}$CO $J=$3--2/1--0 intensity ratios and {a} comparison {to} the Spitzer 8 $\>\mu$m images}
{Figures \ref{G30.5_ratio} (a)-(d)} show $R^{13}_{\rm 3-2/1-0}$ maps of the 88 $\>$km s$^{-1}$, 93 $\>$km s$^{-1}$, 103 $\>$km s$^{-1}$, and 113 $\>$km s$^{-1}$ clouds, respectively.
The clipping levels are adopted {at} 3$\sigma$ {resulting in} 0.8 K $\>$km s$^{-1}$, 1.1 K $\>$km s$^{-1}$, 1.0 K $\>$km s$^{-1}$, and 1.1 K $\>$km s$^{-1}$ for the $^{13}$CO $J=$3--2 emissions, respectively.
{Figures \ref{G30.5_ratio}} (e)-(h) show the $^{13}$CO $J=$3--2 (contours) distribution superposed on the Spitzer 8 $\>\mu$m images. 
The 88 $\>$km s$^{-1}$ cloud ({Figures \ref{G30.5_ratio}a and \ref{G30.5_ratio}e}) indicates a {local} high-intensity ratio ($R^{13}_{\rm 3-2/1-0} > 0.6$) {that} corresponds to the infrared source of IRAS 18445-0222 ({Figures \ref{G30.5_ratio}a, \ref{G30.5_ratio}e}).
The 93 $\>$km s$^{-1}$ cloud ({Figures \ref{G30.5_ratio}b and \ref{G30.5_ratio}f}) has a low ratio ($R^{13}_{\rm 3-2/1-0} < 0.3$), but morphological correspondence of the infrared peak at $(l,b)\sim(\timeform{30.52D}, \timeform{-0.22D})$.
The 103 $\>$km s$^{-1}$ cloud ({Figures \ref{G30.5_ratio}c, \ref{G30.5_ratio}g}) indicates high $R^{13}_{\rm 3-2/1-0} > 0.6$ at G030.213-00.516, G030.404-00.238, IRAS 18447-0229, and G030.489-00.364, which {have counter parts in the 8 $\>\mu$m image}.
The 113 $\>$km s$^{-1}$ cloud ({Figures \ref{G30.5_ratio}d, \ref{G30.5_ratio}h}) has enhanced $R^{13}_{\rm 3-2/1-0} > 0.6$ at the western and eastern side.
The 8 $\>\mu$m image also corresponds to the high $R^{13}_{\rm 3-2/1-0}$ at the eastern side of G30.5.
To summarize {the results {of} G30.5}, these four clouds are also likely to be partially associated with G30.5 because of high-intensity ratios or morphological correspondence with the infrared images.

\subsection{W43 South (G29.96-0.02)}
\label{sec:W43South}

\subsubsection{$^{13}$CO and C$^{18}$O spatial distributions and velocity structures}
We find two different velocity clouds (93 $\>$km s$^{-1}$ and 102 $\>$km s$^{-1}$) for W43 South. 
{Figure \ref{W43South_integ}} shows the integrated intensity maps of $^{13}$CO and C$^{18}$O $J=$1--0 emissions, where crosses indicate the radio continuum sources obtained by NVSS (Condon et al. 1998). These sources correspond to the OB-type stars {identified} by radio continuum {observations}, where the numbering is the same as {in} Beltr\'an et al. (2013).
The 93 $\>$km s$^{-1}$ cloud ({Figures  \ref{W43South_integ}a and  \ref{W43South_integ}c}) has clumpy structures and a peak intensity at the source {number 2}. This {$^{13}$CO cloud} is distributed over 10 pc {in} the direction of the Galactic longitude. 
The 102 $\>$km s$^{-1}$ cloud ({Figures  \ref{W43South_integ}b and  \ref{W43South_integ}d}) has peak intensities at {sources nummber 4, 5, 8, and 9. These extend} from the northern to the southern side. 
 We also demonstrate the velocity channel maps of W43 South in the Appendix {Figures} \ref{W43South_ch1}}-\ref{W43South_ch3}.
We made the position-velocity diagram toward W43 South. {Figure  \ref{W43South_lv}} shows the Galactic longitude-velocity diagram of  (a) $^{13}$CO, and (b) C$^{18}$O $J=$1--0, respectively. 
{The two clouds can be identified on the position-velocity diagram {(Figure  \ref{W43South_lv})}.}
The black box shows the RRL velocity (96.7 $\>$km s$^{-1}$) at $(l,b)\sim(\timeform{29.944D}, \timeform{-0.042D})$ obtained by Lockman et al (1989), which coincides {with} the intermediate velocity of two clouds. The two clouds converge around $l\sim \timeform{29.9D}$. 
We also find a “V-shape"-like structure at the center of the RRL velocity (blue dotted lines). {From the C$^{18}$O emission in Figure  \ref{W43South_lv}b, the two clouds and “V-shape"-like structure can be seen more clearly.} {We point out that the V-shape exists slightly removed from the overlapping region of the two clouds. We argue that this velocity structure shows a signpost of cloud-cloud collision (e.g., Fukui et al. 2018b). More detailed discussion is described in Section \ref{sec:bridge}}

\subsubsection{$^{13}$CO $J=$3--2/1--0 intensity ratios and {a} comparison {to} the Spitzer 8 $\>\mu$m images}
{Figures  \ref{W43South_ratio}(a) and \ref{W43South_ratio}(b)} show the $R^{13}_{\rm 3-2/1-0}$ maps of the 93 $\>$km s$^{-1}$ and 102 $\>$km s$^{-1}$ clouds, respectively. The clipping levels are adopted {at} 3$\sigma$ (0.9 K $\>$km s$^{-1}$, and 1.1 K $\>$km s$^{-1}$) for the $^{13}$CO $J=$3--2 emissions. 
{Figures \ref{W43South_ratio} (c) and  \ref{W43South_ratio}(d)} indicate the $^{13}$CO $J=$3--2 distributions superposed on the Spitzer 8 $\>\mu$m images. 
The 93 $\>$km s$^{-1}$ cloud ({Figures \ref{W43South_ratio}a and \ref{W43South_ratio}c}) has high $R^{13}_{\rm 3-2/1-0}>0.6$ around {sources number 1, 2, 6, and 7}, which correspond to the 8 $\>\mu$m peaks.
The 102 $\>$km s$^{-1}$ cloud ({Figures \ref{W43South_ratio}b and \ref{W43South_ratio}d}) also shows high $R^{13}_{\rm 3-2/1-0}>0.6$ at {sources number 1, 5, 8, and 9}, which correspond to the  8 $\>\mu$m peaks.
These $R^{13}_{\rm 3-2/1-0}$ results coincide with the positions of the {high dust temperature ($ > 70 $K)} obtained from the results of {Herschel} 70-350 $\>\mu$m (see Figure 7 of Beltr\'an et al. 2013). 
To summarize {the} results of W43 South, these two clouds are also likely to be physically associated with W43 South {because of their} high-intensity ratios and morphological correspondence with the radio continuum sources.

\section{Discussion}
\label{sec:discussion}

{{This section is structured} as follows: {subsection} \ref{sec:property} gives a comparison {of} {the brightness temperature} {with those in} other GMCs (W51 and M17); {subsection} \ref{sec:SFE} shows the star formation efficiency (SFE) of GMCs; {subsection \ref{sec:multiple_clouds} gives interpretation of the multiple clouds and velocity structures in W43 Main, G30.5, and W43 South}; {in} {subsection} {\ref{sec:bridge}} -- \ref{sec:scenario} {we propose} the dense gas and high-mass star formation scenario produced by cloud-cloud collisions in the GMC complex in turbulent conditions. {{Finally}, subsection \ref{sec:collision_freq} demonstrates the frequency of cloud-cloud collisions in the W43 GMC complex.}} 

\subsection{{Comparison {of} the brightness temperature {with} the W51 and M17 GMCs}}
\label{sec:property}

{We carried out the analysis {using} histograms of the molecular gas and of the W43, W51, and M17 GMC, aiming to reveal the molecular gas properties of the W43 GMC complex.}
{Figure \ref{integ_all}} shows the integrated intensity maps of $^{12}$CO,  $^{13}$CO, and C$^{18}$O $J=$1--0 for the W51 and M17 GMCs obtained by the FUGIN project. 

\begin{itemize}
\item The W51 GMC \\
W51 is a GMC including the high-mass star-forming regions of W51A and W51B (e.g., Carpenter \& Sanders 1998; Parsons et al. 2012; Ginsburg et al. 2015, and reference therein). The parallactic distance is estimated to be 5.4 kpc near the tangential direction of the Sagittarius Arm (Sato et al. 2010), which is {about the} same distance as the W43 GMC complex. 
The {CO} gas of {three isotopes} has a peak intensity at W51A, and the low-density gas traced by $^{12}$CO is distributed {over} 100 pc  ({Figures \ref{integ_all} a-c}). 
{A} more detailed analysis of the W51 GMC using the FUGIN data is presented by Fujita et al. (2019b).
\item The M17 GMC\\
M17 is also {a GMC, composed} of M17 SW and M17 SWex. M17 SW includes the massive cluster NGC 6618 (e.g., Povich et al. 2007; Hoffmeister et al. 2008; Chini \& Hoffmeister 2008 and {references} therein), and M17 SWex has {an} infrared dark cloud, which is {the location} of onset of massive star formation (Povich \& Whitney 2010). 
The parallactic distance is derived to be 2.0 kpc, which corresponds to the {distance to the} Sagittarius Arm (Chibueze et al. 2016). 
All isotopes of CO have a strong peak at M17 SW, while the diffuse gas is distributed {over} 30 pc near M17 SWex ({Figures \ref{integ_all} d-f}).
Detailed {properties of the M17 cloud} using the FUGIN data are described by Yamagishi et al. (2016) and Nishimura et al. (2018).
\end{itemize}



{Figures \ref{BDF_hist}(a), (b), and (c) show the histograms of the brightness temperatures of $^{12}$CO, $^{13}$CO, and C$^{18}$O, which correspond to the Brightness Distribution Function (BDF) 
introduced by Sawada et al. (2012a, 2012b, 2018). We smoothed the data of M17 to $\timeform{110"}$ in order to achieve the same spatial resolution ($\sim$ 1 pc) {as} W43 and W51.}
{We normalized the histograms at  (a) 7.3-9.2 K,  (b) 5.0-5.8 K, and  (c) 1.9-2.1 K.} 
{The slope for M17 in $^{12}$CO and $^{13}$CO ({Figures  \ref{BDF_hist}a and  \ref{BDF_hist}b}) is shallower than {those} for W43 and W51. M17 has excess voxels with high $T_{\rm mb}$ in the BDF. This is consistent with the trend reported by Sawada et al. (2012a, 2012b) that regions containing active star-forming regions show shallower distributions. If the detected emission within the beam is optically thick, the line becomes saturated with $T_{\rm mb}${, reflecting} the gas excitation temperature. Hence the excess with higher $T_{\rm mb}$ than $\sim$ 30 K {indicate} the extended warm gas component due to H\,\emissiontype{II} regions with respect to the cloud extent, as is prominent for M17.}

{However, the histogram of C$^{18}$O, which is optically thinner than other isotopes, has a different shape to others, and W43 has a high proportion of dense gas with high brightness temperature ({see a black arrow} in {Figure \ref{BDF_hist}c}). This can be interpreted that the dense gas in the W43 GMC has locally a higher column density than the W51 and M17 GMCs.}


\subsection{{Comparison of the star formation efficiency}}
\label{sec:SFE}

{We estimated the star formation efficiency (SFE) of these GMCs with  the aim of clarifying the present star formation activity of the W43 GMC complex. The SFE
is given by
\begin{eqnarray}
{\rm SFE} &= {M_{*} \over {M_{*}+M_{\rm cloud}}},
\label{eq:75}
\end{eqnarray}
where $M_{\rm cloud}$ and $M_{\rm *}$ {are} the cloud mass and stellar mass, respectively. We {derive} the molecular mass from the FUGIN data. A detailed explanation of our methods is given in Appendix 1.
 Table 6 shows our results for {the} SFEs of each GMC complex. The total stellar masses were estimated from the number of earliest O-type star(s) by assuming the number distribution of the Salpeter IMF (${dN \over dM} \propto M^{-2.35}$: Salpeter 1955). 
 The summation range is adopted from {$0.08 M_{\odot}$ stars}. The earliest spectral types are O3.5III in W43 Main, O5III in W43 South (G29.96-0.02), O4V in W51, and O4V in M17, respectively (see Table 1 in Binder \& Povich 2018). The masses of these earliest-type stars are adopted by referring to observational O-star parameters (see Tables 4 and 5 in Martins et al. 2005). {The total stellar mass of} W43 is derived from summing W43 Main and W43 South, and {that of} W51 is consistent with the results of infrared spectroscopic observations by Bik et al. (2019).}

{In Table 6, we see that W43 has a low SFE compared to} W51 and M17 {calculated using} all isotopes. This indicates that W43 has a large proportion of molecular to stellar mass. When we estimate the SFE from the dense gas (C$^{18}$O), W43 shows $\sim 4 \%$, which is consistent with other nearby star-forming regions ($3\%$ - $6\%$: Evans et al. 2009), and Orion GMC ($\sim 4\%$: Nishimura et al. 2015). 
{Figure \ref{W43_C18Oratio} presents the C$^{18}$O/$^{12}$CO $J=$1--0 intensity ratio map, which shows the relatively high-density regions in the W43 GMC complex. }
This map shows high ratios ($\sim$0.07-0.08) around W43 Main and W43 South, which correspond to the present active star-forming region. 
Moreover, the W43 GMC complex has locally-high density regions including G30.5.
These results indicate that future star formation in W43 is likely to be more active.  
{Therefore, we suggest that while {“the current} SFE" of the whole W43 GMC is $\sim 4 \%$, the star formation activity in W43 will possibly increase in the future.} 
This possibility is also proposed by Nguyen-Luong et al. (2011) and Motte et al. (2017). }
We discuss the origin of the dense gas with multiple O-type stars in subsection {\ref{sec:multiple_clouds} - \ref{sec:scenario},} based on {the} properties of the W43 GMC complex.

\subsection{{Interpretation of the multiple clouds in W43 Main, G30.5, and W43 South}}
\label{sec:multiple_clouds}

{We find multiple $^{13}$CO clouds with a velocity difference of $\sim$ 10-20 $\>$km s$^{-1}$ for W43 Main, G30.5, and W43 South, which have {locally-high} column density molecular gas.}
If we assume a cloud separation of $r= 20$-30 pc and velocity difference of $v=$15 km s$^{-1} \times \sqrt{2} \sim 20$ $\>$km s$^{-1}$, adopting a projection angle of \timeform{45D},
the total mass required to gravitationally bind can be estimated as
\begin{eqnarray}
M &= {r v^2 \over 2G} \sim 10^6\ M_{\odot}.
\label{eq:75}
\end{eqnarray}
This {mass} is a factor of two larger than the averaged mass of each $^{13}$CO cloud as $5 \times 10^5\ M_{\odot}$.
We suggest that the it is difficult to interpret the $^{13}$CO clouds as a gravitationally bound system like the two-body system. 
{Therefore, we hypothesize that the clouds in W43 Main, G30.5, and W43 South collided with each other by chance, and the collisions have produced the dense gas and local mini-starburst in the W43 GMC complex.}

\subsection{{Spectral features indicative of cloud-cloud collisions}}
\label{sec:bridge}

{When two clouds collide, as a natural consequence of momentum exchange, an intermediate velocity component, so-called “the bridge feature" is expected to appear between the two clouds (e.g., Haworth et al. 2015a, 2015b). 
We thus carried out a detailed spectral analysis to investigate the bridge features in W43 Main. 
{Figure \ref{bridge_all}(a) and (b) illustrate an example $^{13}$CO spectrum and an excess emission as a residual of two Gaussian subtraction from the $^{13}$CO $J =$3-2 spectrum, respectively. Figure \ref{bridge_all}(c) shows their spatial distributions of $^{13}$CO $J =$3-2, and (d) shows the distribution of the $J =$ 3-2/1-0 intensity ratio for the bridge component defined as the excess emission. We defined the bridge features by the following procedures using the $^{13}$CO data sets of $J=$3-2 and 1-0 lines.
\begin{enumerate}
\item We applied the least-square fittings of the red and blue-shifted components to the Gaussian functions (Figure \ref{bridge_all} (a)).
\item We limit the center velocity ($V_{\rm cent}$) and tried the fittings changing the range of $V_{\rm cent}$ to see the most successful condition. As a result, we set the criteria of subtracting the Gaussian functions if $V_{\rm cent}$ falls in the range of $77 \leq V_{\rm cent} < 83$ km s$^{-1}$ and $92 \leq V_{\rm cent} < 100$ km s$^{-1}$ for each component from the original data.
\item The bridge features are thus defined as the residual excess emission with the integrated velocity range from 82.3 to 88.2 km s$^{-1}$ as shown in the green area of figure  \ref{bridge_all}(b).
\end{enumerate}
The bridge features (green contours) are almost exclusively distributed in the region where the two clouds (blue and red contours) overlap in W43 Main. }
Furthermore, {the intensity ratio of the bridge features ($R^{13}_{\rm 3-2/1-0} \gtrsim 0.7$) is enhanced from the typical value ($R^{13}_{\rm 3-2/1-0} \sim 0.3$) without high-mass star formation in W43. 
This result shows that the bridge features have high density and/or high temperature around W43 Main.} 

{We also tried to apply the same procedures to W43 South. The spectra are, however, more blended and technically difficult to be decomposed due to relatively large velocity dispersions of the two components, and also to small velocity separation possibly because of the viewing angle with respect to the relative motion. Hence the bridge features are not identified clearly. The position-velocity diagram better indicate the two velocity components and the bridge feature connecting them (see Figure \ref{W43South_lv}). Higher resolution, higher sensitivity, and more optically thin data may help the decomposition to reveal more obvious evidence of the collisional interaction in W43 South. }

{According to the cloud-cloud collision model, the interface layer of two clouds is decelerated and have intermediate velocity, and hence corresponds to the bridge feature that is localized in the overlapping region of two cloud{s} on the plane of the sky. The bridge features are also expected to have high density and/or temperature by the collisional shock compression. Therefore, {we suggest that {the bridge feature in W43 Main} almost originate from cloud-cloud collisions, and support our hypothesis.}


\subsection{{Interpretation of velocity distributions}}
\label{sec:interpretation_velocity}

{We also propose  that a cloud-cloud collision hypothesis can explain the origin of the V-shape structure on the position velocity diagram (Figure \ref{W43South_lv}). When two clouds having different sizes collide and merge, the cavity is created in the large cloud.  
If we make the position-velocity diagram toward colliding spots in this stage, we can expect to observe a V-shaped velocity distribution (see Figure 5 middle panel of Haworth et al. 2015a). }
Indeed, the observations of the isolated compact cluster GM 24 also demonstrated that a cloud-cloud collision can produce 
the V-shape velocity structure from a comparison to the numerical simulation (see Figure 7 of Fukui et al. 2018b).
We note that the V-shape in W43 South exists away from the overlapping region of the 93 $\>$km s$^{-1}$ and 102 $\>$km s$^{-1}$ clouds. 
We suggest that this shows another collision {occurring} to the 102 $\>$km s$^{-1}$ cloud. 
Fujita et al. (2019b) indeed argue that the multiple cloud collisions trigger massive star formation in the W51 GMC.}


The stellar feedback might be an alternative interpretation of the complex velocity structures, instead of a cloud-cloud collision. However, we cannot find elliptical velocity distributions on the position-velocity diagram expected for an expanding-shell (e.g., Figure 12 of Fukui et al. 2012). Moreover, the ionization effects in W43 South are likely to be small because of the young high-mass star-forming region ($\sim$ 0.1 Myr: Watson \& Hanson 1997). 
{Indeed, most  H\emissiontype{II} regions in W43 South have a compact size ($< $1 pc: Table 6 of Beltran et al. 2013), and are embedded in the molecular clouds.} Dale et al. (2013) reported that the influence of the stellar winds from OB-type stars on molecular clouds is less effective compared to expanding H\,\emissiontype{II} regions. {Therefore, we conclude that the velocity structures of the clouds in the W43 GMC complex is explained better by the cloud-cloud collision rather than the feedback effect.}


\subsection{{Dense gas and multiple O-type star formation scenario in the W43 GMC complex}}
\label{sec:scenario}

{Figure \ref{W43_scenario} (a)} shows our proposed model of the top-view schematic images of the W43 GMC complex. 
The light blue represents the low-density gas traced by $^{12}$CO of the Scutum Arm. 
Our proposed model shows that more small-scale clouds ($\sim$ 10-20 pc) collide with each other, and produce the local dense gas. 
{If we roughly {assume} the colliding cloud size {as 20-30 pc and the velocity difference of the cloud as 10-20 km s$^{-1}$ , the collisional time scale is derived as 20-30 pc/(10-20) $\>$km s$^{-1}$ $\sim\ 1-3$ Myr.} This value is consistent with the age of O-type stars in W43 Main (1-6 Myr: Motte et al. 2003; Bally et al . 2010). The stellar age of W43 South is $\sim$ 0.1 Myr (Watson \& Hanson 1997), which is shorter than the collisional timescale. This might imply that the star formation in the initial phase is triggered {by} a cloud-cloud collision.

{{Figure \ref{W43_scenario}(b)} presents the Galactic-scale top view of the W43 GMC complex from the North Galactic Pole based on Figure 11 of Sofue et al ({2019}). 
The red arrows of {Figure \ref{W43_scenario}} show the continuous {converging} flow along the spiral arm and the long bar in the Milky Way.  
This mechanism, proposed by previous studies, is a possible theory to explain the origin of cloud-cloud collisions in W43 GMC complex at the bar-end region (e.g., Nguyen-Luong et al. 2011, Motte et al. 2014, Renaud et al. 2015), and common with external barred spiral galaxies (e.g., M 83: Kenney \& Lord 1991; NGC 3627: Beuther et al. 2017).
 We will describe them in the separate paper since it extends the scope of this paper.}


{We propose that the cloud-cloud collision scenario in the GMC complex might be able to explain the origin of the CMF {being} shallower than IMF in W43-MM1 (Motte et al. 2018b).
Another super star cluster, NGC 3603, triggered by a cloud-cloud collision (Fukui et al. 2014), was reported to have {a} top-heavy stellar mass function (Harayama et al. 2008).
The Galactic center 50 $\>$km s$^{-1}$ cloud is also suggested to have a similar tendency  (e.g., Tsuboi et al. 2015; Uehara et al. 2019).
{This suggestion is also supported by the MHD simulation (Fukui et al. 2019)}}

\subsection{{The frequency of cloud-cloud collisions in the W43 GMC complex}}
\label{sec:collision_freq}

{We estimated the frequency of cloud-cloud collisions in the W43 GMC complex from the mean free paths, assuming random cloud motion. The collision frequency ($\tau_{\rm col}$) is given by 
\begin{eqnarray}
\tau_{\rm col} = {1 \over n_{\rm cl} \sigma_{\rm cl} v_{\rm rel}} \sim 7\ {\rm Myr},
\label{eq:75}
\end{eqnarray}
where $n_{\rm cl}$, $\sigma_{\rm cl}$, and $v_{\rm rel}$ are the number density, collisional cross-section, and relative velocity. 
The number of clouds is calculated as the ratio of the $^{13}$CO total and average mass as $1.1\times 10^7\ M_{\odot}/4.6 \times 10^5\ M_{\odot} \sim 24$.
The number density is derived as $n_{\rm cl} \sim 2 \times 10^{-5}$pc$^{-3}$, assuming a GMC volume of $4\pi/3 \times$
75 pc $\times$ 50 pc $\times$ 75 pc. 
The collisional cross-section is estimated as $\sigma_{\rm cl} = \pi (10\ {\rm pc})^2$ assuming a $^{13}$CO cloud {radius} of 10 pc. 
The relative velocity is adopted as $v_{\rm rel} = 15 \times \sqrt{2} \sim 20\ {\rm km s^{-1}}$ assuming a projection angle of  \timeform{45D}.
The results for 7 Myr is factor of 3-4 shorter than the typical age of GMCs, which is 20-30 Myr (e.g., Kawamura et al. 2009). 
This indicates that $^{13}$CO clouds are likely to experience 3-4 collisional events in their {life-time,} if we assume the random motion of clouds.
The Galactic-scale numerical simulation also shows the high merger rate of $\sim$ 2-3 Myr in the massive GMC like W43 (Fujimoto et al. 2014a).}

\section{Conclusions}
\label{sec:conclusion}

The conclusions of this paper are summarized as follows.
{
\begin{enumerate}
\item We carried out new large-scale CO $J=$ 1--0 observations of the W43 GMC complex {in} the tangential direction of the Scutum Arm as a part of the FUGIN legacy survey.
We revealed the spatial distribution of molecular gas and velocity structures. 
\item The low-density gas traced by $^{12}$CO is distributed {in a region of} 150 pc $\times$ 100 pc ($l \times b$), and has {a} large velocity dispersion ({20-30 $\>$km s$^{-1}$}).
The dense gas traced by C$^{18}$O {reveals} three high-mass star-forming regions (W43 Main, G30.5, and W43 South), with clumpy structures.
\item We found 2-4 velocity components with velocity difference of $\sim$ 10-20 $\>$km s$^{-1}$ in each region.
At least two clouds are likely to be physically associated with high-mass star-forming regions from their common high-intensity ratios {of $R^{13}_{\rm 3-2/1-0}$} {and their similar distribution to} infrared {nebulae}.
\item {We made brightness temperature histograms of W43, W51, and M17 GMC, revealing the molecular gas properties of the W43 GMC complex. 
We showed that localized dense gas has a high brightness temperature compared to the W51 and M17 GMC.}
\item “The current SFE" of the entire W43 GMC has low value ($\sim 4\%$) compared to the W51 and M17 GMCs. {On the other hand, 
the W43 GMC complex has locally-high density C$^{18}$O gas from the brightness distribution function. Therefore, we suggest that the star formation activity in W43 will possibly increase in the future.}
\item The velocity separation of these clouds in W43 Main, G30.5, and W43 South is too large {for each cloud} to be gravitationally bound. 
{We also find bridging features and a V-shape structure on the position-velocity diagram. These features are a signpost of cloud-cloud collision, suggested by previous studies.}
Therefore, we {propose} that {the} supersonic cloud-cloud collision {hypothesis} can explain the origin of dense gas and the local mini-starburst in the W43 GMC complex.
\item  We argued {that} the {{converging} gas flow from} the Scutum Arm and the long bar {causes} the highly turbulent condition, which makes frequent cloud-cloud collisions in the W43 GMC complex. 
\end{enumerate}
}

\section*{Acknowledgements}
We are grateful to Professor Shu-ichiro Inutsuka of Nagoya University for a useful discussion. 
{The authors are grateful to the referee for thoughtful comments on the paper.
We would like to thank Dr. Tom J. L. C. Bakx of Nagoya University for English language editing.}
{We also grateful to Ms. Kisetsu Tsuge and Mr. Rin Yamada of Nagoya University for useful comments about Figures and texts.}
The Nobeyama 45-m radio telescope is operated by Nobeyama Radio Observatory, a branch of the National Astronomical Observatory of Japan. 
Data analysis was carried out on the Multi-wavelength Data Analysis System operated by the Astronomy Data Center (ADC), National Astronomical Observatory of Japan.
The work is financially supported by a Grant-in-Aid for Scientific Research (KAKENHI, No. 15K17607, 15H05694, 17H06740, 18K13580) from MEXT (the Ministry of Education, Culture, Sports, Science and Technology of Japan) and JSPS (Japan Society for the Promotion of Science).

 This work is based on observations made with the Spitzer Space Telescope, which is operated by the Jet Propulsion Laboratory, California Institute of Technology under a contract with NASA. Support for this work was provided by NASA.

The James Clerk Maxwell Telescope is operated by the East Asian Observatory on behalf of The National Astronomical Observatory of Japan; Academia Sinica Institute of Astronomy and Astrophysics; the Korea Astronomy and Space Science Institute; Center for Astronomical Mega-Science (as well as the National Key R\&D Program of China with No. 2017YFA0402700). Additional funding support is provided by the Science and Technology Facilities Council of the United Kingdom and participating universities in the United Kingdom and Canada.

We would like to thank Editage (www.editage.com) for English language editing.

Software: We utilized Astropy, a community-developed core Python package for astronomy (Astropy Collaboration et al. 2013, 2018), NumPy  (Van Der Walt et al. 2011), Matplotlib (Hunter 2007), IPython (Pe\'rez et al. 2007), and Montage \footnote{http://montage.ipac.caltech.edu} software.

\appendix
\section{Procedures of the physical parameters of the molecular clouds derived from the FUGIN data}
In this section, we introduce the calculation method of the physical parameters of the molecular gas, assuming local thermal equilibrium (LTE) (e.g., Wilson et al. 2009; Mangum \& Shirley 2015). 

\begin{itemize}

\item The radiative transfer equation{: [$T_{\rm mb}$]}\\
The observed brightness temperature [$T_{\rm mb} (v)$] at radial velocity [$v$] is given by
\begin{eqnarray}
T_{\rm mb} (v) = (J(T_{\rm ex})- J(T_{\rm bg}))(1- \exp(-\tau (v))),
\label{eq:75}
\end{eqnarray}
where {$J(T)$,} $T_{\rm ex} $, $T_{\rm bg}$, and $\tau (v)$ are {the Rayleigh-Jeans equivalent temperature}, the excitation temperature, the background continuum emissions, and the optical depth, respectively.
We adopted $T_{\rm bg}=2.73$ K, which corresponds to the cosmic microwave background emission.

\item The excitation temperature: [$T_{\rm ex}$]\\
If we assume that the peak temperature of $^{12}$CO $J=$1--0 is {optically} thick ($\tau \to \infty$), the excitation temperature is given by
\begin{eqnarray}
T_{\rm ex} &=& 5.5 \bigg/ \ln \left(1+ {5.5  \over T_{\rm mb}(\rm ^{12}CO peak)/{\rm K} + 0.82 }\right) {\    [\rm K]}.
\label{eq:75}
\end{eqnarray}
\item The optical depth: $[\tau_{13} (v)]$ and $[\tau_{18} (v)]$\\
If we assume the same excitation temperatures in $^{12}$CO, $^{13}$CO, and C$^{18}$O, the optical depths are derived at each voxel from the $^{13}$CO and C$^{18}$O brightness temperatures, and the excitation temperature by using the following equations:
\begin{eqnarray}
\tau_{13} (v) &=& -\ln \left[1-{T_{\rm mb}({\rm ^{13}CO}) \over 5.3\ {\rm K}} \left\{ {1 \over \exp({5.3\ {\rm K} \over T_{\rm ex}})-1}-0.16 \right\}^{-1} \right],\\
\tau_{18} (v) &=& -\ln \left[1-{T_{\rm mb}({\rm C^{18}O}) \over 5.3\ {\rm K}} \left\{ {1 \over \exp({5.3\ {\rm K} \over T_{\rm ex}})-1}-0.17 \right\}^{-1} \right].
\label{eq:75}
\end{eqnarray}

\item The CO column density: $N ({\rm ^{13}CO})$ and $N ({\rm C^{18}O})$\\
The $^{13}$CO and C$^{18}$O column densities are calculated by summing all velocity channels from the following equations:
\begin{eqnarray}
N ({\rm ^{13}CO}) &= 2.4 \times 10^{14}  \sum_v {T_{{\rm ex}}\ {/\rm K}\  \tau_{13} (v)\ \Delta v ({\rm ^{13}CO}) {/\rm km\ s^{-1}} \over 1-\exp \left(-{5.3\ {\rm K} \over T_{\rm ex} } \right) } {\   [\rm cm^{-2}]}{,}\\
N ({\rm C^{18}O}) &= 2.5 \times 10^{14}  \sum_v {T_{{\rm ex}}\ {/\rm K}\  \tau_{18} (v)\ \Delta v ({\rm C^{18}O}) {/\rm km\ s^{-1}} \over 1-\exp \left(-{5.3\ {\rm K} \over T_{\rm ex} } \right) } {\   [\rm cm^{-2}]}{,}
\label{eq:75}
\end{eqnarray}
where $\Delta v$ ($^{13}$CO) and $\Delta v$ (C$^{18}$O) are adopted as velocity channel resolution of 0.65 $\>$km s$^{-1}$.

\item The H$_2$ column density\\
The H$_2$ column density is calculated using the following methods, using {an} X-factor ($X({\rm ^{12}CO})$) and abundance ratios of the isotopes. 
\begin{eqnarray}
N^{12}_{\rm X} ({\rm H_2}) &= X({\rm ^{12}CO}) \times W({\rm ^{12}CO})\\
N^{13}_{\rm LTE} ({\rm H_2}) &= X[{\rm ^{13}CO}] \times N ({\rm ^{13}CO}) \\
N^{18}_{\rm LTE} ({\rm H_2}) &= X[{\rm C^{18}O}] \times N ({\rm C^{18}O}) 
\label{eq:75}
\end{eqnarray}
We adopted {an} X-factor of $2.0\times 10^{20}$ [(K kms$^{-1}$)$^{-1}$ cm$^{-2}$] with having a $\pm 30\%$ uncertainty (Bolatto et al. 2012).
When we converted to the H$_2$ column densities from the $^{13}$CO and C$^{18}$O emission, we utilized the conversion factor of $X[{\rm ^{13}CO}]= 7.7 \times 10^5$ and $X[{\rm C^{18}O}]=5.6 \times 10^6$. These factors are derived from the isotope abundance ratio of [$^{12}$C]/[$^{13}$C] $= 77$, [$^{16}$O]/[$^{18}$O] $= 560$ (Wilson \& Rood 1994), and the CO and H$_2$ abundance ratio of $[^{12}$CO]/[H$_2$] $= 10^{-4}$ (e.g., Frerking et al. 1982; Pineda et al. 2010).

\item Total molecular mass\\
The total molecular mass is given by
\begin{eqnarray}
M =  \mu_{\rm H_2} m_{\rm H} D^2 \sum_i {\Omega} N_i(\rm H_2),
\label{eq:75}
\end{eqnarray}
\end{itemize}
where $\mu_{\rm H_2} \sim 2.8$ is mean molecular weight {per hydrogen molecule including the contribution of helium (e.g., Appendix A.1. of Kauffmann et al. 2008)}, $m_{\rm H}=1.67 \times 10^{-24}$ g is the proton mass, $D$ is the distance to each GMC,
 and {$\Omega$ is the solid angle of pixel $i$.}


\begin{figure*}[h]
\begin{center} 
 \includegraphics[width=18cm]{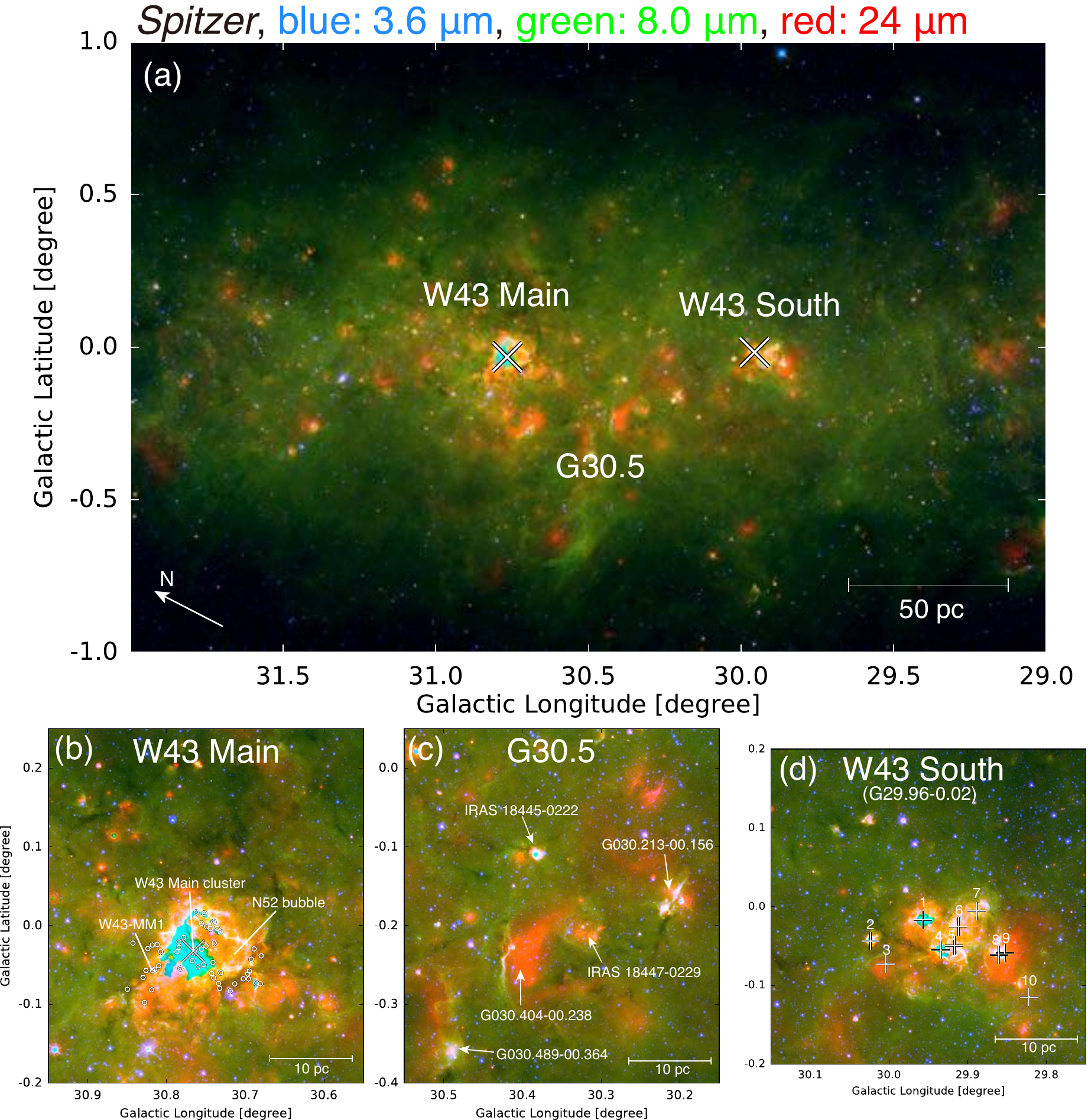}
\end{center}
\caption{(a) Spitzer three-color composite image of W43. Blue, green, and red {represent} the Spitzer/IRAC 3.6 $\>\mu$m, Spitzer/IRAC 8 $\>\mu$m (Benjamin et al. 2003), and Spitzer/MIPS 24 $\>\mu$m (Carey et al. 2009) {distributions}. The X marks indicate W43 Main (Blum et al. 1999) and W43 South (Wood \& Churchwell 1989). (b) Close-up image of W43 Main. The white circles indicate the 51 protocluster candidates (W43 MM1-MM51) cataloged by Motte et al. (2003).  (c) Close-up image of G30.5. (d) Close-up image of W43 South. The white crosses indicate the radio continuum sources identified by Condon et al. (1998).}
\label{W43_spitzer}
\end{figure*}

\clearpage
\begin{table*}
\tbl{Properties of W43 Main and W43 South.}{
\begin{tabular}{cccccccccc}
\hline
\multicolumn{1}{c}{Name} & Total luminosity   &  \# of O-type stars & Age & References \\
\hline
W43 Main & 7--10 $\times 10^6 L_{\odot}$    & $\sim$ 14 -- 50  & 1--6 Myr & [1,2,3,4,5]\\
W43 South & 2--6 $\times 10^6 L_{\odot}$   & 6 & $\sim$0.1 Myr & [2,6,7]\\
\hline
\end{tabular}}\label{pro_W43Main_W43South}
\begin{tabnote}
References [1] Hattori et al. (2016), [2] Lin et al. (2016), [3] Deharveng et al. (2010), [4] Motte et al. (2003), [5] Bally et al. (2010), [6] Beltr\'an et al. (2013), [7] Watson \& Hanson (1997)\\
\end{tabnote}
\end{table*}

\begin{table*}
\tbl{Spectral types of O-type stars in the W43 Main cluster.}{
\begin{tabular}{cccccccccc}
\hline
\multicolumn{1}{c}{Name} & Galactic Longitude  & Galactic Latitude & Spectral Type  \\
\hline
W43 \#1a & \timeform{30.766995D} & \timeform{-0.034752D} &WN7\\
W43 \#1b  & \timeform{30.766884D} & \timeform{-0.034631D} &O4 III\\
W43 \#2 & ---  & ---  &O3.5 III\\
W43 \#3a &  \timeform{30.766215D} &  \timeform{-0.035086D} & O3.5 V/O4 V \\
W43 \#3b &\timeform{30.766167D}  &  \timeform{-0.034975D}& O3.5 V/O4 V\\
\hline
\end{tabular}}\label{Ostar_W43Main}
\begin{tabnote}
References: Blum et al. (1999); Luque-Escamilla et al. (2011); Binder \& Povich (2018)\\
\end{tabnote}
\end{table*}

\begin{table*}
\tbl{Spectral types of OB-type stars in W43 South.}{
\begin{tabular}{cccccccccc}
\hline
\multicolumn{1}{c}{Number} & Galactic Longitude  & Galactic Latitude & Spectral Type  \\
\hline
\#1 & \timeform{29.957D}  & \timeform{-0.0170D} & O6\\
\#2 & \timeform{30.023D}  & \timeform{-0.0438D}  & B0\\
\#3 &  \timeform{30.004D}  & \timeform{-0.0730D}   & B0\\
\#4 &  \timeform{29.934D}  & \timeform{-0.0555D}  & O5\\
\#5 &  \timeform{29.917D}  & \timeform{-0.0497D}   & O6.5\\
\#6 &  \timeform{29.912D}  & \timeform{-0.0252D}   & O9.5\\
\#7 &  \timeform{29.889D}  & \timeform{-0.0056D}   & O8.5\\
\#8 &  \timeform{29.860D}  & \timeform{-0.0614D}   & O9.5\\
\#9 &  \timeform{29.853D}  & \timeform{-0.0591D}  & B0 \\
\#10 &  \timeform{29.822D}  & \timeform{-0.1150D}   & B0.5 \\
\hline
\end{tabular}}\label{Ostar_W43South}
\begin{tabnote}
References: Beltr\'an et al. (2013)\\
\end{tabnote}
\end{table*}

\clearpage

\begin{table*}
\tbl{Observational properties of data sets.}{
\begin{tabular}{cccccccccc}
\hline
\multicolumn{1}{c}{Telescope/Survey} & Line & Receiver &Effective   &  Velocity & RMS noise$^{\dag}$ & References \\
& && Resolution & Resolution & level &\\
\hline
Nobeyama 45-m/FUGIN &$^{12}$CO $J=$ 1--0 \footnotemark[]  & FOREST &\timeform{20"}  & 1.3 $\>$km s$^{-1}$ & $\sim 1.0$ K  & [1,2,3]\\
 &$^{13}$CO $J=$ 1--0\footnotemark[] & FOREST & \timeform{21"}  &  1.3 $\>$km s$^{-1}$& $\sim 0.35$ K  &  [1,2,3]\\
&C$^{18}$O $J=$ 1--0\footnotemark[] &FOREST &\timeform{21"}  &  1.3 $\>$km s$^{-1}$& $\sim 0.35$ K &  [1,2,3]\\
JCMT/CHIMPS &$^{13}$CO $J=$ 3--2\footnotemark[] &HARP  & \timeform{15"}  &  0.5 $\>$km s$^{-1}$& $\sim 0.14$ K & [4,5]\\
\hline
\hline
Telescope/Survey & Band  & Detector &  Resolution & References & & \\
\hline
{\it Spitzer}/GLIMPSE   & 3.6 $\>\mu$m & IRAC & $\sim$\timeform{2"} & [6,7,8] & \\
{\it Spitzer}/GLIMPSE   &  8.0 $\>\mu$m & IRAC &$\sim$\timeform{2"} & [6,7,8]  &\\
{\it Spitzer}/MIPSGAL   & 24 $\>\mu$m  & MIPS & \timeform{6"} &  [9,10]  &\\
\hline
\end{tabular}}\label{obs_param}
\begin{tabnote}
\footnotemark[$\dag$] The value of rms noise levels are after smoothing (space and/or velocity) data sets. \\
References [1] Minamidani et al. (2015), [2] Minadamidani et al. (2016), [3] Umemoto et al. (2017), [4] Rigby et al. (2017), [5] Buckle et al. (2009), [6] Benjamin et al. (2003), [7] Fazio et al. 2004, [8] Churchwell et al. (2009), [9] Carey et al. (2009), [10] Rieke et al. (2004) \\
\end{tabnote}
\end{table*}

\begin{figure*}[h]
\begin{center} 
 \includegraphics[width=10cm]{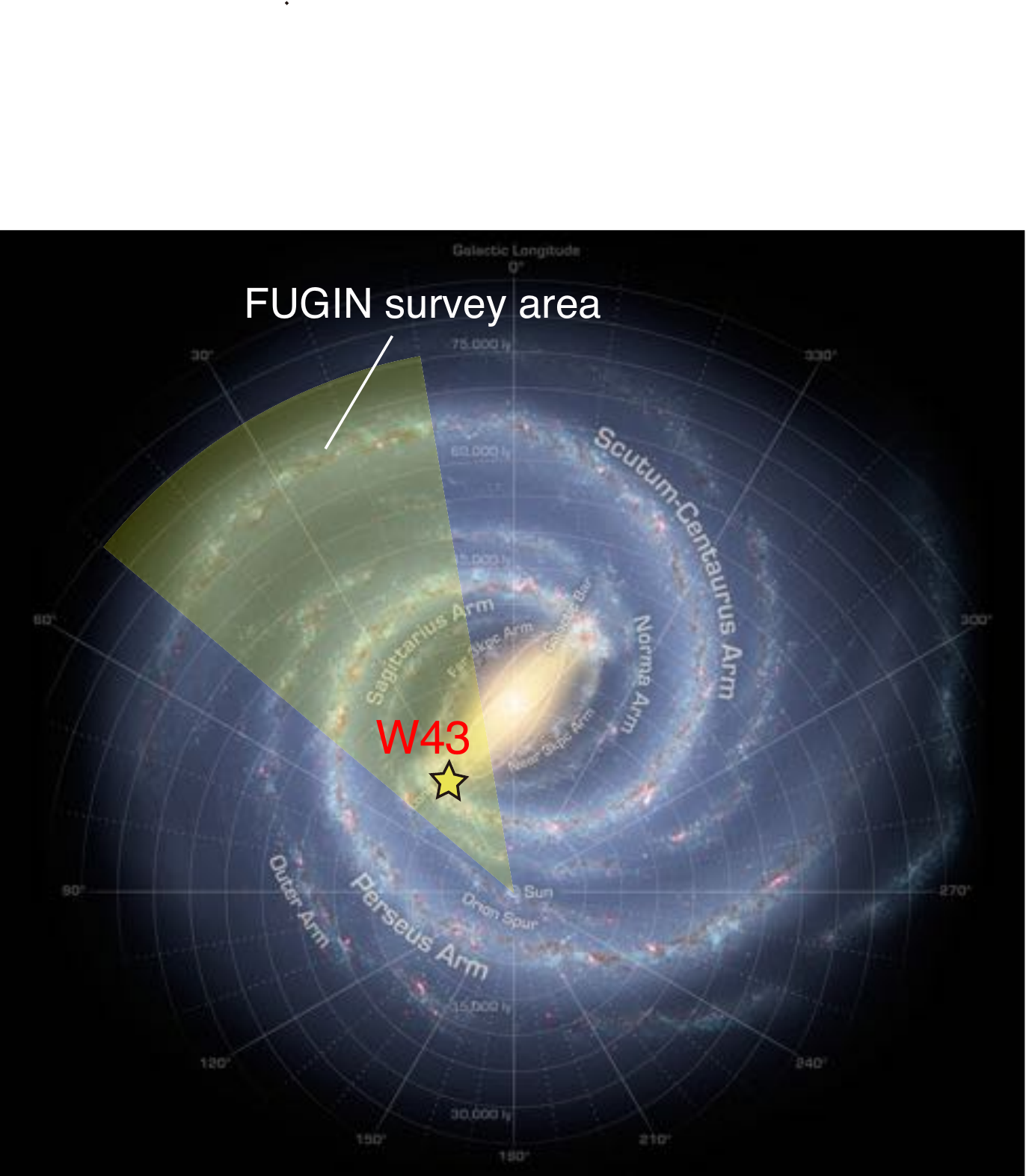}
\end{center}
\caption{{The top-view of the Milky Way (NASA/JPL-Caltech/ESO/R. Hurt). The star symbol indicates the position of the W43 GMC complex. The yellow shadow shows the inner survey area of the FUGIN project.} }
\label{MW_large}
\end{figure*}

\begin{figure*}[h]
\begin{center} 
 \includegraphics[width=18cm]{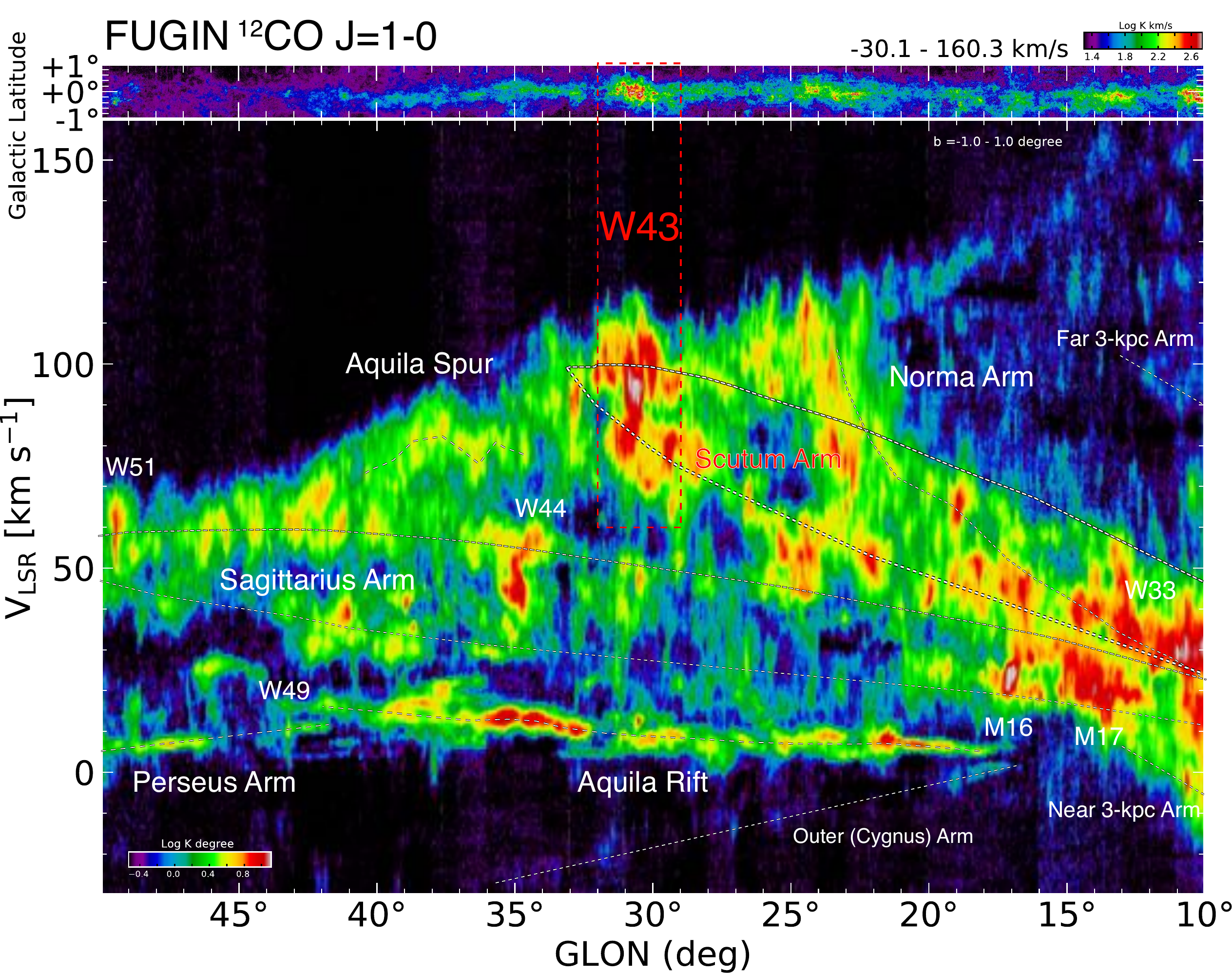}
\end{center}
\caption{Upper panel : Integrated intensity map of $^{12}$CO $J=$1--0 with the {integrated} velocity range of $-30$ $\>$km s$^{-1}$-$160$ $\>$km s$^{-1}$. Lower panel: Longitude-velocity diagram of $^{12}$CO $J=$1--0 with the integrated latitude range of $\timeform{-1D}$-$\timeform{+1D}$. The white dotted lines indicate the spiral arm at the first quadrant from Reid et al. (2016). 
{The near/far 3kpc arm is also referred to Dame \& Thaddeus( 2008).} The color {is} adopted {in the} logarithmic scale.}
\label{FUGIN_large_12CO}
\end{figure*}

\begin{figure*}[h]
\begin{center} 
 \includegraphics[width=18cm]{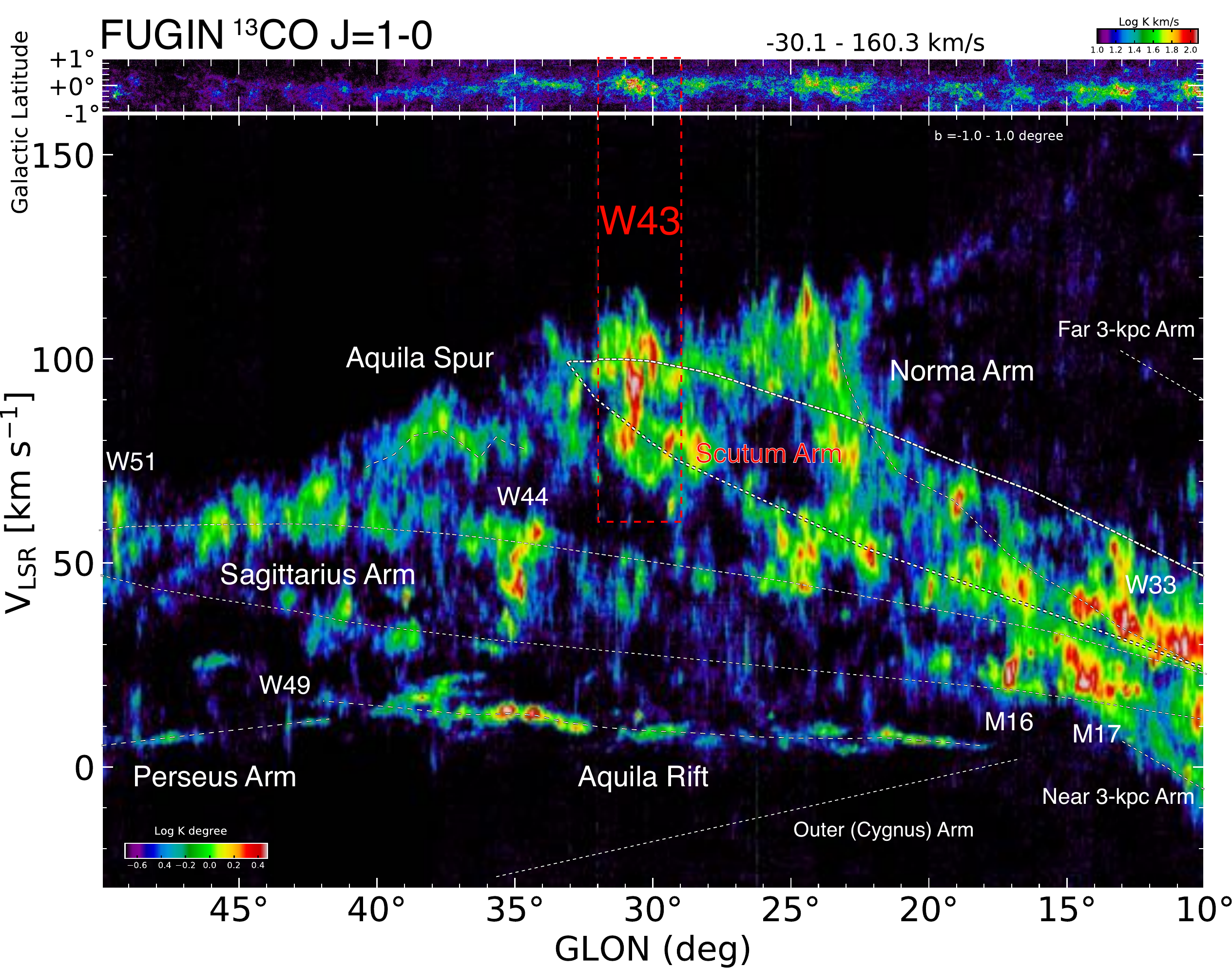}
\end{center}
\caption{{Same as Figure \ref{FUGIN_large_12CO}, but for $^{13}$CO $J=$1--0}}
\label{FUGIN_large_13CO}
\end{figure*}

\begin{figure*}[h]
\begin{center} 
 \includegraphics[width=18cm]{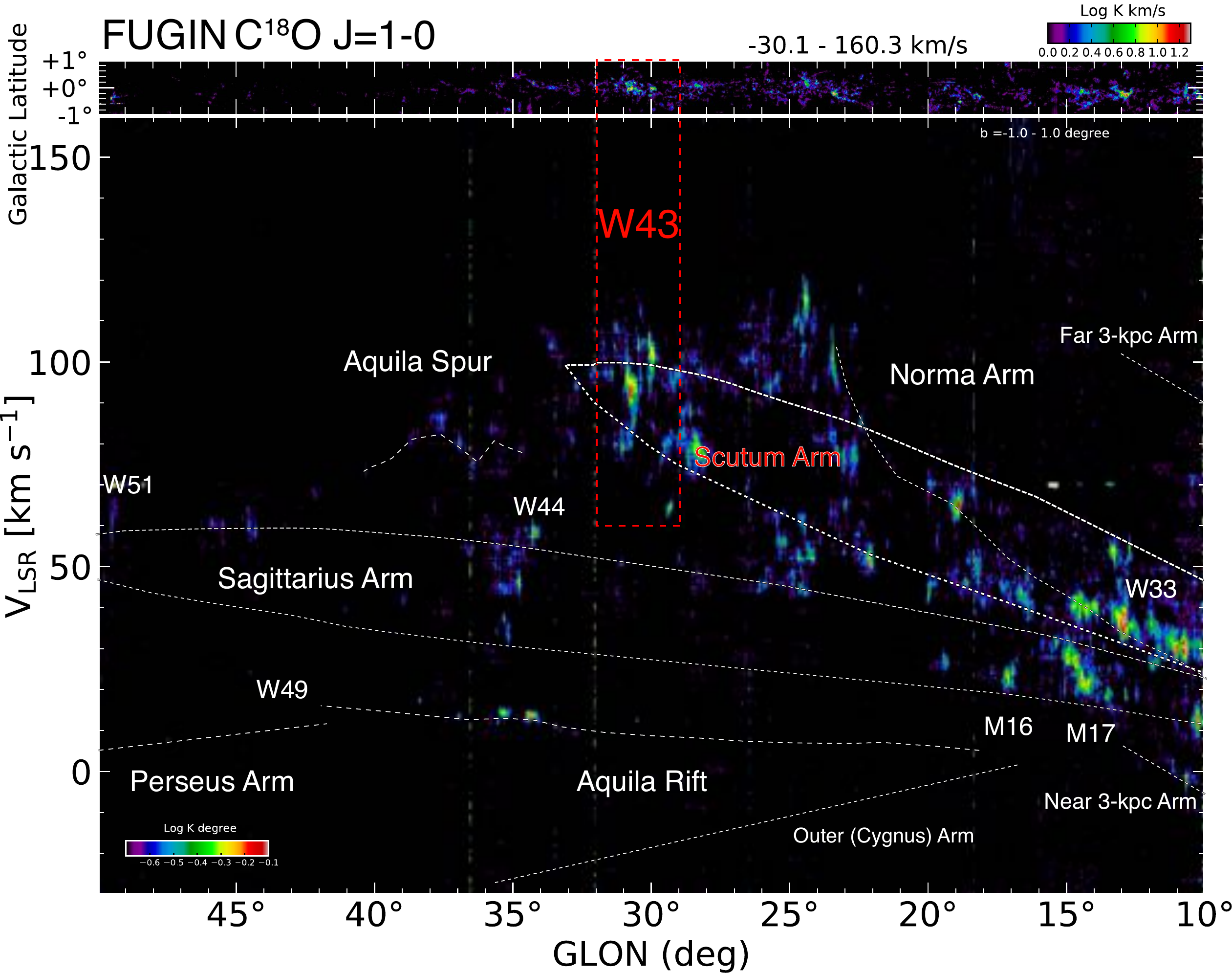}
\end{center}
\caption{{Same as Figure \ref{FUGIN_large_12CO}, but for C$^{18}$O $J=$1--0}. The data is smoothed to be $\sim \timeform{100"}$. }
\label{FUGIN_large_C18O}
\end{figure*}

\begin{figure*}[h]
\begin{center} 
 \includegraphics[width=13cm]{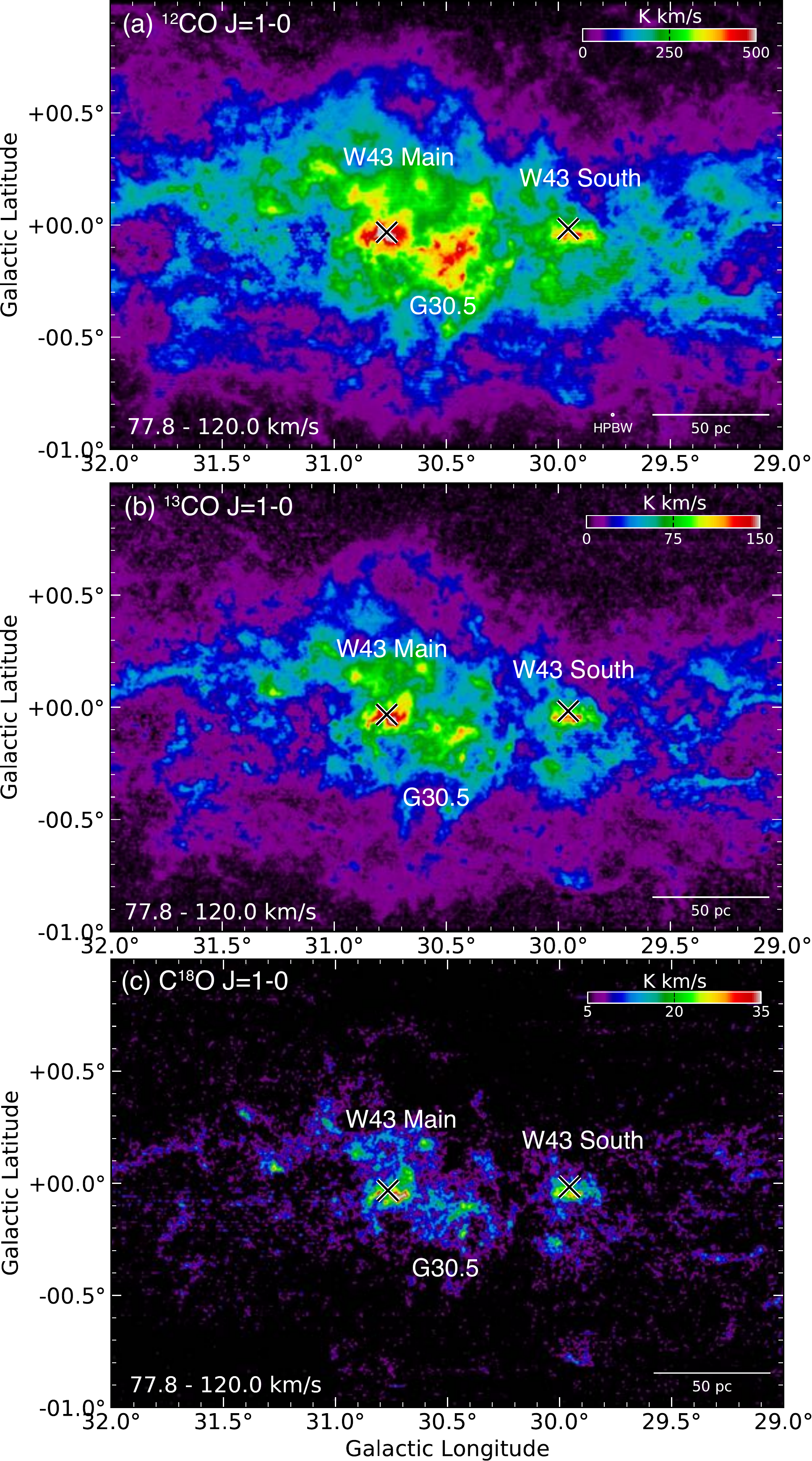}
\end{center}
\caption{Integrated intensity maps of  (a) $^{12}$CO,  (b) $^{13}$CO, and (c) C$^{18}$O $J=$1--0 for the W43 GMC complex. The integrated velocity range is from $78$ $\>$km s$^{-1}$ to $120$ $\>$km s$^{-1}$. The final beam size after convolution is \timeform{40"}. The white crosses indicate W43 Main (Blum et al. 1999) and W43 South (Wood \& Churchwell 1989).}
\label{W43_integ}
\end{figure*}

\begin{figure*}[h]
\begin{center} 
 \includegraphics[width=18cm]{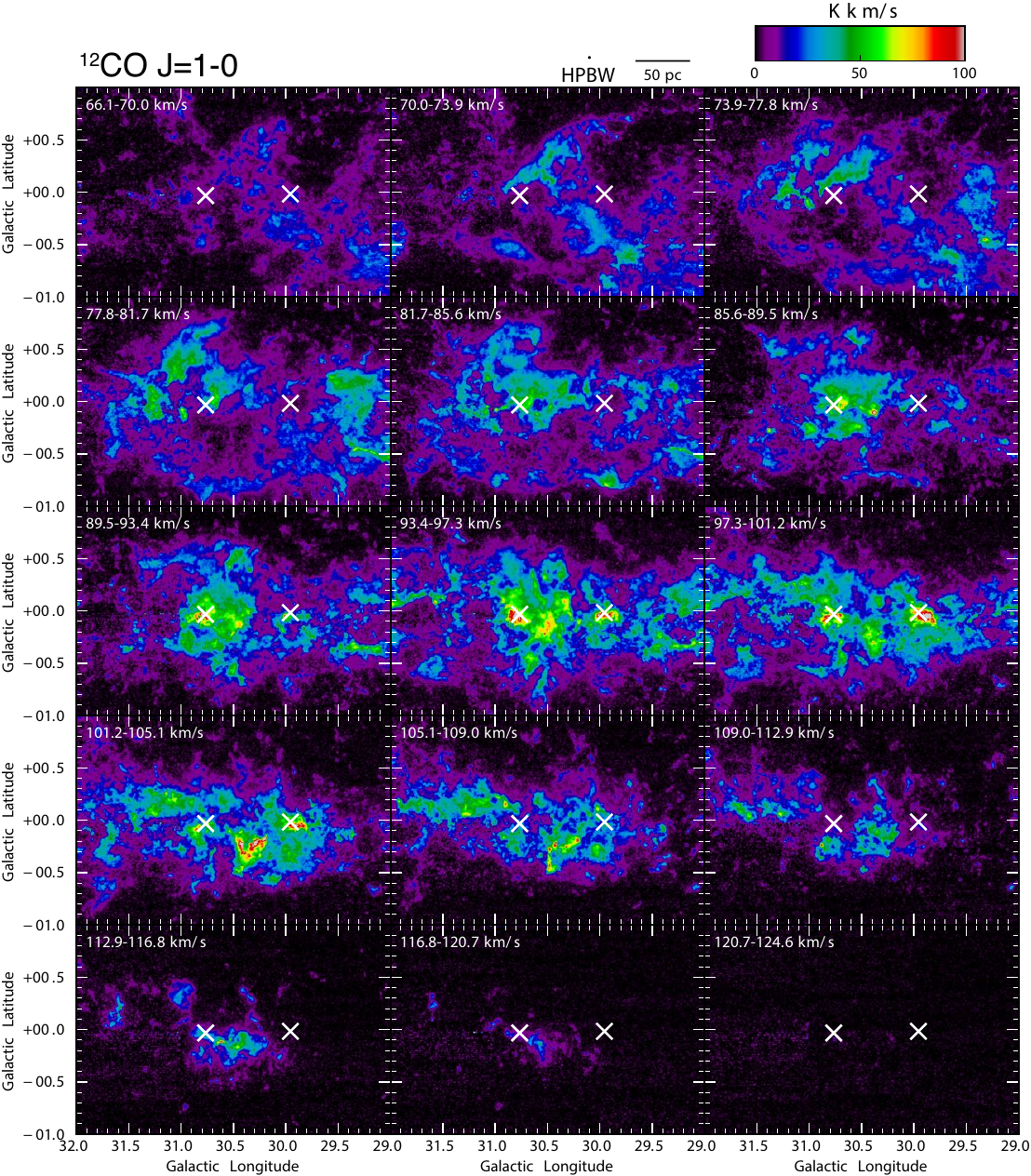}
\end{center}
\caption{Velocity channel map of the $^{12}$CO $J=$ 1--0 emission with a velocity step of  3.9 $\>$km s$^{-1}$. The final beam size after convolution is \timeform{40"}. The white crosses indicate W43 Main (Blum et al. 1999) and W43 South (Wood \& Churchwell 1989).}
\label{W43_12COch}
\end{figure*}

\begin{figure*}[h]
\begin{center} 
 \includegraphics[width=18cm]{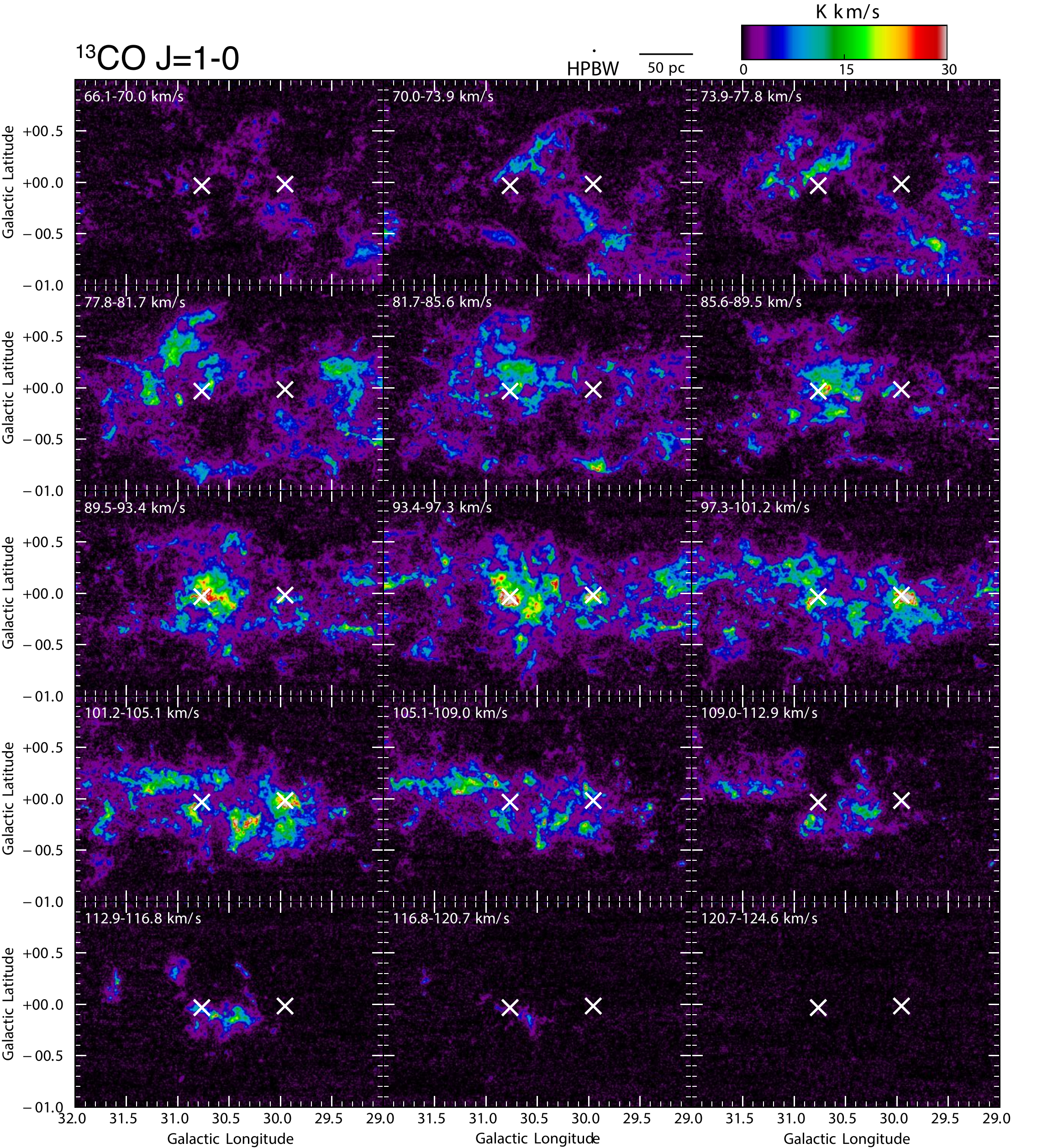}
\end{center}
\caption{{Same as Figure \ref{W43_12COch}, but for $^{13}$CO $J=$1--0.}}
\label{W43_13COch}
\end{figure*}

\begin{figure*}[h]
\begin{center} 
 \includegraphics[width=18cm]{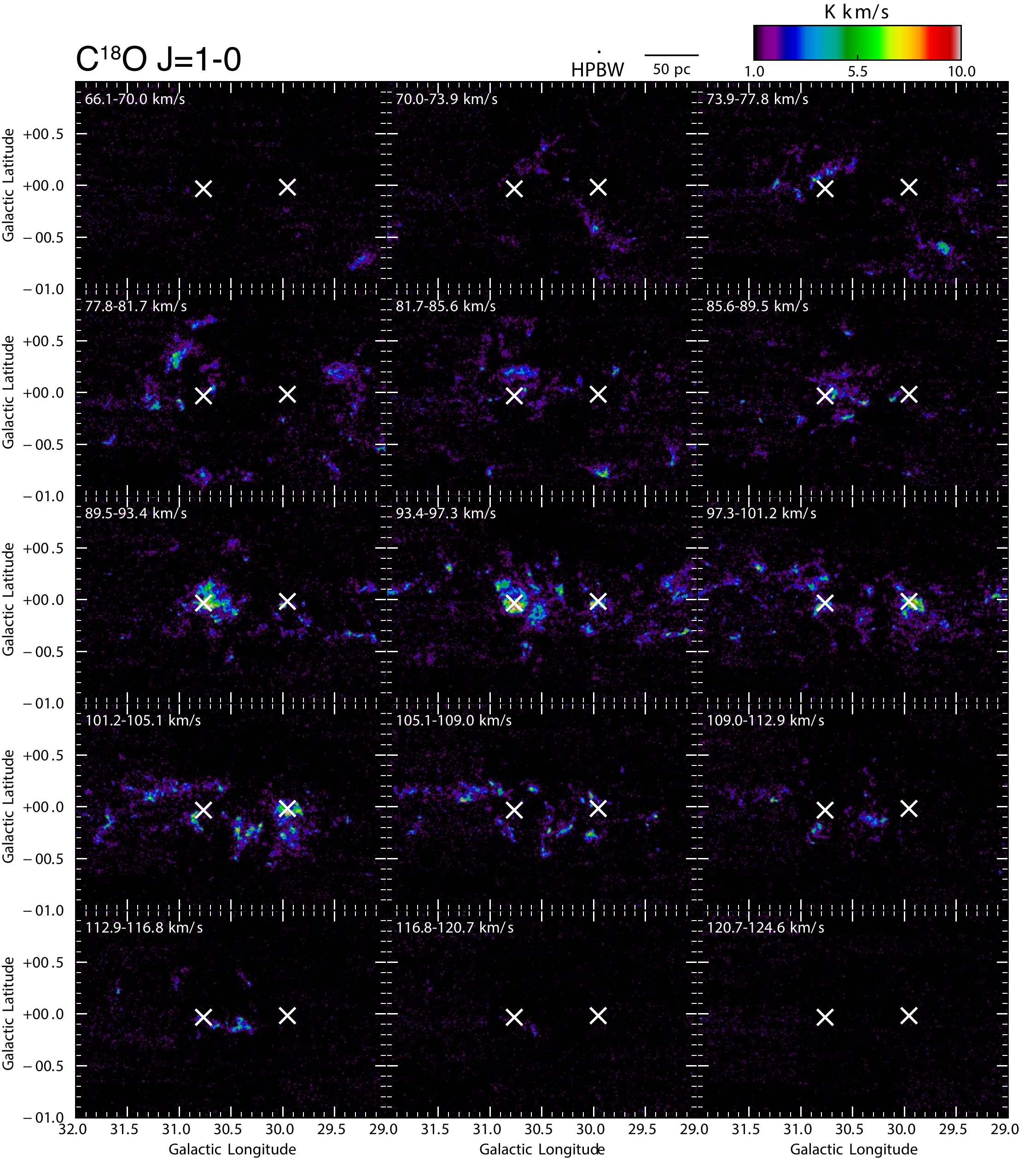}
\end{center}
\caption{{Same as Figure \ref{W43_12COch}, but for C$^{18}$O $J=$1--0.}}
\label{W43_C18Och}
\end{figure*}

\begin{figure*}[h]
\begin{center} 
 \includegraphics[width=10cm]{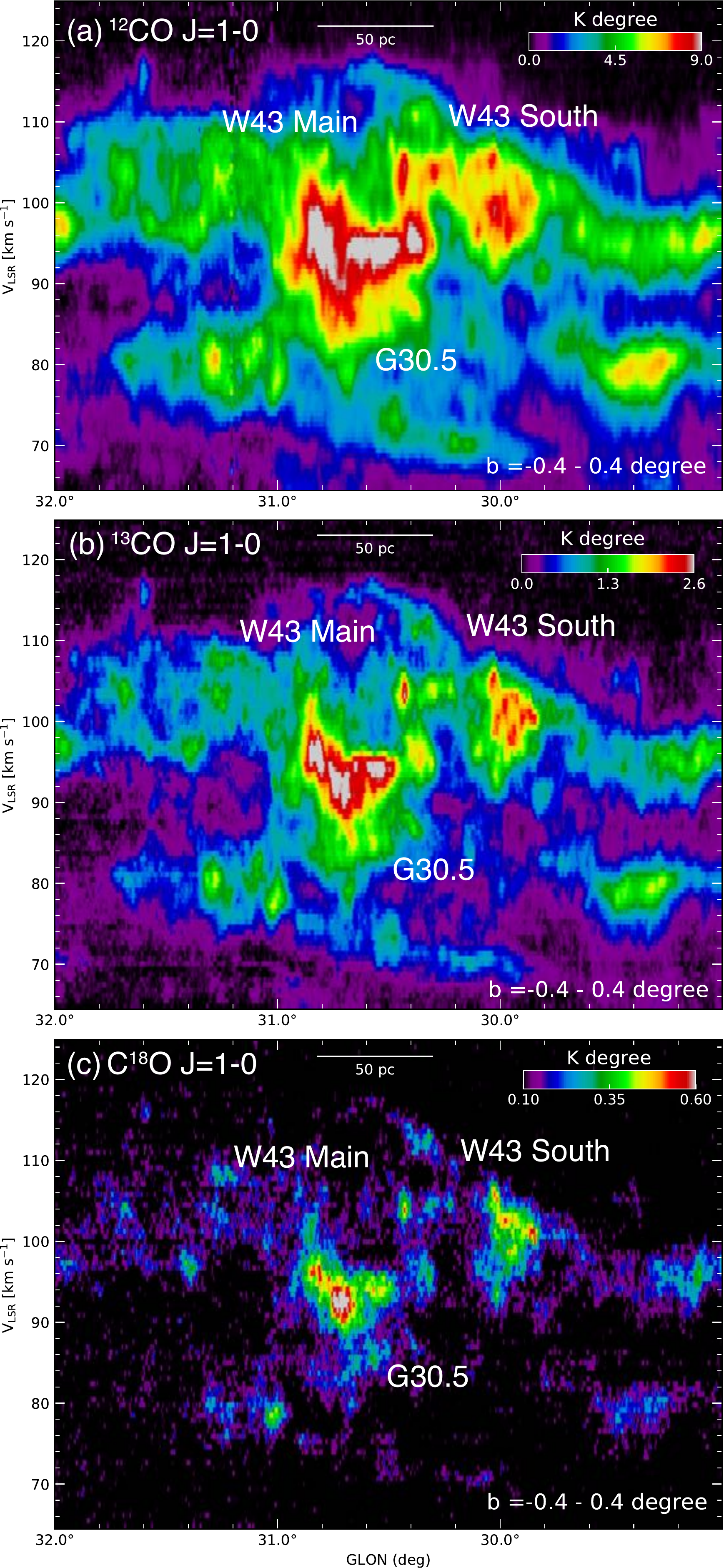}
\end{center}
\caption{Longitude-velocity diagram of the (a) $^{12}$CO, (b) $^{13}$CO, and  (c) C$^{18}$O $J=$ 1--0 with the integrated latitude range of $b=-\timeform{0.4D}$-- $+\timeform{0.4D}$.}\label{W43_lv}
\end{figure*}

\clearpage
\begin{table*}[h]
\tbl{Physical properties of the molecular clouds}{
\begin{tabular}{cccccccccccccc}
\hline
\multicolumn{1}{c}{Name}&  $N^{12}_{\rm X\ peak}$ & $N^{12}_{\rm X\ mean}$ & $N^{13}_{\rm LTE\ peak}$ & $N^{13}_{\rm LTE\ mean}$& $N^{18}_{\rm LTE\ peak}$ & $N^{18}_{\rm LTE\ mean}$&  $M^{12}_{\rm X}$ & $M_{\rm LTE}^{13}$ & $M_{\rm LTE}^{18}$\\
 & [cm$^{-2}$] & [cm$^{-2}$] & [cm$^{-2}$] & [cm$^{-2}$] & [cm$^{-2}$] & [cm$^{-2}$] & [$M_{\odot}$]  & [$M_{\odot}$] &  [$M_{\odot}$]\\
(1) & (2) &(3)& (4) & (5) & (6) & (7) & (8) & (9) & (10)\\
\hline\hline
W43 GMC complex &$3.3 \times 10^{23}$&$1.1 \times 10^{22}$& $3.0 \times 10^{23}$ &$1.4 \times 10^{22}$ & $3.2 \times 10^{23}$& $2.2 \times 10^{21}$ &$1.4 \times 10^7$ & $1.1 \times 10^7$ & $1.7\times10^6$\\
\hline
W43 Main (Total) &$1.2 \times 10^{23}$&$5.5 \times 10^{22}$& $2.6 \times 10^{23}$ &$6.4 \times 10^{22}$ & $3.1 \times 10^{23}$ &$1.4 \times 10^{22}$&$2.0 \times 10^6$&$2.4 \times 10^6$&$5.3 \times 10^5$\\
82 $\>$km s$^{-1}$ cloud & $2.2 \times 10^{22}$&$8.0 \times 10^{21}$& $4.6 \times 10^{22}$ &$8.0 \times 10^{21}$ & $6.0 \times 10^{22}$ &$2.6 \times 10^{20}$&$2.9 \times 10^5$&$2.6 \times 10^5$&$8.2 \times 10^3$\\
94 $\>$km s$^{-1}$ cloud & $8.1 \times 10^{22}$&$3.3\times 10^{22}$& $2.2 \times 10^{23}$ &$4.4 \times 10^{22}$ & $3.1 \times 10^{23}$ &$1.2 \times 10^{22}$&$1.2 \times 10^6$&$1.6 \times 10^6$&$4.5 \times 10^5$\\
103 $\>$km s$^{-1}$ cloud & $2.8 \times 10^{22}$&$1.0\times 10^{22}$& $7.9 \times 10^{22}$ &$8.9 \times 10^{21}$ & $8.6 \times 10^{22}$ &$1.3 \times 10^{21}$&$3.8 \times 10^5$&$3.2 \times 10^5$&$4.7 \times 10^4$\\
115 $\>$km s$^{-1}$ cloud & $2.5 \times 10^{22}$&$3.7\times 10^{21}$& $5.7 \times 10^{22}$ &$4.6 \times 10^{21}$ & $4.8 \times 10^{22}$ &$7.4 \times 10^{20}$&$1.4 \times 10^5$&$1.1 \times 10^5$&$1.8 \times 10^4$\\
\hline
G30.5 (Total) &$9.7 \times 10^{22}$&$4.5 \times 10^{22}$&$2.0 \times 10^{23}$&$3.9 \times 10^{22}$&$2.3 \times 10^{23}$&$6.1 \times 10^{21}$&$1.7 \times 10^6$&$1.4 \times 10^6$&$2.2 \times 10^5$\\
88 $\>$km s$^{-1}$ cloud & $3.2 \times 10^{22}$&$6.8 \times 10^{21}$&$1.1 \times 10^{23}$&$7.7 \times 10^{21}$&$1.4 \times 10^{23}$&$1.2 \times 10^{21}$&$2.5 \times 10^5$&$1.8 \times 10^5$&$2.7 \times 10^4$\\
93 $\>$km s$^{-1}$ cloud & $3.7 \times 10^{22}$&$1.5 \times 10^{22}$&$7.2 \times 10^{22}$&$1.3 \times 10^{22}$&$6.2 \times 10^{22}$&$6.5 \times 10^{20}$&$5.7 \times 10^5$&$4.6 \times 10^5$&$2.3 \times 10^4$\\
103 $\>$km s$^{-1}$ cloud & $4.7 \times 10^{22}$&$1.5 \times 10^{22}$&$1.7 \times 10^{23}$&$1.5 \times 10^{22}$&$2.3 \times 10^{23}$&$3.6 \times 10^{21}$&$5.7 \times 10^5$&$5.3 \times 10^5$&$1.3 \times 10^5$\\
113 $\>$km s$^{-1}$ cloud & $2.4 \times 10^{22}$&$7.4 \times 10^{21}$&$4.0 \times 10^{22}$&$6.1 \times 10^{21}$&$5.9 \times 10^{22}$&$1.3 \times 10^{21}$&$2.7 \times 10^5$&$1.9 \times 10^5$&$4.1 \times 10^4$\\
\hline
W43 South (Total) &$9.3 \times 10^{22}$&$3.0 \times 10^{22}$&$3.0 \times 10^{23}$&$3.4 \times 10^{22}$&$3.2 \times 10^{23}$&$1.3 \times 10^{22}$&$9.7 \times 10^5$&$1.1 \times 10^6$& $4.0 \times 10^5$\\
93 $\>$km s$^{-1}$ cloud & $3.9 \times 10^{22}$&$7.9 \times 10^{21}$&$1.3 \times 10^{23}$&$9.3 \times 10^{21}$&$1.9 \times 10^{23}$&$2.2 \times 10^{21}$&$2.6 \times 10^5$&$2.5 \times 10^5$& $6.1 \times 10^4$ \\
102 $\>$km s$^{-1}$ cloud & $6.2 \times 10^{22}$&$1.7 \times 10^{22}$&$2.4 \times 10^{23}$&$2.5 \times 10^{22}$&$3.0 \times 10^{23}$&$1.1 \times 10^{22}$&$5.6 \times 10^5$&$7.4 \times 10^5$& $3.3 \times 10^5$\\
\hline\hline
\end{tabular}
}
\label{tab:first}
\begin{tabnote}
The clipping level is adopted as 5 $\sigma$ ($\sim$ 5 K of $^{12}$CO and 1.75 K of $^{13}$CO,  C$^{18}$O$J=$1--0). The integrated velocity range of the total molecular gas is $78$ to $120$ $\>$km s$^{-1}$. The velocity clouds of W43 Main, G30.5, and W43 South are the same as in Figures \ref{W43Main_integ},\ref{G30.5_integ}, and \ref{W43South_integ}, respectively.\\
*Columns: (1) Name. (2) Peak H$_2$ column density from $^{12}$CO assuming the X-factor. (3) Average H$_2$ column density of the cloud from $^{12}$CO assuming the X-factor. (4) Peak H$_2$ column density from $^{13}$CO assuming the LTE. (5) Average H$_2$ column density of the cloud from $^{13}$CO assuming the LTE. (6) Peak H$_2$ column density from C$^{18}$O assuming the LTE. (7) Average H$_2$ column density of the cloud from C$^{18}$O assuming the LTE. (8) Total H$_2$ mass from $^{12}$CO assuming the X-factor.  (9) Total H$_2$ mass from $^{13}$CO assuming the LTE. (10) Total H$_2$ mass from C$^{18}$O assuming the LTE. 
\end{tabnote}
\end{table*}

\begin{figure*}[h]
\begin{center} 
 \includegraphics[width=18cm]{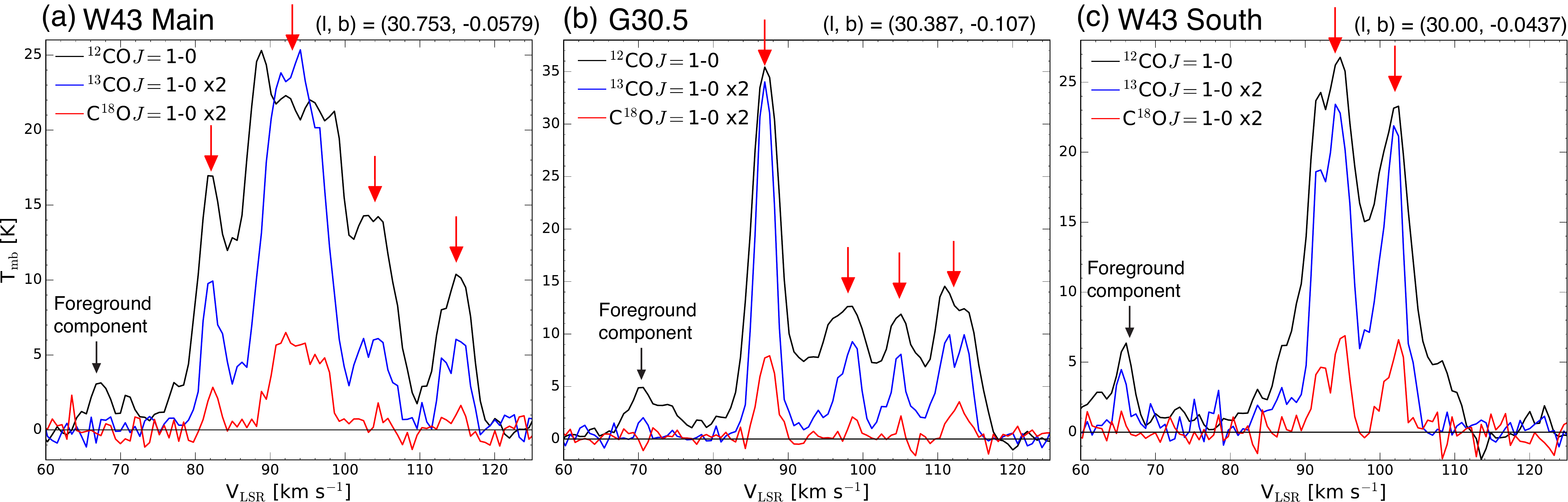}
\end{center}
\caption{{Spectra for $^{12}$CO, $^{13}$CO, and C$^{18}$O $J=$1--0 obtained at (a) W43 Main, (b) G30.5, and (c) W43 South. The red arrows indicate the velocity components associated with the W43 GMC complex. The position of each spectrum shows the upper right of each panel. }}
\label{W43_spectra}
\end{figure*}

\begin{figure*}[h]
\begin{center} 
 \includegraphics[width=18cm]{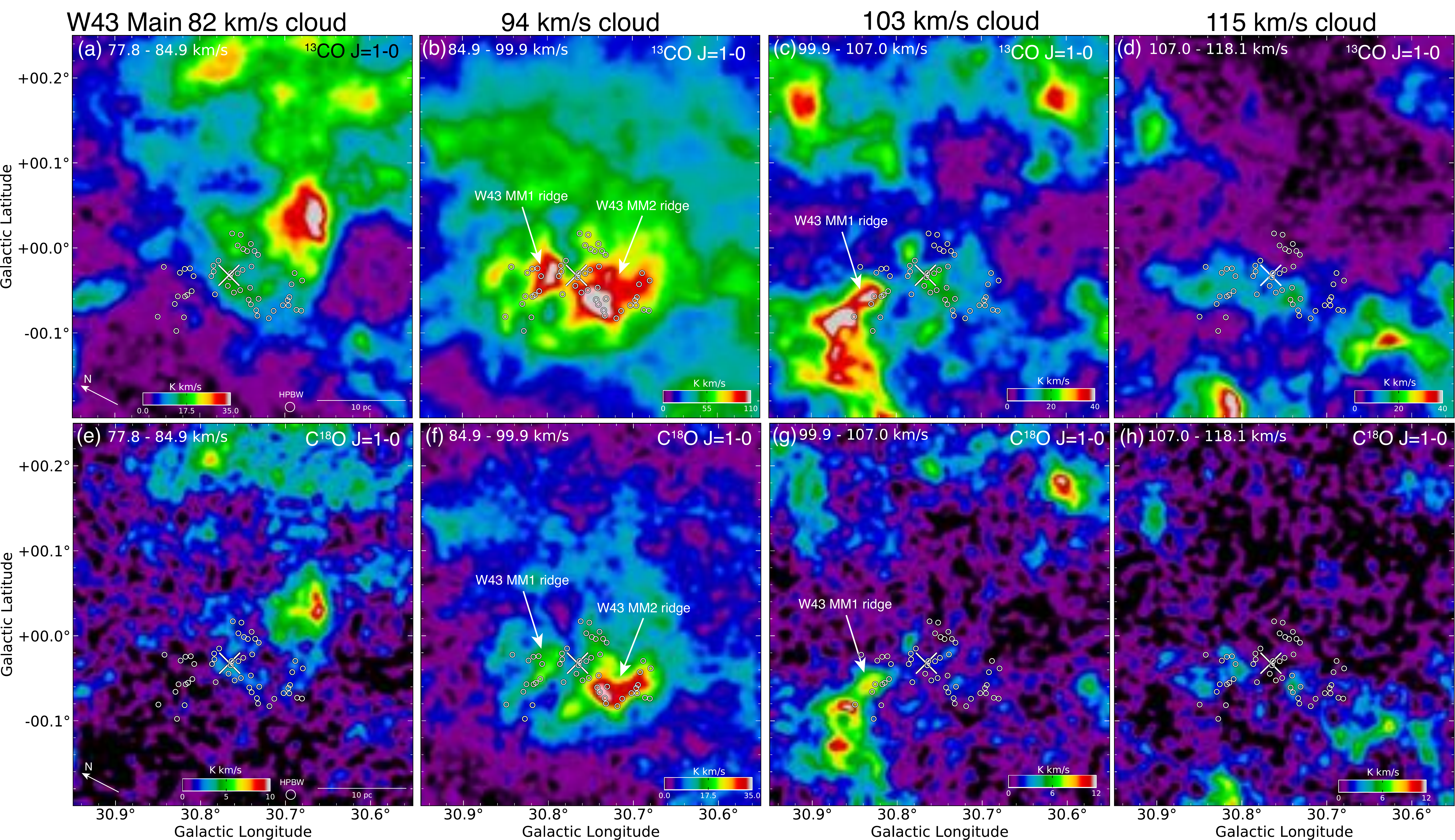}
\end{center}
\caption{ (a), (b), (c), (d) Integrated intensity map of $^{13}$CO $J=$ 1--0 for the (a) 82 $\>$km s$^{-1}$,  (b) 94 $\>$km s$^{-1}$,  (c) 103 $\>$km s$^{-1}$, and (d) 115 $\>$km s$^{-1}$ cloud. The Xs indicate the W43 Main cluster (Blum et al. 1999). The white circles present the 51 {protocluster candidates} (W43 MM1-MM51) cataloged by Motte et al. (2003). The final beam size after convolution is $\sim \timeform{40"}$. (e), (f), (g), (h) Integrated intensity map of C$^{18}$O $J=$ 1--0 obtained for the (e) 82 $\>$km s$^{-1}$ ,  (f) 94 $\>$km s$^{-1}$, (g) 103 $\>$km s$^{-1}$, and (h) 115 $\>$km s$^{-1}$ cloud.}
\label{W43Main_integ}
\end{figure*}

\begin{figure*}[h]
\begin{center} 
 \includegraphics[width=17cm]{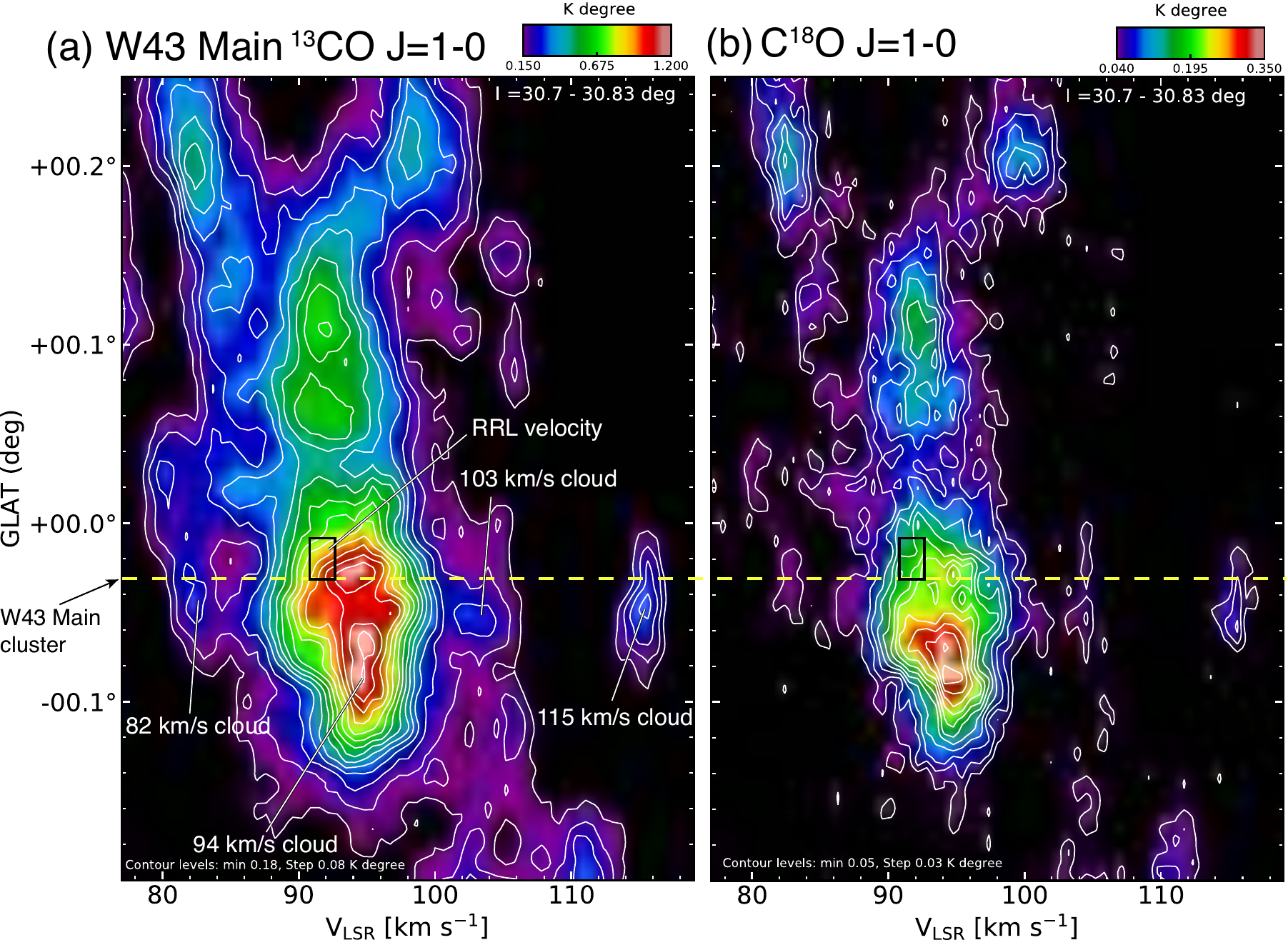}
\end{center}
\caption{ Galactic latitude-velocity diagram of  (a) $^{13}$CO and (b) C$^{18}$O $J=$ 1--0 integrated over the latitude range from \timeform{30.7D} to \timeform{30.83D}. The contour levels and intervals are 0.18 K degree and 0.08 K degree of (a), 0.05 K degree and 0.03 K degree of (b), respectively. The black boxes show the radio recombination line velocity (91.7 $\>$km s$^{-1}$) at $(l, b) = (\timeform{30.780D}, \timeform{-0.020D})$ from Luisi et al (2017), where their resolution is $\sim$ 1.86 $\>$km s$^{-1}$ $\times$ \timeform{82"}. The yellow dashed line indicates the position of the W43 Main cluster (Blum et al. 1999). }
\label{W43Main_lv}
\end{figure*}

\begin{figure*}[h]
\begin{center} 
 \includegraphics[width=18cm]{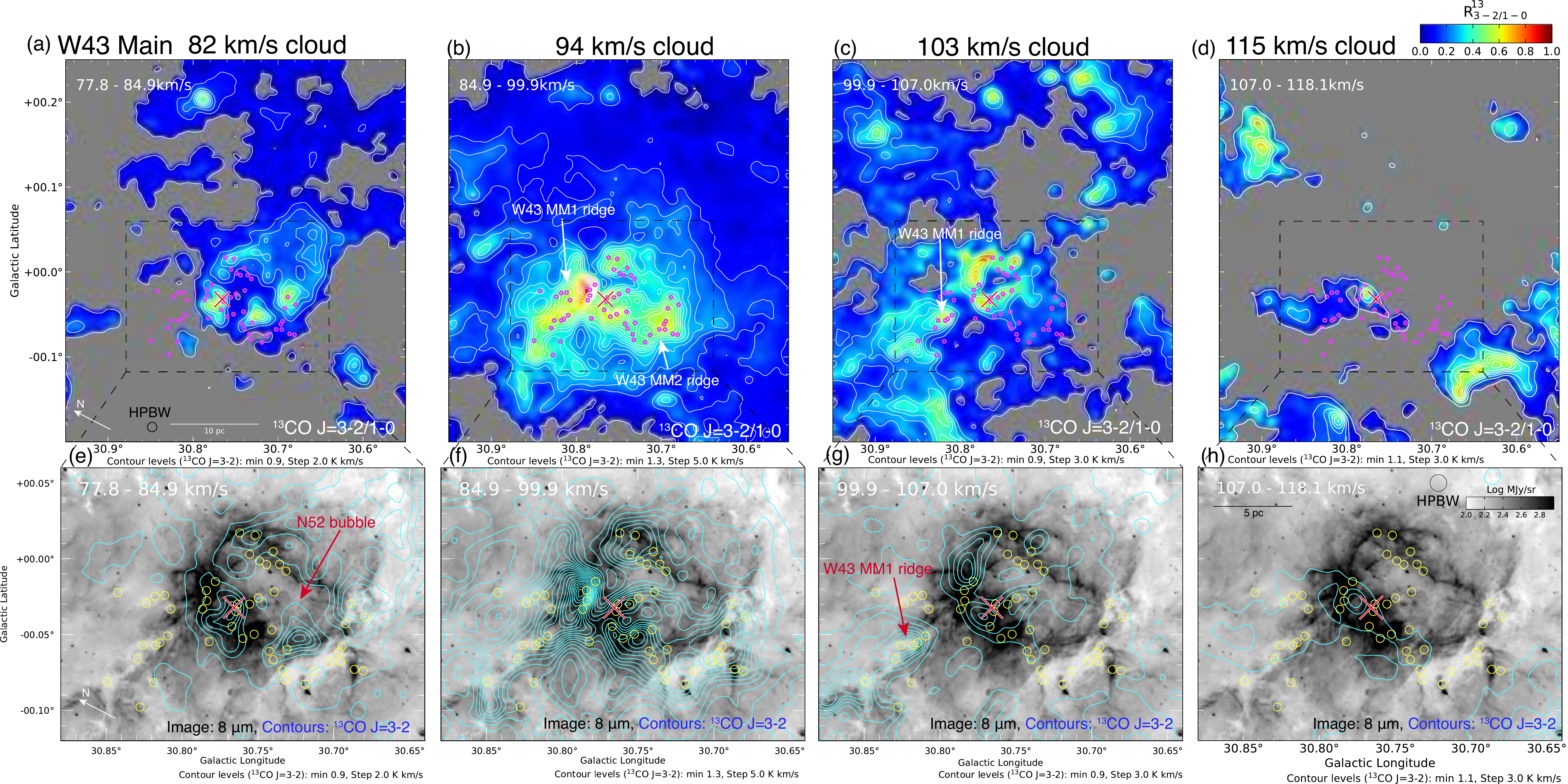}
\end{center}
\caption{ (a), (b), (c), (d) Intensity ratio map of $^{13}$CO $J=$ 3--2/$^{13}$CO $J=$ 1--0 from JCMT and {FUGIN} for the  (a) 82 $\>$km s$^{-1}$,  (b) 94 $\>$km s$^{-1}$, (c) 103 $\>$km s$^{-1}$, and (d) 115 $\>$km s$^{-1}$ cloud. The Xs indicate W43 Main cluster (Blum et al. 1999). The circles indicate the 51 compact fragments (W43 MM1-MM51) cataloged by Motte et al. (2003). The final beam size after convolution is $\sim \timeform{40"}$ . The clipping levels are adopted as $3\sigma$ ($\sim 0.9$ K $\>$km s$^{-1}$$\sim 1.3$ K $\>$km s$^{-1}$, $\sim 0.9$ K $\>$km s$^{-1}$, and $\sim 1.1$ K $\>$km s$^{-1}$) of each integrated velocity range. (e), (f), (g), (h) Integrated intensity map of $^{13}$CO $J=$ 3--2 (contours) obtained with JCMT (CHIMPS) superposed on the Spitzer 8 $\>\mu$m image for the (e) 82 $\>$km s$^{-1}$, (f) 94 $\>$km s$^{-1}$, (g) 103 $\>$km s$^{-1}$, and (h) 115 $\>$km s$^{-1}$ cloud.}
\label{W43Main_ratio}
\end{figure*}

\begin{figure*}[h]
\begin{center} 
 \includegraphics[width=17cm]{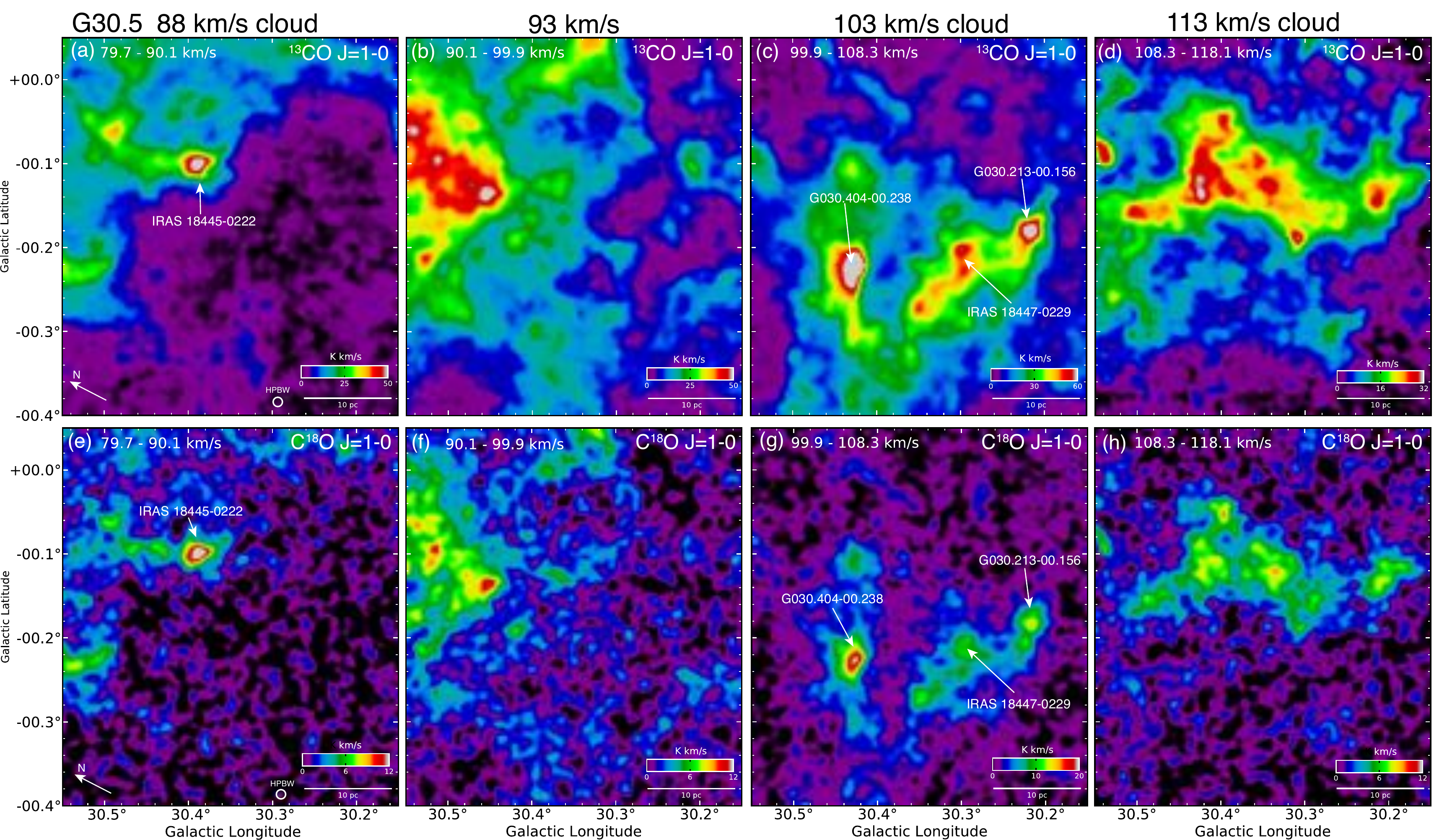}
\end{center}
\caption{ (a), (b), (c), (d) Integrated intensity map of $^{13}$CO $J=$ 1--0 for the (a) 88 $\>$km s$^{-1}$, (b) 93 $\>$km s$^{-1}$, (c) 103 $\>$km s$^{-1}$ and  (d) 113 $\>$km s$^{-1}$. The final beam size after convolution is $\sim \timeform{40"}$. (e), (f), (g), (h) Integrated intensity map of C$^{18}$O $J=$ 1--0 for the (e) 88 $\>$km s$^{-1}$, (f) 93 $\>$km s$^{-1}$, (g) 103 $\>$km s$^{-1}$  and (h) 113 $\>$km s$^{-1}$ cloud .}
\label{G30.5_integ}
\end{figure*}

\begin{figure*}[h]
\begin{center} 
 \includegraphics[width=17cm]{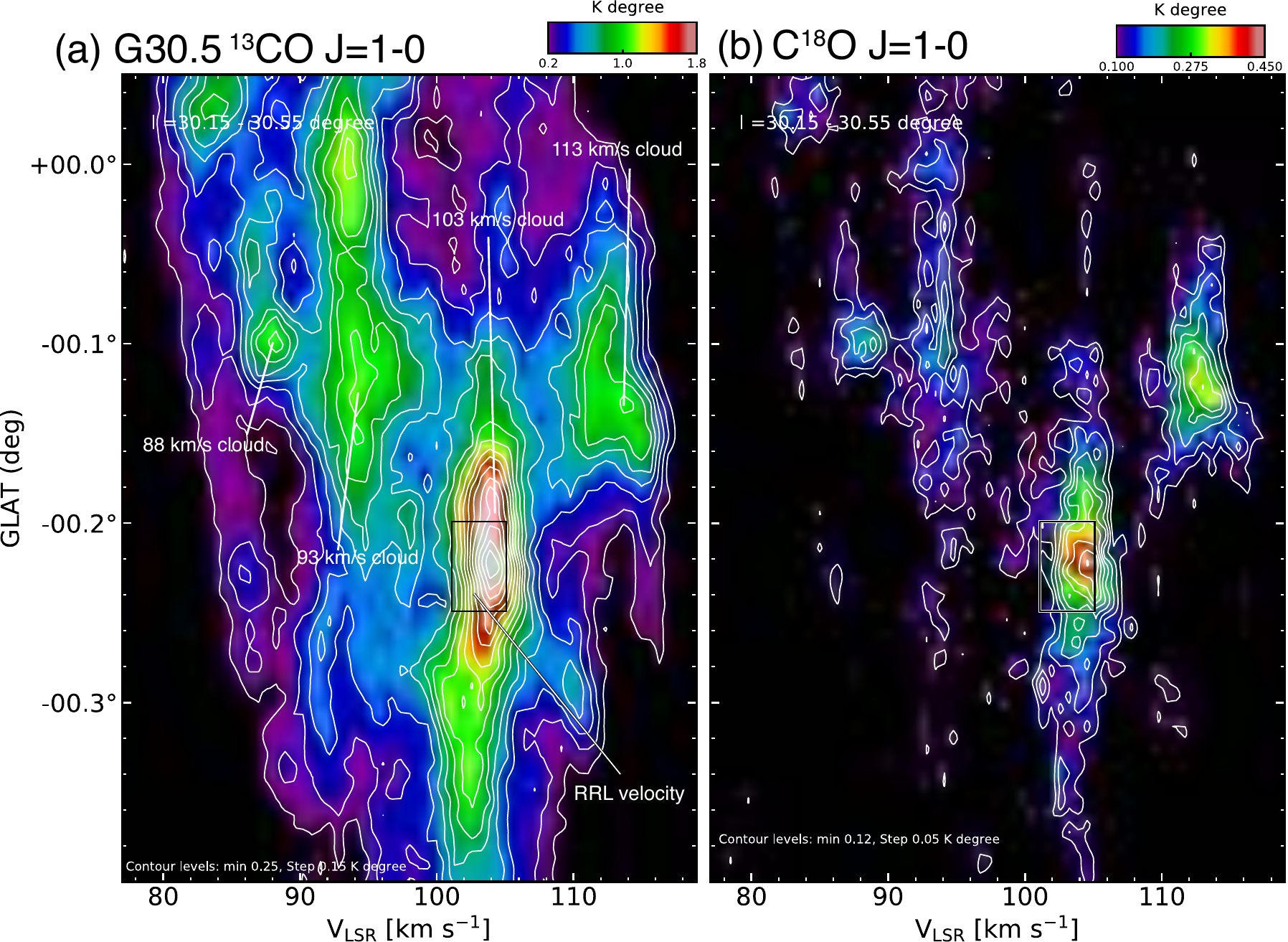}
\end{center}
\caption{Galactic latitude-velocity diagram of (a) $^{13}$CO and (b) C$^{18}$O $J=$ 1--0 integrated over the latitude range from \timeform{30.15D} to \timeform{30.55D}. The black boxes show the radio recombination line velocity ($\sim 102.5$ $\>$km s$^{-1}$) at $(l,b)\sim (\timeform{30.404D}, \timeform{-0.238D})$ from Lockman (1989), where their resolution is $\sim$ 4 $\>$km s$^{-1}$ $\times$ \timeform{3'}. The contour levels and intervals are 0.25 K degree and 0.15 K degree of (a), 0.12 K degree and 0.05 K degree of (b), respectively}
\label{G30.5_lv}
\end{figure*}

\begin{figure*}[h]
\begin{center} 
 \includegraphics[width=17cm]{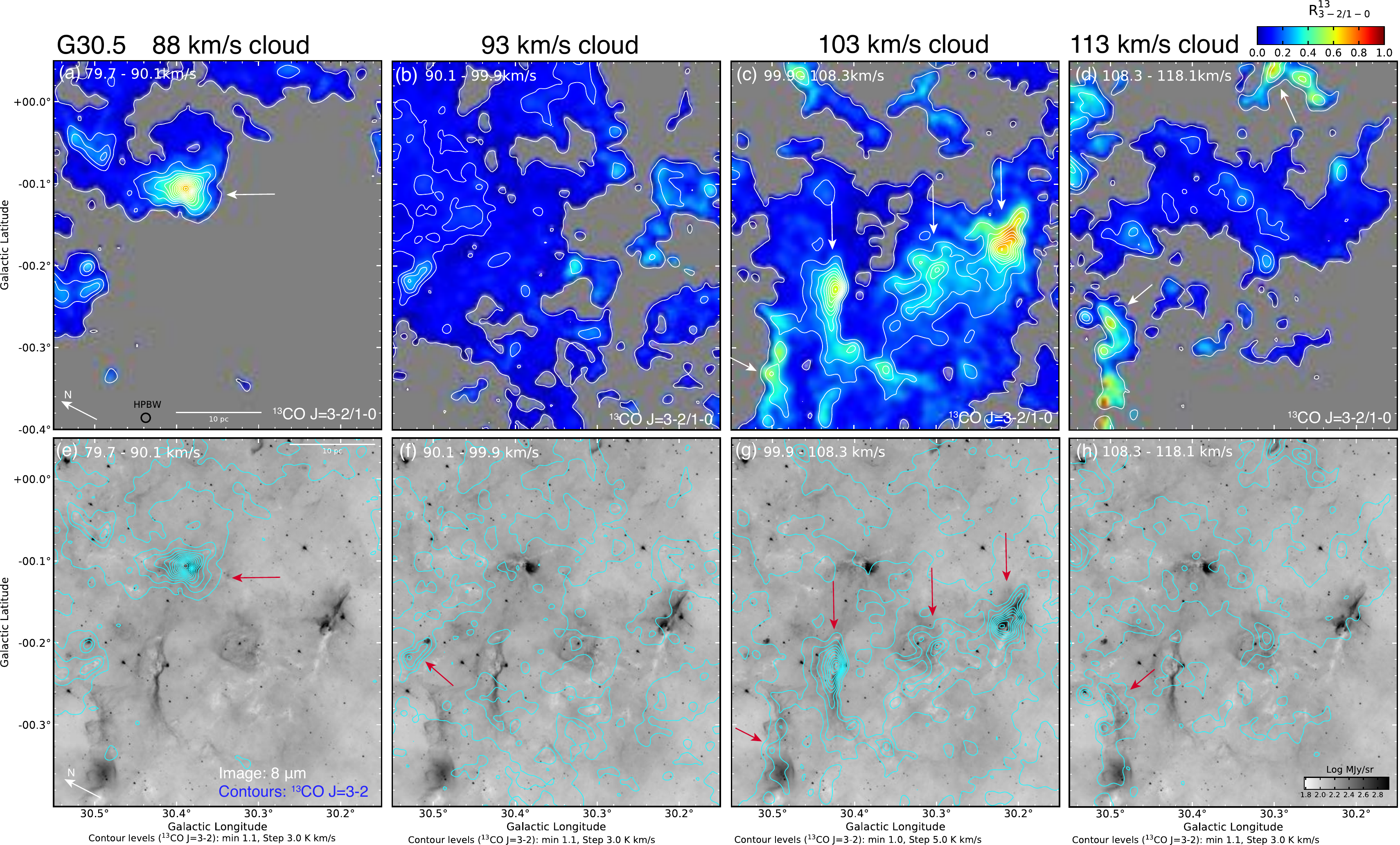}
\end{center}
\caption{ (a), (b), (c), (d) Intensity ratio map of $^{13}$CO $J=$ 3--2/$^{13}$CO $J=$ 1--0 from JCMT and {FUGIN} for the (a) 88 $\>$km s$^{-1}$ cloud,  (b) 93 $\>$km s$^{-1}$ cloud,  (c) 103 $\>$km s$^{-1}$ cloud and (d) 113 $\>$km s$^{-1}$ cloud. The final beam size after convolution is $\sim \timeform{40"}$. The clipping levels are adopted as $3\sigma$ ($\sim 0.8$ K $\>$km s$^{-1}$, $\sim 1.1$ K $\>$km s$^{-1}$, $\sim 1.0$ K $\>$km s$^{-1}$, and $\sim 1.1$ K $\>$km s$^{-1}$) of each integrated velocity range. (e), (f), (g), (h) Integrated intensity map of $^{13}$CO $J=$ 3--2 (contours) obtained {by JCMT} superposed on the Spitzer 8 $\>\mu$m image for the (e) 88 $\>$km s$^{-1}$ cloud, (f) 93 $\>$km s$^{-1}$ cloud, (g) 103 $\>$km s$^{-1}$ cloud and (h) 113 $\>$km s$^{-1}$ cloud.}
\label{G30.5_ratio}
\end{figure*}

\begin{figure*}[h]
\begin{center} 
 \includegraphics[width=17cm]{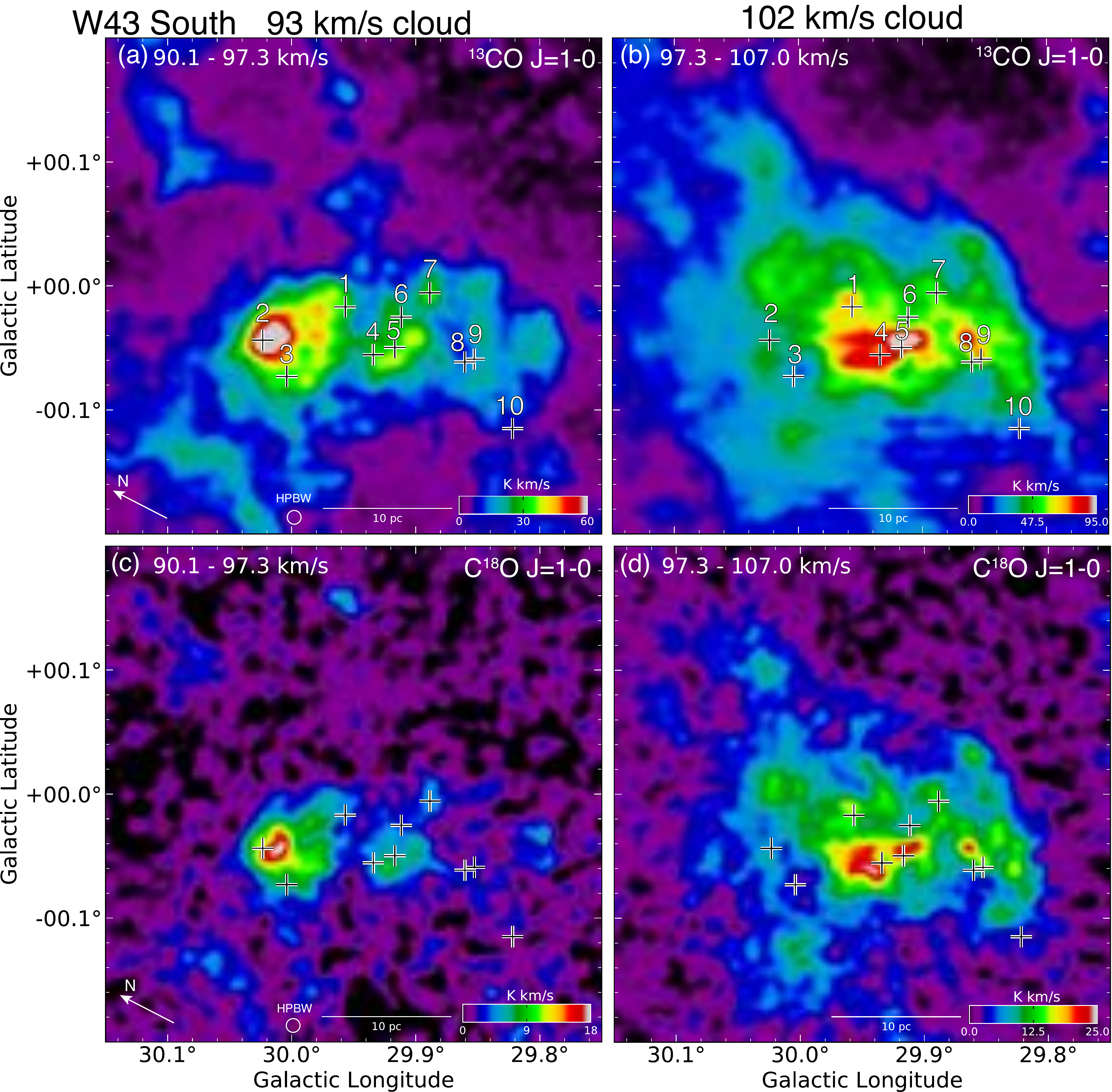}
\end{center}
\caption{(a), (b) Integrated intensity map of $^{13}$CO $J=$ 1--0 for the (a) 93 $\>$km s$^{-1}$ cloud  and (b) the 102 $\>$km s$^{-1}$ cloud. The crosses indicate radio continuum sources identified by {the} NVSS survey (Condon et al. 1998), and the numbering is the same as in Beltr\'an et al. (2013). The final beam size after convolution is $\sim \timeform{40"}$ . (c), (d) Integrated intensity map of C$^{18}$O $J=$ 1--0 for (c) the 93 $\>$km s$^{-1}$ cloud and (d) the 102 $\>$km s$^{-1}$ cloud. }
\label{W43South_integ}
\end{figure*}

\begin{figure*}[h]
\begin{center} 
 \includegraphics[width=18cm]{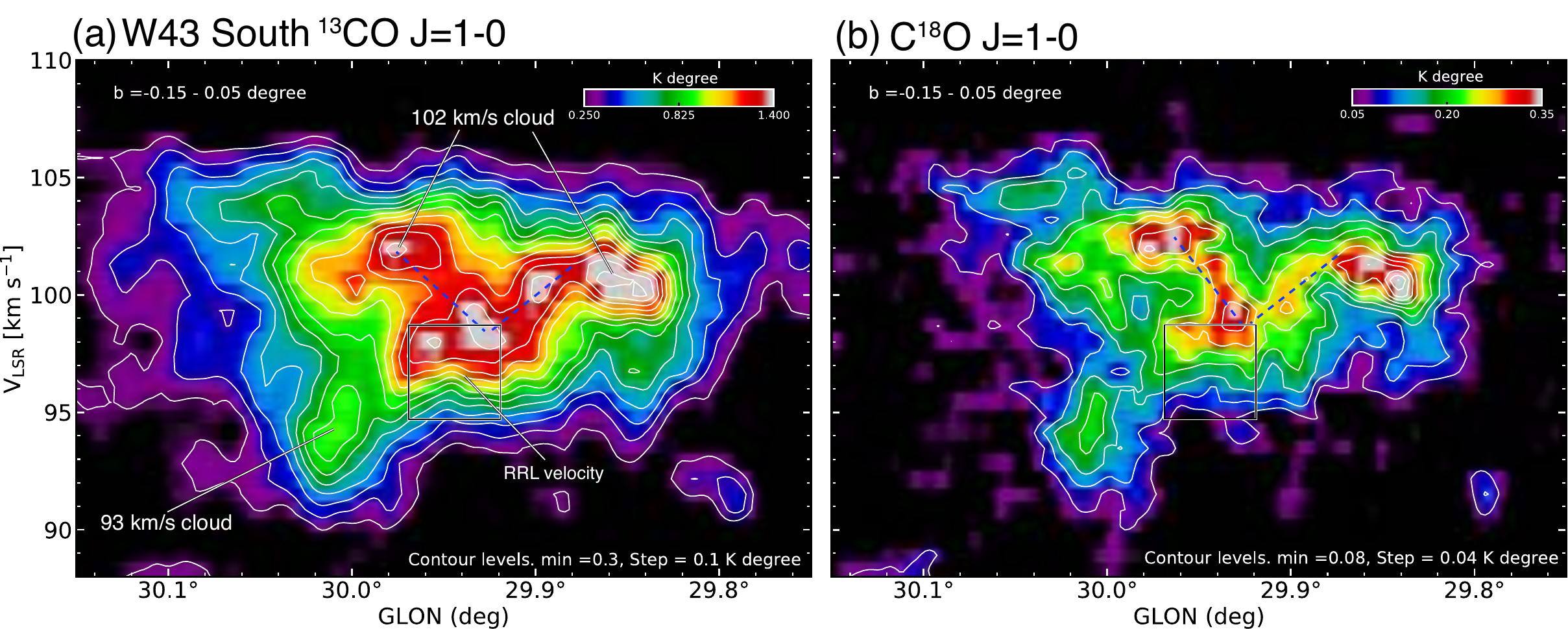}
\end{center}
\caption{Galactic longitude-velocity diagram of (a) $^{13}$CO and (b) C$^{18}$O $J=$ 1--0 integrated over the latitude range from \timeform{-0.15D} to \timeform{0.05D}. The contour levels and intervals are 0.3 K degree and 0.1 K degree of (a), 0.08 K degree and 0.04 K degree of (b), respectively. The black boxes show the H110$\alpha$ radio recombination line velocity ($\sim 96.7$ $\>$km s$^{-1}$) at $(l,b)\sim (\timeform{29.944D}, \timeform{-0.042D})$ from Lockman (1989), where their resolution is $\sim$ 4 $\>$km s$^{-1}$ $\times$ \timeform{3'}. The blue dotted lines present the V-shape like structure.}
\label{W43South_lv}
\end{figure*}

\begin{figure*}[h]
\begin{center} 
 \includegraphics[width=17cm]{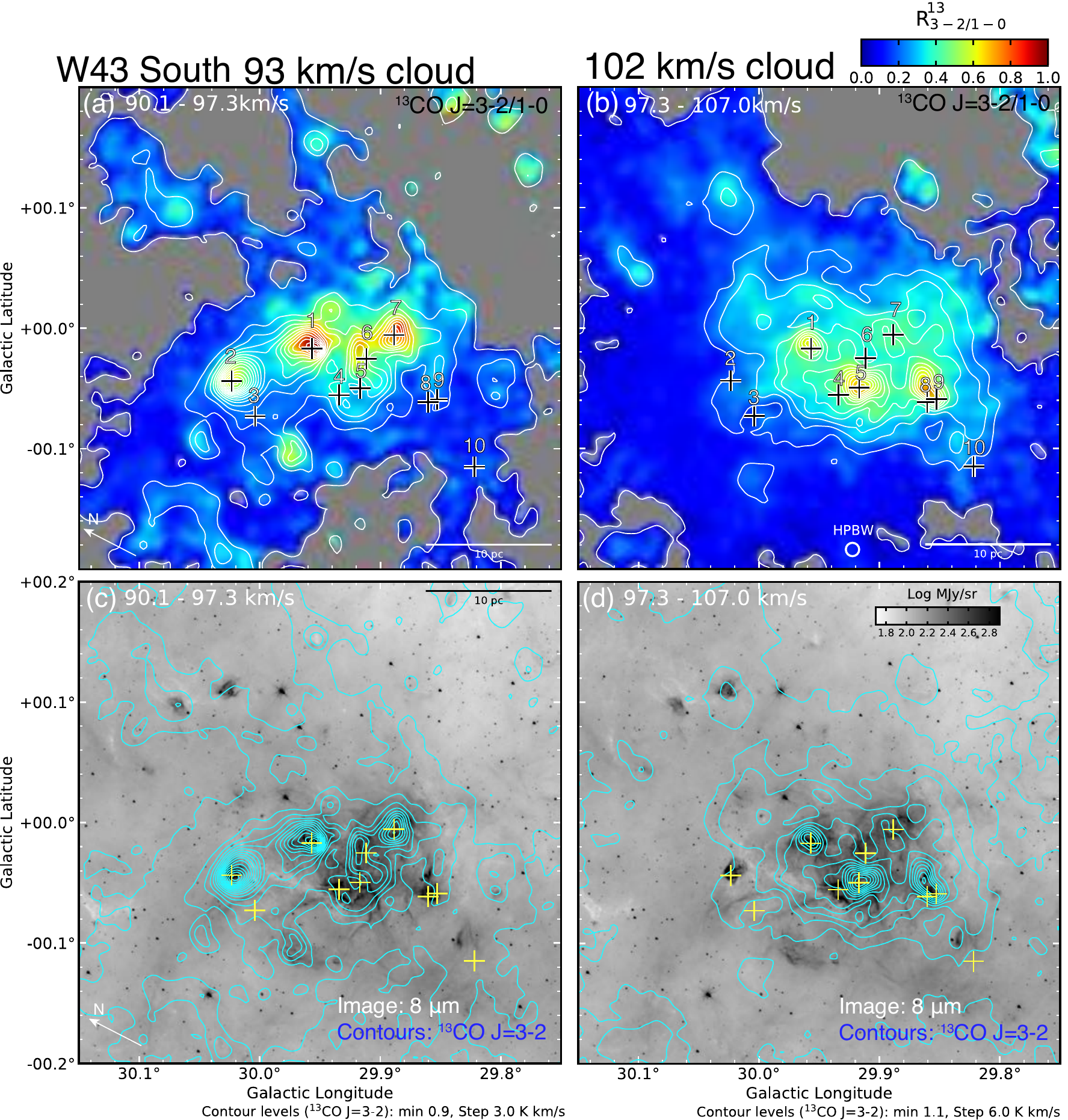}
\end{center}
\caption{(a), (b) Intensity ratio map of $^{13}$CO $J=$ 3--2/$^{13}$CO $J=$ 1--0 from JCMT and {FUGIN} for the (a) 93 $\>$km s$^{-1}$ cloud and (b) 102 $\>$km s$^{-1}$ cloud. The crosses indicate radio continuum sources identified by {the} NVSS survey (Condon et al. 1998), and the numbering is the same as in Beltr\'an et al. (2013). The final beam size after convolution is $\sim \timeform{40"}$ . The clipping levels are adopted as $3\sigma$ ($\sim 0.9$ K $\>$km s$^{-1}$ and $\sim 1.1$ K $\>$km s$^{-1}$) of each integrated velocity range. (c), (d) Integrated intensity map of $^{13}$CO $J=$ 3--2 (contours) obtained with JCMT superposed on the Spitzer 8 $\>\mu$m image for the (c) 93 $\>$km s$^{-1}$ cloud and (d) 102 $\>$km s$^{-1}$ cloud.}
\label{W43South_ratio}
\end{figure*}

\begin{figure*}[h]
\begin{center} 
 \includegraphics[width=15cm]{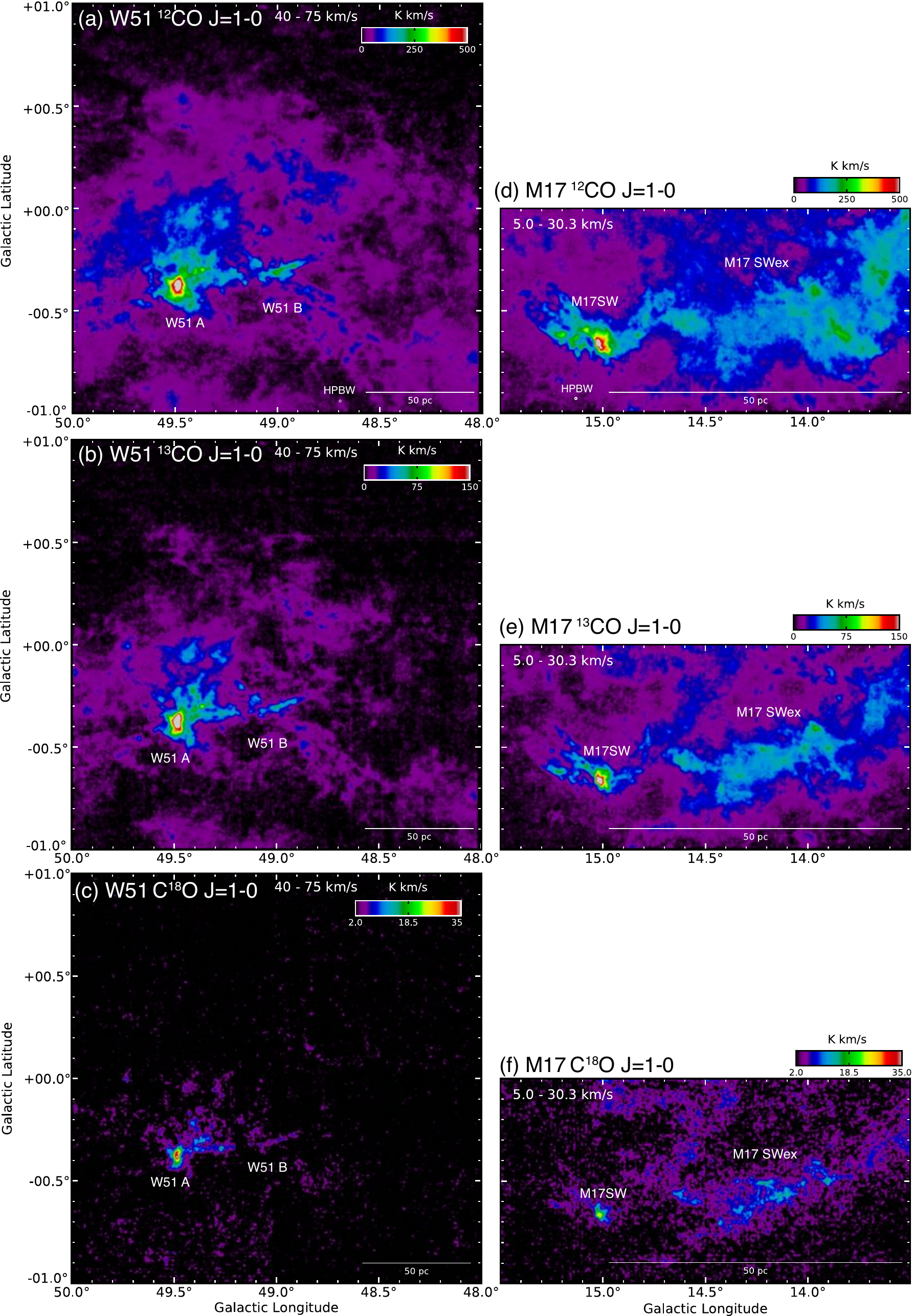}
\end{center}
\caption{Integrated intensity maps of (a,d) $^{12}$CO, (b,e) $^{13}$CO, and (c,f) C$^{18}$O $J=$1--0 for the W51 and M17 GMC obtained with {FUGIN}. The integrated velocity range of W51 and M17 is from $40$ $\>$km s$^{-1}$ to $75$ $\>$km s$^{-1}$ and from $5$ $\>$km s$^{-1}$ to $30$ $\>$km s$^{-1}$, respectively.}
\label{integ_all}
\end{figure*}

\clearpage

\begin{figure*}[h]
\begin{center} 
 \includegraphics[width=18cm]{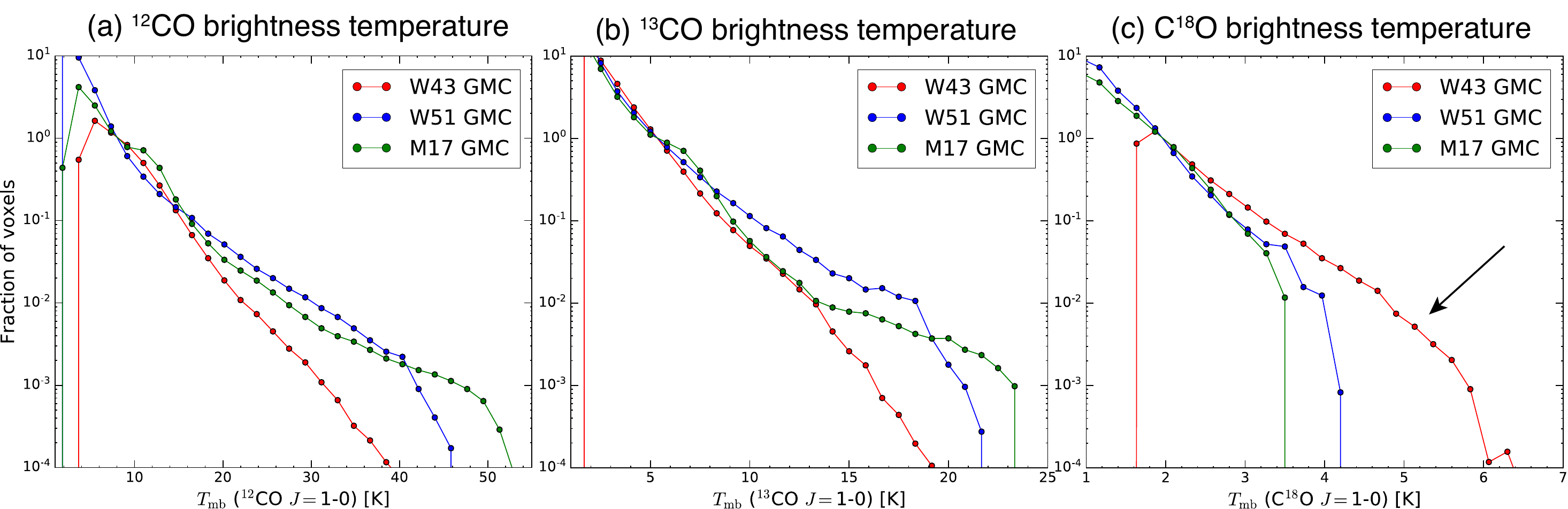}
\end{center}
\caption{Histogram of the brightness temperature of the voxels in the W43, W51, and M17 GMC from  (a) $^{12}$CO, (b) $^{13}$CO, and (c) C$^{18}$O $J=$1--0, respectively. The frequency is normalized by 7.3-9.2 K, 5.0-5.8 K, and 1.9-2.1 K, respectively. The clipping levels are adopted as 5$\sigma$ of each region.}
\label{BDF_hist}
\end{figure*}

\begin{table*}
\tbl{Physical properties and Star Formation Efficiency (SFE) of W43, W51, and M17}{
\begin{tabular}{cccccccccccc}
\hline
\multicolumn{1}{c}{Name} & $M^{12}_{\rm x}$  & $M^{13}_{\rm LTE}$ & $M^{18}_{\rm LTE}$  &  Stellar mass& SFE$_{\rm ^{12}CO}$ & SFE$_{\rm ^{13}CO}$  &  SFE$_{\rm C^{18}O}$  \\
&[$M_{\odot}$]  & [$M_{\odot}$] &  [$M_{\odot}$]  &   [$M_{\odot}$]& [$\%$] & [$\%$] & [$\%$]\\
(1) & (2) &(3)& (4) & (5) & (6) & (7) & (8) \\
\hline
W43 GMC  & $1.4\times 10^7$  & $1.1\times 10^7$ & $1.7\times 10^6 $  &$\sim 8 \times 10^{4}$$^{\dag}$ & $\sim 0.5$ & {$\sim 0.7$} & $\sim 4$ \\
W51 GMC  & $2.9\times 10^6$  & $1.9\times 10^6$ & $2.9\times 10^5$  &$\sim 4 \times 10^{4}$ & $\sim 1$ & $\sim 2$ & $\sim 12$\\
M17 GMC &  $6.8\times 10^5$ & $6.4\times 10^5$ & $1.6\times 10^5$  & $\sim 5 \times 10^{4}$ &$\sim 7$  & $\sim 7$ & $\sim 24$\\
\hline
\end{tabular}}\label{tab:first}
\begin{tabnote}
$\dag$ The total stellar mass of W43 is calculated by summing up W43 Main and W43 South (G29.96-0.02).\\
Columns: (1) Name. (2) Total H$_2$ mass from $^{12}$CO assuming the X-factor.  (3) Total H$_2$ mass from $^{13}$CO assuming the LTE. (4) Total H$_2$ mass from C$^{18}$O assuming the LTE. (5) The total stellar mass derived by assuming the IMF from the earliest spectral type star(s) referring to Binder \& Povich (2018). The masses of O-type star(s) are adopted from the observational stellar parameters of Tables 4 and 5 in Martins et al. (2005). (6) Star formation efficiency derived by the total H$_2$ mass from $^{12}$CO. (7) Star formation efficiency derived by the total H$_2$ mass from $^{13}$CO. (8) Star formation efficiency derived by the total H$_2$ mass from C$^{18}$O.
\end{tabnote}
\end{table*}

\begin{figure*}[h]
\begin{center} 
 \includegraphics[width=18cm]{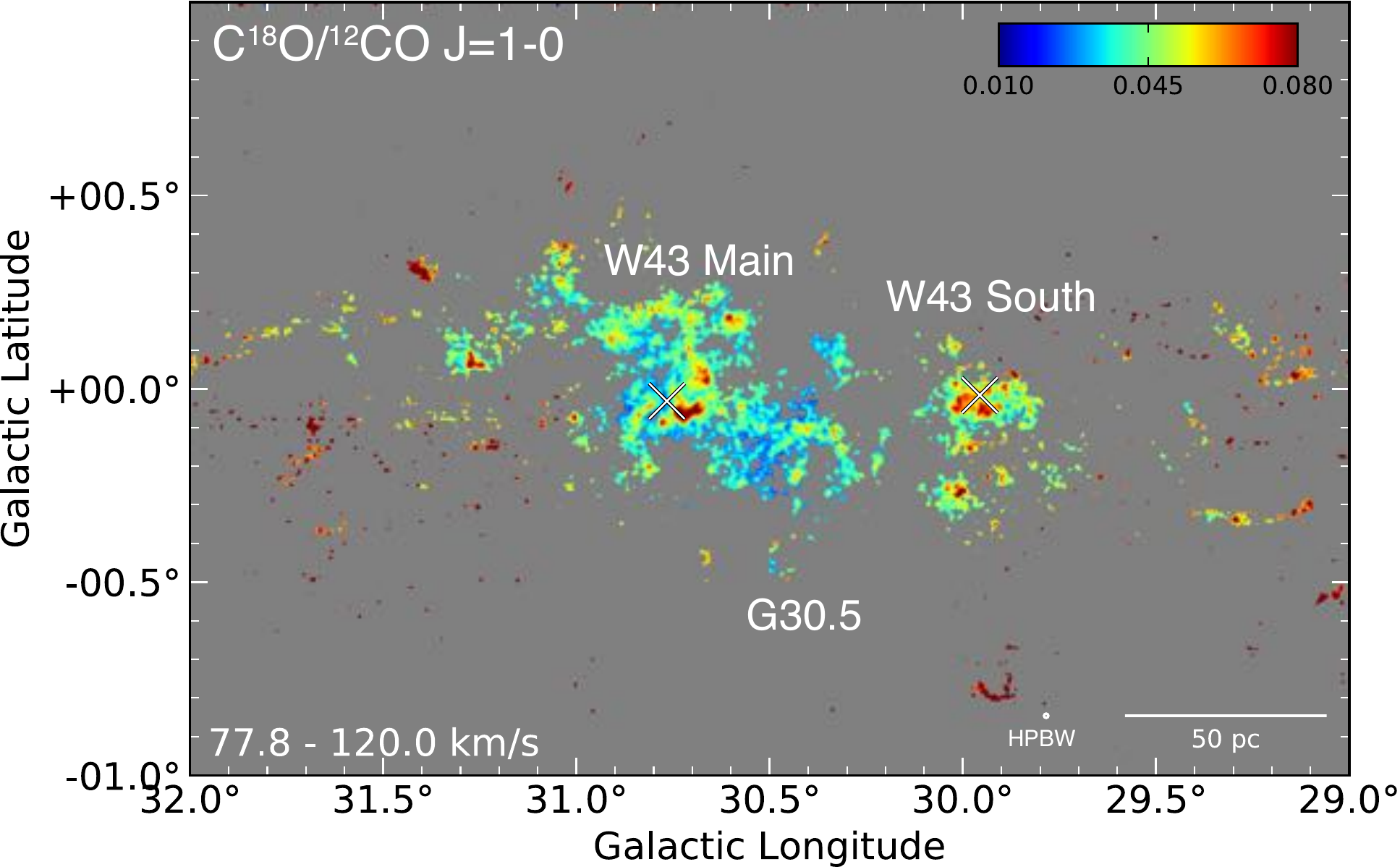}
\end{center}
\caption{{Intensity ratio map of C$^{18}$O/$^{12}$CO $J=$ 1--0 for the W43 GMC complex. The white crosses indicate W43 Main (Blum et al. 1999) and W43 South (Wood \& Churchwell 1989). The clipping levels are adopted as 5$\sigma$.}}
\label{W43_C18Oratio}
\end{figure*}

\begin{figure*}[h]
\begin{center} 
 \includegraphics[width=18cm]{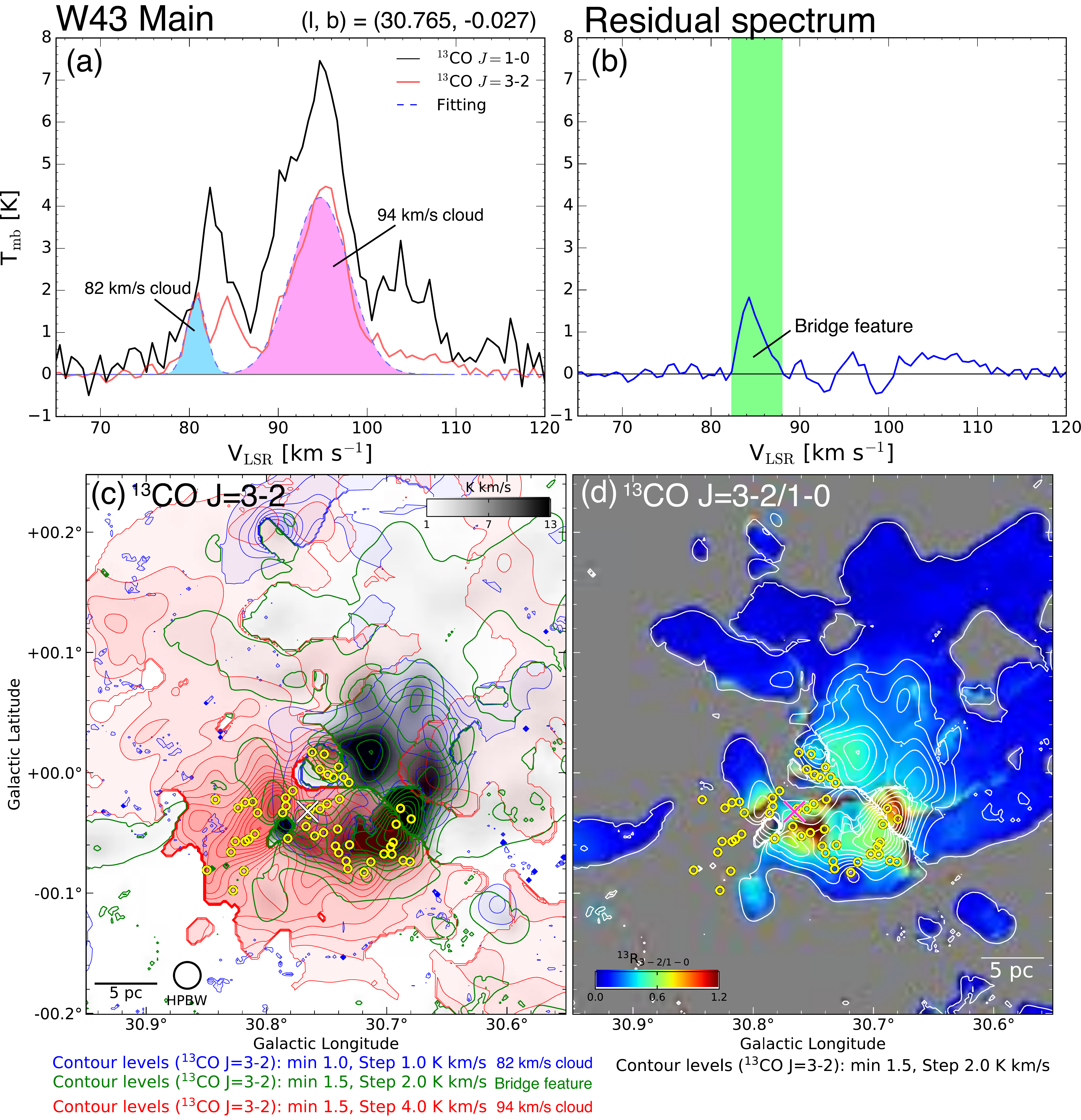}
\end{center}
\caption{{(a) The spectra and the Gaussian fitting results of the $^{13}$CO $J=$3-2 emission in W43 Main. (b) The residual spectrum of $^{13}$CO $J=$3-2.  The green area indicates the integrated velocity range of the bridge feature. (c) The spatial distributions of bridges (green contours) superposed on the two clouds (blue and red contours) in W43 Main. (d) The $R^{13}_{\rm 3-2/1-0}$ map of bridges in W43 Main. {The clipping level  is adopted as {1.5} K km s$^{-1}$. The data was smoothed to be a spatial resolution of $\sim$ \timeform{80"}.}
Plots and symbols are the same as Figure \ref{W43_spitzer} (b).}}
\label{bridge_all}
\end{figure*}

\begin{figure*}[h]
\begin{center} 
 \includegraphics[width=18cm]{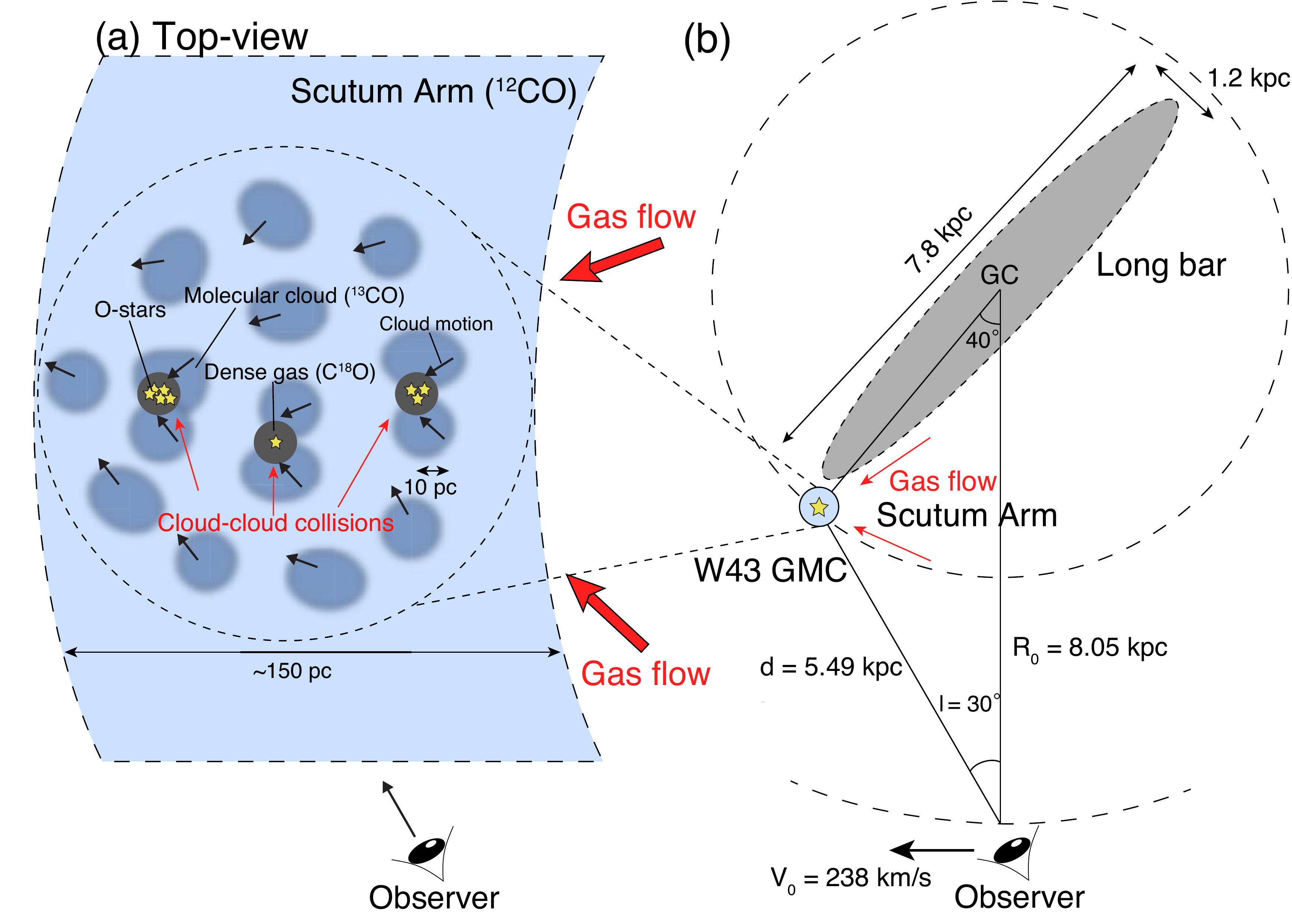}
\end{center}
\caption{Schematic picture of our proposed dense gas and O-type star formation scenario of the {W43 GMC complex in (a) $\sim$ 100 pc scale} and (b) {the} Galactic-scale.
 The Galactic-scale illustration is based on Figure 11 of Sofue et al ({2019}). The distance to the Galactic center ($R_{0}$) and W43 ($d$) are adopted from the VLBI astrometry results obtained by Honma et al. (2012) and Zhang et al. (2014), respectively.  {The solar constant ($V_{0}$) is adopted from Honma et al. (2012)}. The long-bar parameters are adopted as 7.8 kpc of the major axis, 1.2 kpc of the minor axis, and \timeform{43D} of the position angle from the Galactic center-Sun axis based on L\'opez-Corredoira et al. (2007). }
 \label{W43_scenario}
\end{figure*}

\clearpage
\section{CO velocity channel maps of W43 Main, G30.5, and W43 South with FUGIN}
We show the velocity channel maps of the $^{12}$CO, $^{13}$CO, C$^{18}$O $J =$ 1--0 emissions toward W43 Main, G30.5, and W43 South, respectively. The velocity range is between 66.1 and 124.6 $\>$km s$^{-1}$.

\begin{figure*}[h]
\begin{center} 
 \includegraphics[width=13cm]{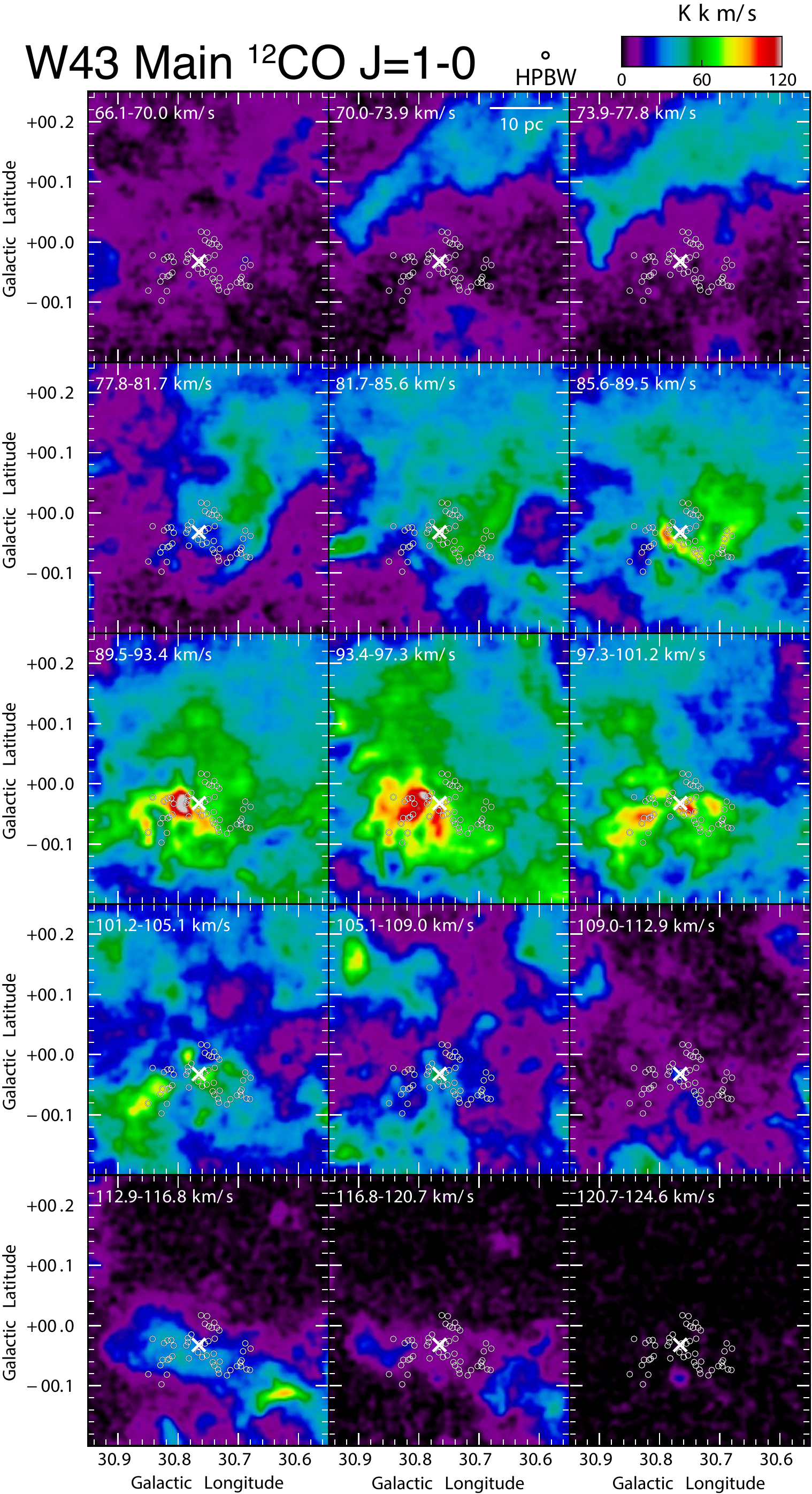}
\end{center}
\caption{Velocity channel map of the $^{12}$CO $J=$ 1--0 emission focused on W43 Main with a velocity step of  3.9 $\>$km s$^{-1}$. The white crosses indicate the W43 Main cluster (Blum et al. 1999). The white circles present the 51 compact fragments (W43 MM1-MM51) cataloged by Motte et al. (2003). The final beam size after convolution is \timeform{40"}.}
\label{W43Main_ch1}
\end{figure*}

\begin{figure*}[h]
\begin{center} 
 \includegraphics[width=13cm]{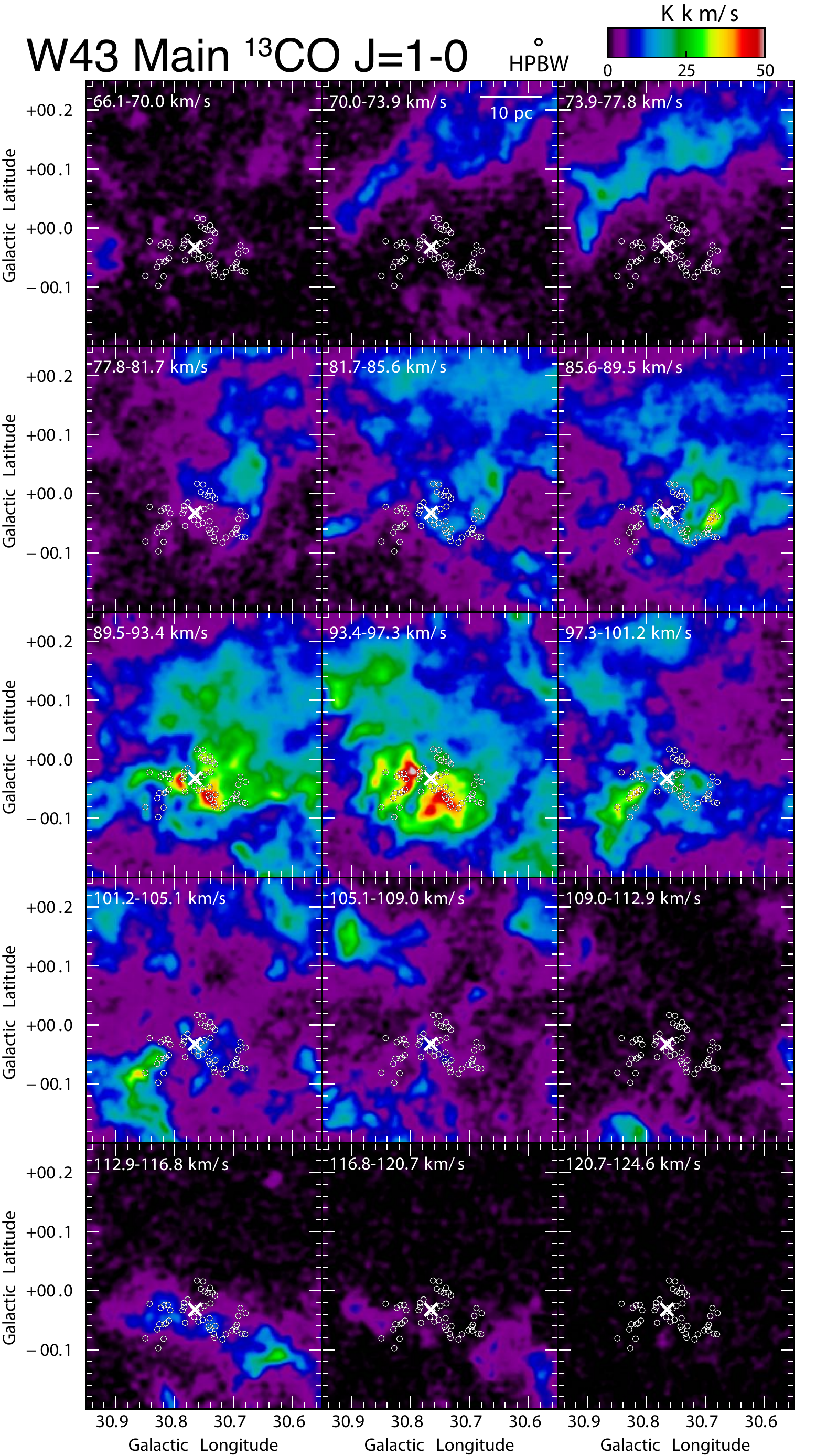}
\end{center}
\caption{{Same as Figure \ref{W43Main_ch1}, but for $^{13}$CO $J=$1--0.}}\label{W43Main_ch2}
\end{figure*}

\begin{figure*}[h]
\begin{center} 
 \includegraphics[width=13cm]{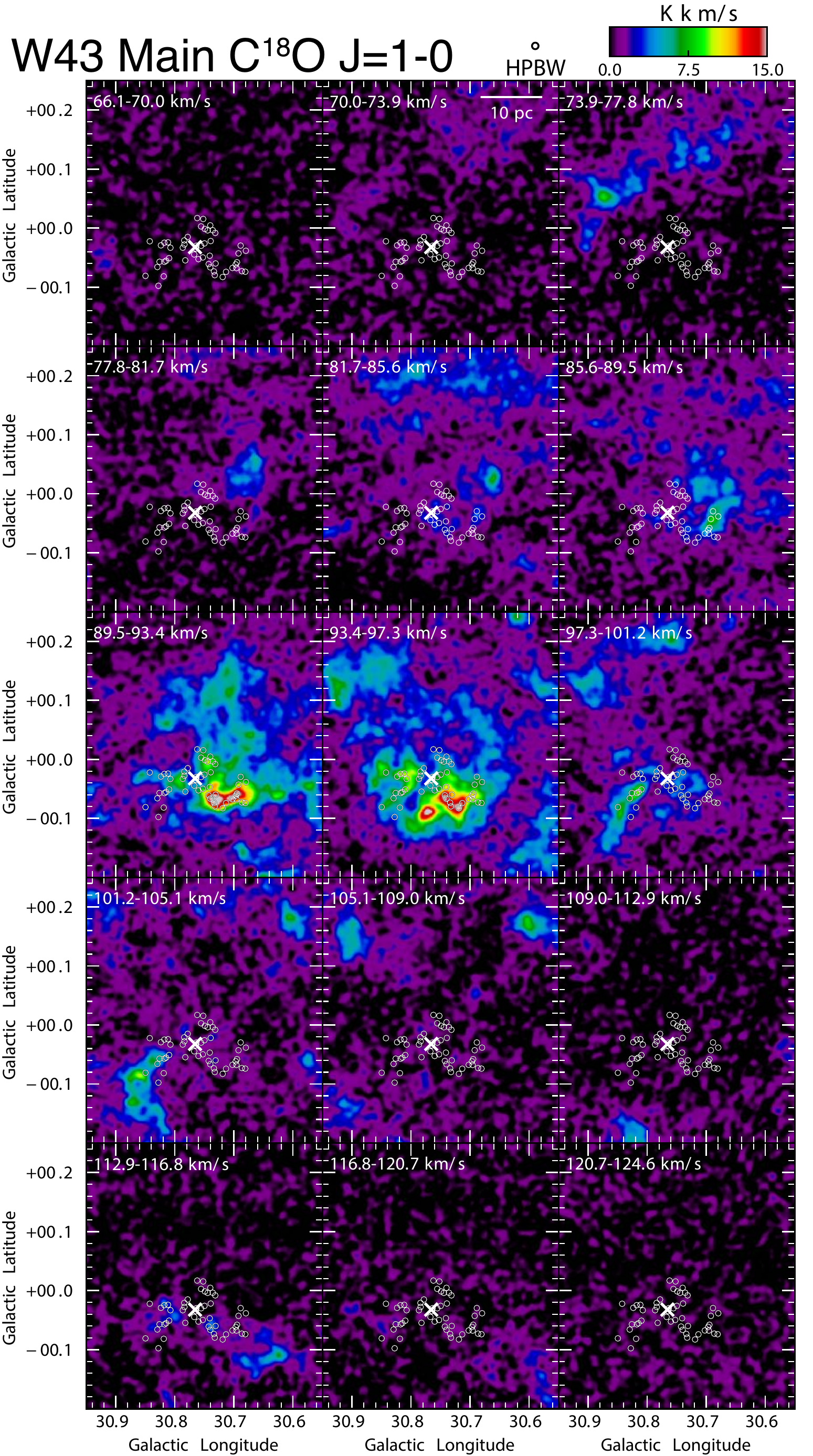}
\end{center}
\caption{{Same as Figure \ref{W43Main_ch1}, but for C$^{18}$O $J=$1--0.}}\label{W43Main_ch3}
\end{figure*}

\begin{figure*}[h]
\begin{center} 
 \includegraphics[width=13cm]{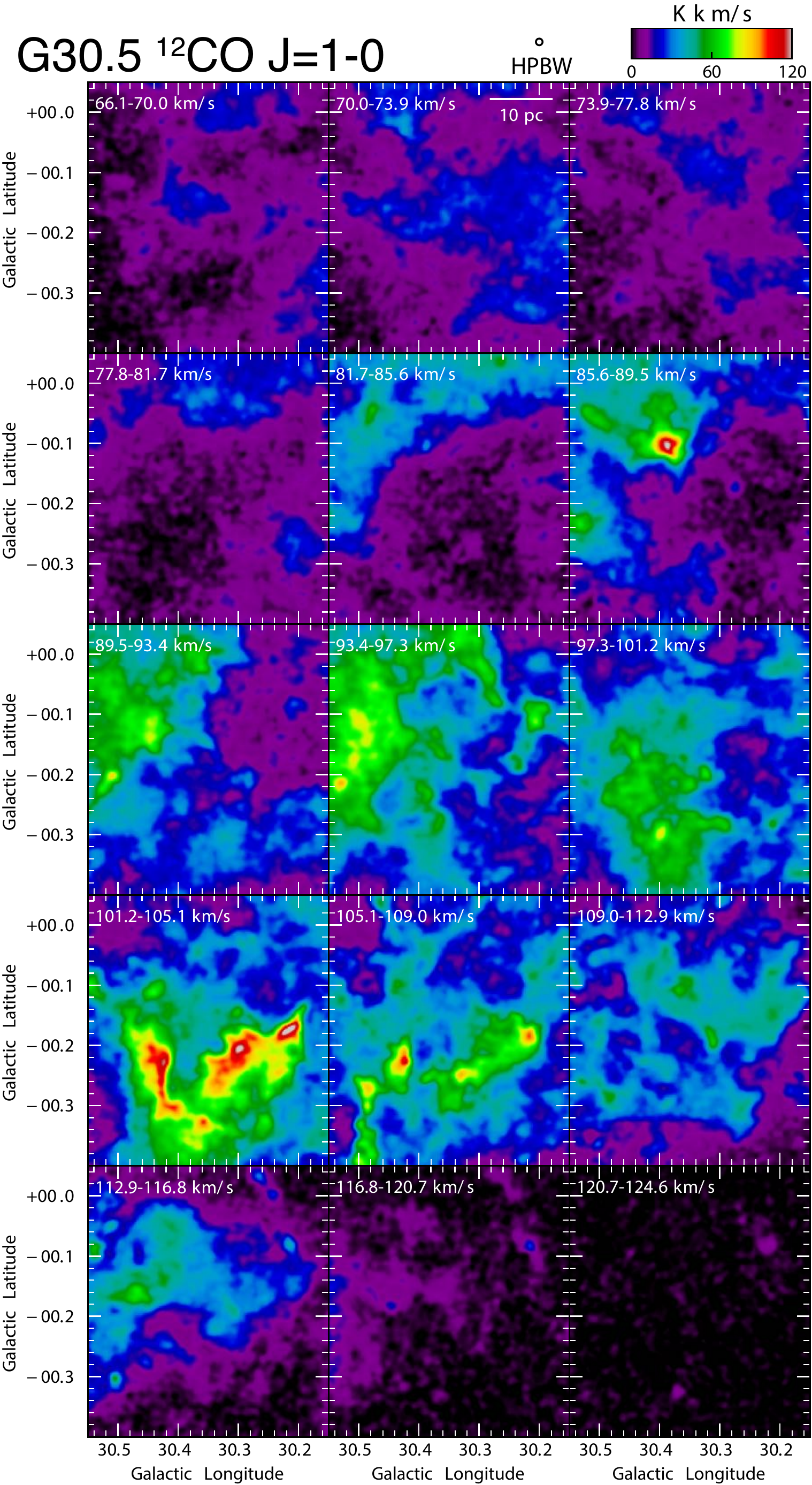}
\end{center}
\caption{Velocity channel map of the $^{12}$CO $J=$ 1--0 emission focused on G30.5 with a velocity step of  3.9 $\>$km s$^{-1}$. The final beam sizes after convolution is \timeform{40"}.}
\label{G30.5_ch1}
\end{figure*}

\begin{figure*}[h]
\begin{center} 
 \includegraphics[width=13cm]{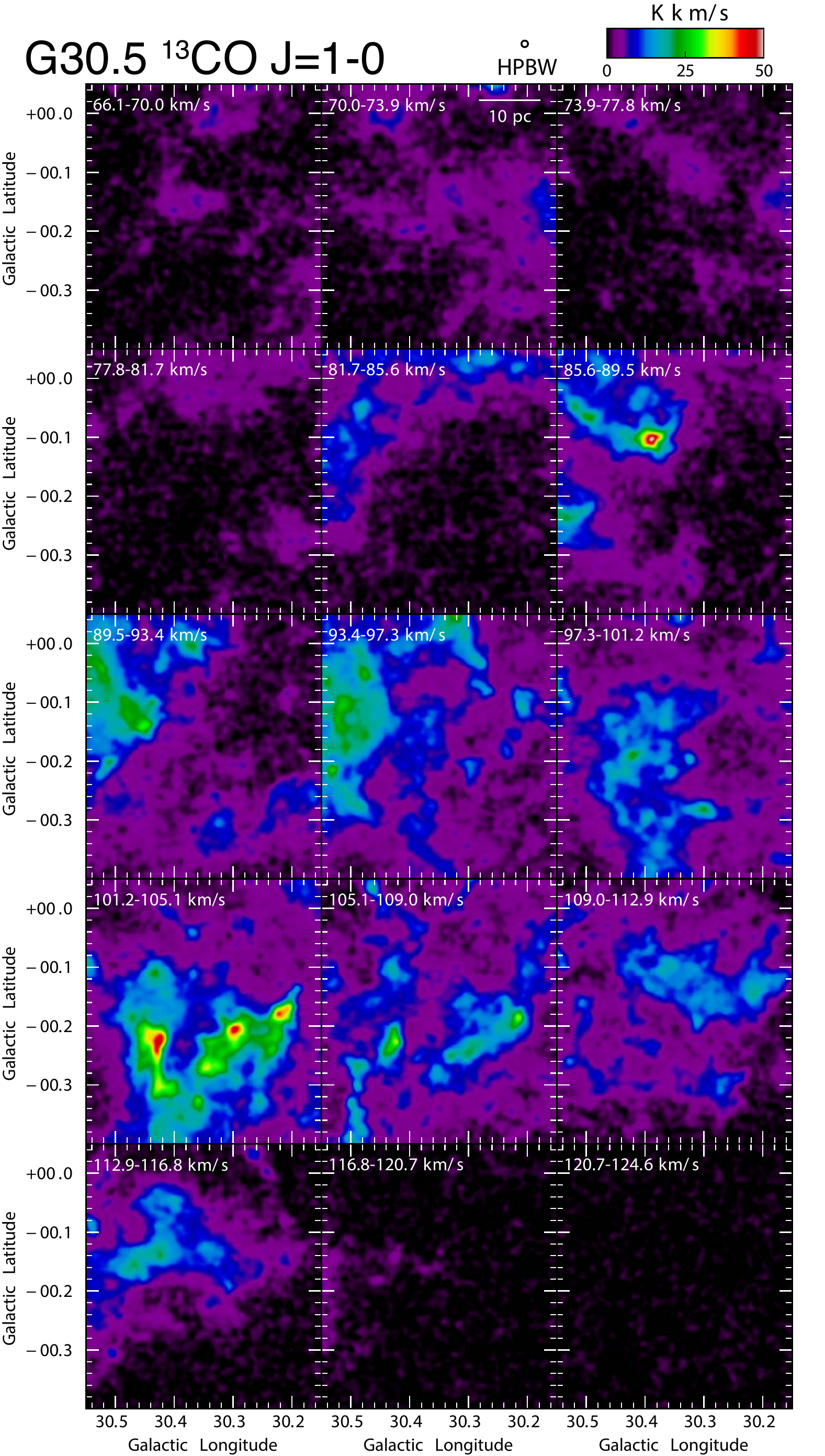}
\end{center}
\caption{{Same as Figure \ref{G30.5_ch1}, but for $^{13}$CO $J=$1--0.}}\label{G30.5_ch2}
\end{figure*}

\begin{figure*}[h]
\begin{center} 
 \includegraphics[width=13cm]{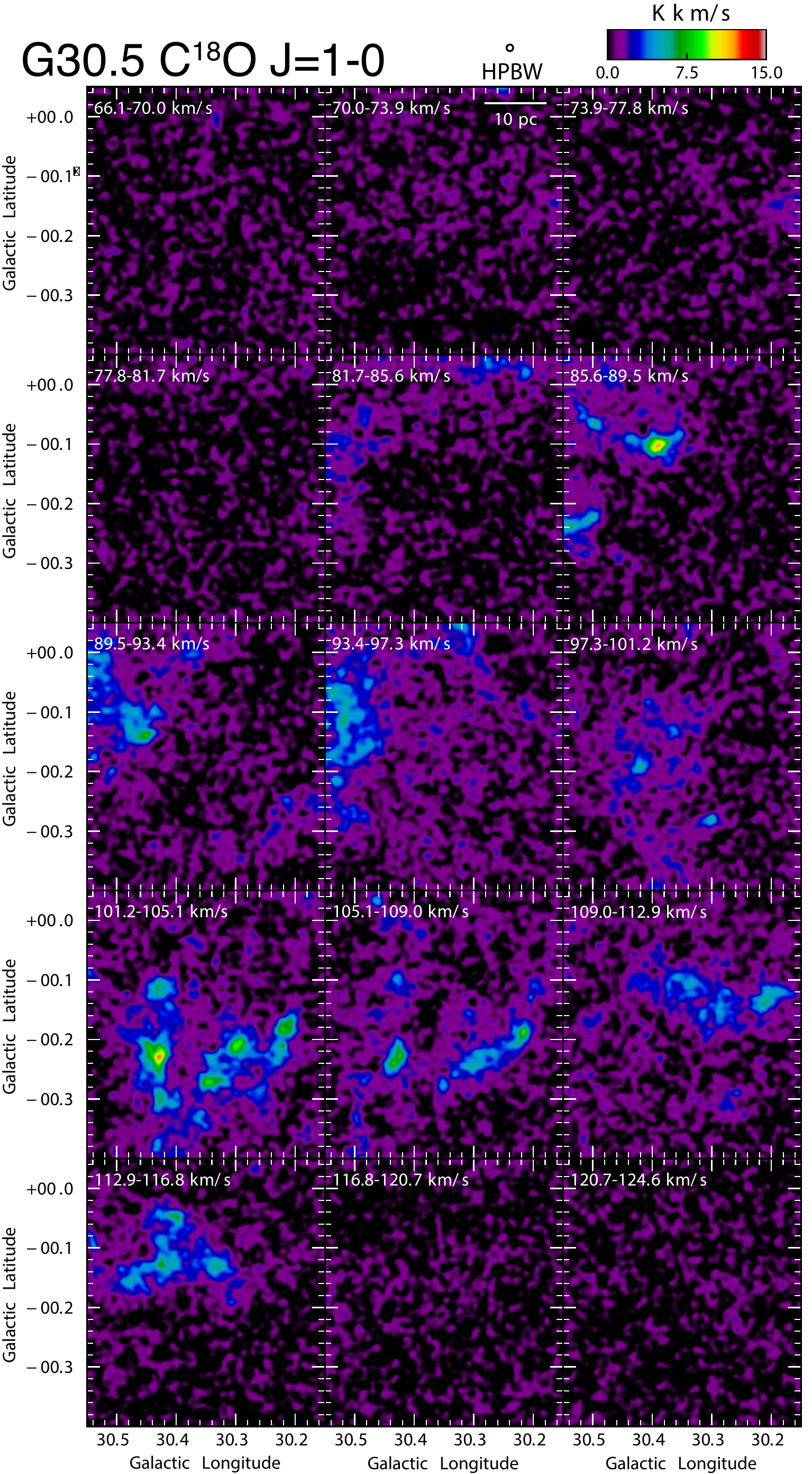}
\end{center}
\caption{{Same as Figure \ref{G30.5_ch1}, but for C$^{18}$O $J=$1--0.}}\label{G30.5_ch3}
\end{figure*}

\begin{figure*}[h]
\begin{center} 
 \includegraphics[width=13cm]{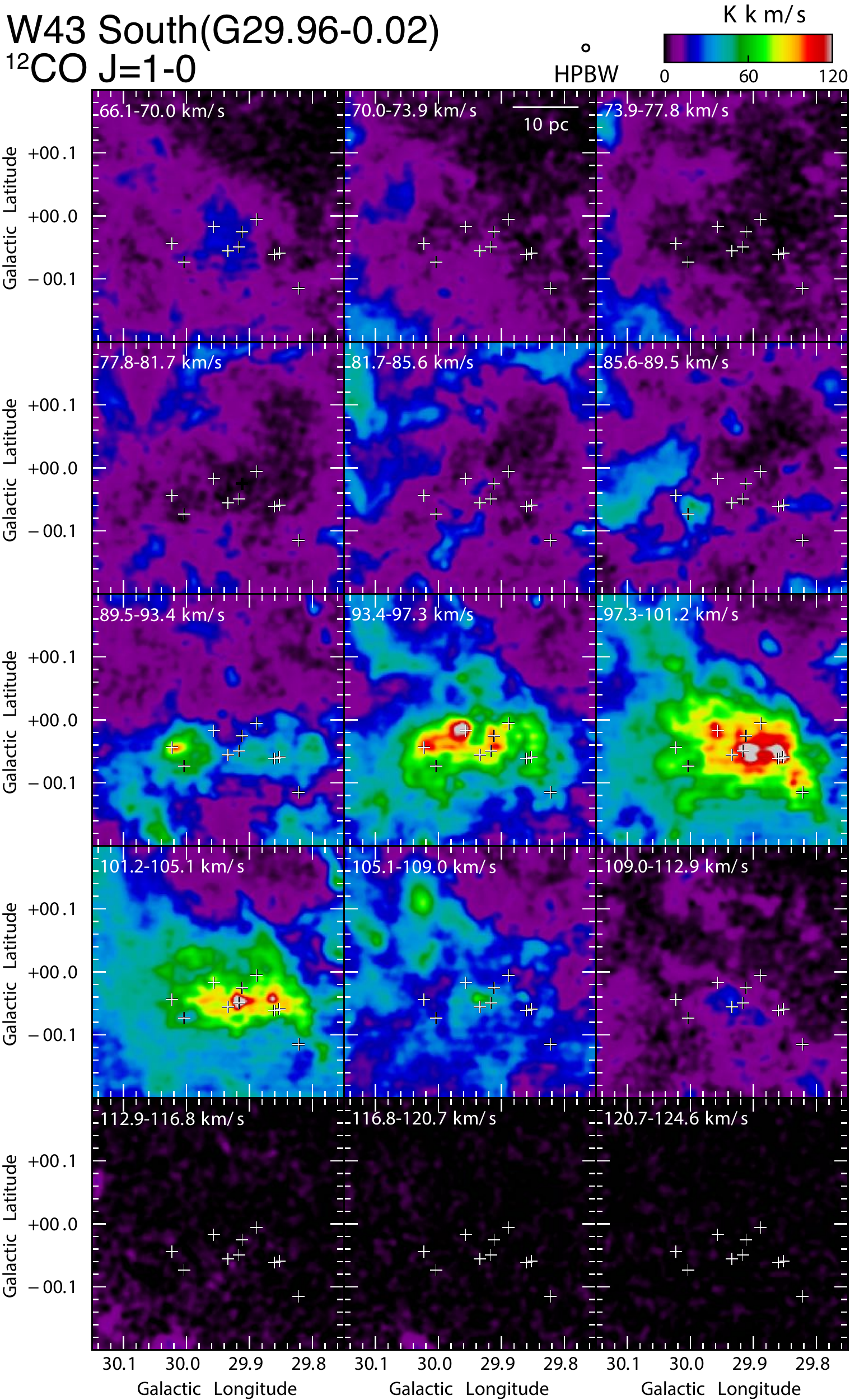}
\end{center}
\caption{Velocity channel map of the $^{12}$CO $J=$ 1--0 emission focused on W43 South with a velocity step of  3.9 $\>$km s$^{-1}$. The crosses indicate the radio continuum sources identified by the NVSS survey (Condon et al. 1998). The final beam size after convolution is \timeform{40"}.}
\label{W43South_ch1}
\end{figure*}

\begin{figure*}[h]
\begin{center} 
 \includegraphics[width=13cm]{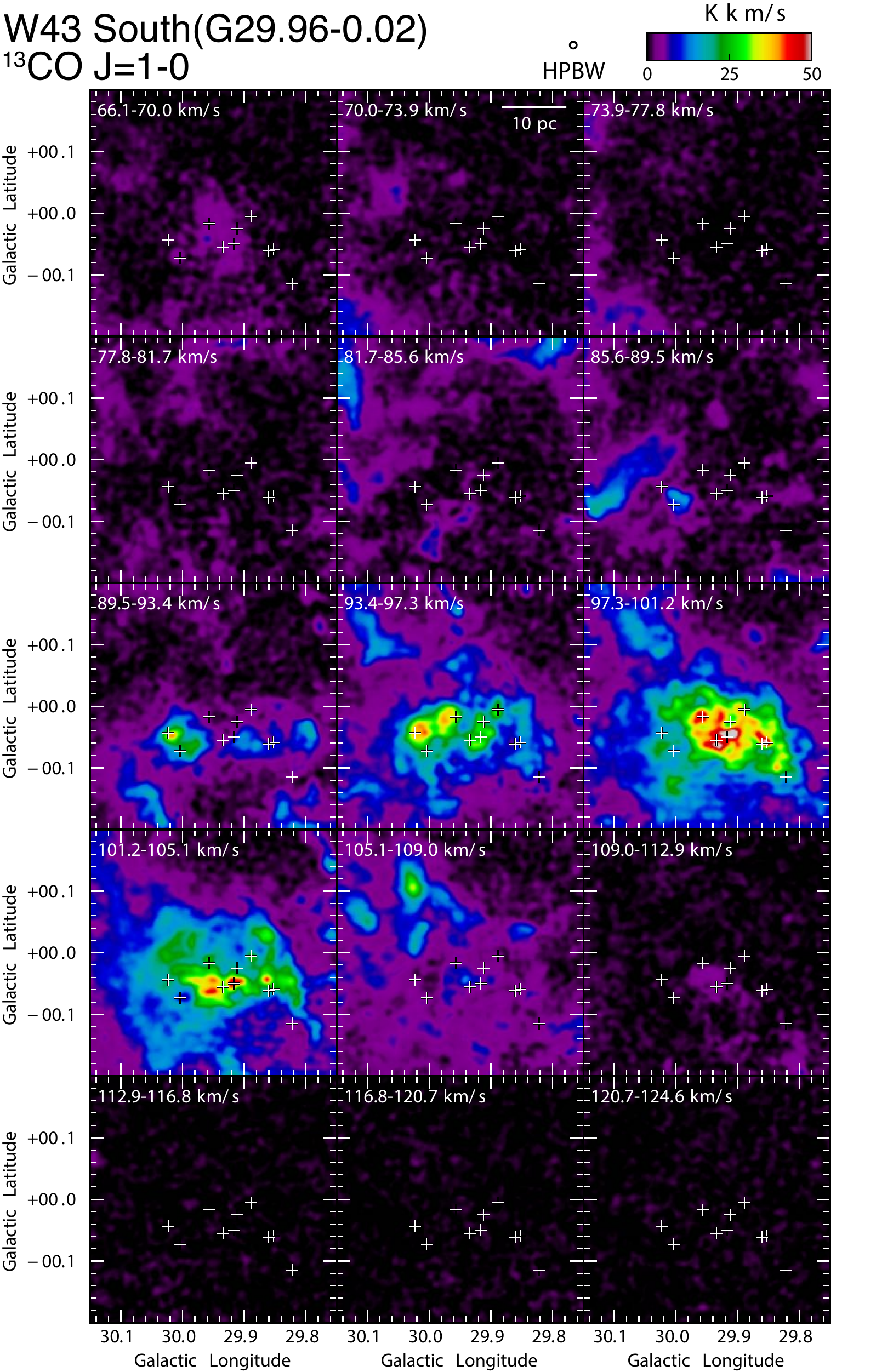}
\end{center}
\caption{{Same as Figure \ref{W43South_ch1}, but for $^{13}$CO $J=$1--0.}}\label{W43South_ch2}
\end{figure*}

\begin{figure*}[h]
\begin{center} 
 \includegraphics[width=13cm]{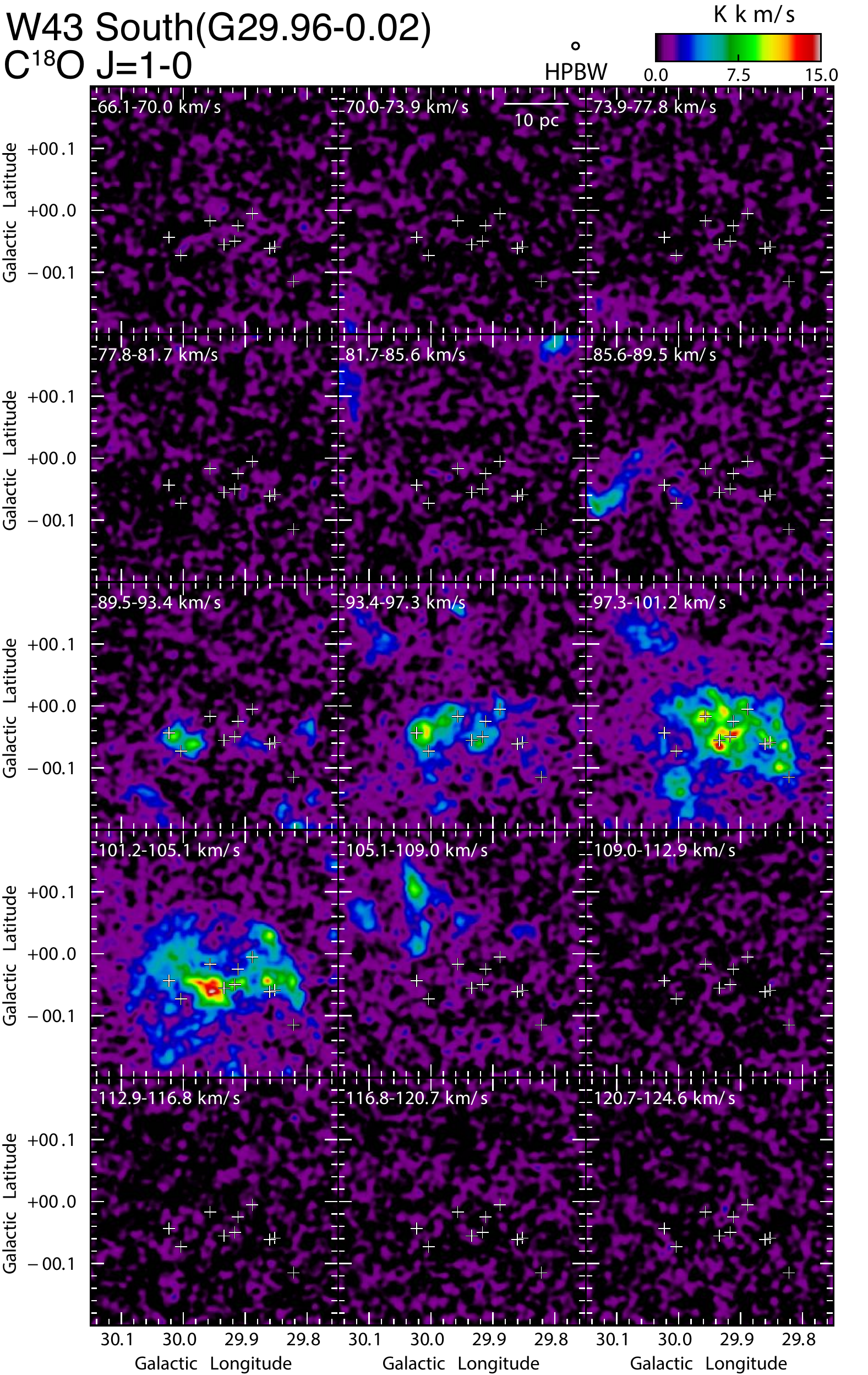}
\end{center}
\caption{{Same as Figure \ref{W43South_ch1}, but for C$^{18}$O $J=$1--0.}}\label{W43South_ch3}
\end{figure*}


\begin{thebibliography}{}
 \bibitem[Aauthor \& Bauthor(2003a)]{key-2}
 Afflerbach, A., Churchwell, E., Hofner, P., \& Kurtz, S., 1994, ApJ, 437, 697
   \bibitem[Aauthor \& Bauthor(2003a)]{key-2}
 Anathpindika, S. V. 2010, MNRAS, 405, 1431
  \bibitem[Aauthor \& Bauthor(2003a)]{key-2}
 Anderson, L. D., Bania, T. M.,  Balser, D. S., Cunningham, V., Wenger, T. V., Johnstone, B. M., \& Armentrout, W. P., 2014, ApJS, 221, 1
 \bibitem[Aauthor \& Bauthor(2003a)]{key-2}
 Astropy Collaboration, et al. 2013, A\&A, 558, A33
  \bibitem[Aauthor \& Bauthor(2003a)]{key-2}
 Astropy Collaboation, et al. 2018, AJ, 156, 123
  \bibitem[Aauthor \& Bauthor(2003a)]{key-2}
 Balfour, S. K., Whitworth, A. P., \& Hubber, D. A. 2017, MNRAS, 465, 3483
 \bibitem[Aauthor \& Bauthor(2003a)]{key-2}
Balfour, S. K., Whitworth, A. P., Hubber, D. A., \& Jaffa, S. E. 2015, MNRAS, 453, 2471
  \bibitem[Aauthor \& Bauthor(2003a)]{key-2}
 Ball, R., Meixner, M. M., Keto, E.,  Arens, J. F., \& Jernigan, J. G., 1996, ApJ, 112, 1645
 \bibitem[Bally, \& Langer(1982)]{1982ApJ...255..143B} 
  {Bally, J., \& Langer, W.~D.\ 1982, \apj, 255, 143}
 \bibitem[Aauthor \& Bauthor(2003a)]{key-2}
Bally, J., et al., 2010, A\&A, 518, L90
 \bibitem[Aauthor \& Bauthor(2003a)]{key-2}
Balser, D. S., Goss, W. M., \& De Pree, C. G., 2001, AJ, 121, 371
 \bibitem[Aauthor \& Bauthor(2003a)]{key-2}
Baug, T., Dewangan, L. K., Ojha, D. K., \& Ninan, J. P., 2016, ApJ, 833, 85
 \bibitem[Aauthor \& Bauthor(2003a)]{key-2}
Beichman, C. A., Neugebauer, G., Habing, H. J., Clegg, P. E., \& Chester, T. J., 1988, Infrared Astronomical Satellite (IRAS) Catalogs and Atlases, Vol. 1, Explanatory Supplement (Washington, D.C.: GPO)
  \bibitem[Benjamin et~al.(2003)]{Benjamin2003}
Beltr\'an, M. T.,Cesaroni, R.,Neri, R., \& Codella, C. 2011, A\&A,525,A151
 \bibitem[Aauthor \& Bauthor(2003a)]{key-2}
Beltr\'an, M. T., et al., 2013, A\&A, 552, A123
  \bibitem[Benjamin et~al.(2003)]{Benjamin2003}
Benjamin, R. A., et al. 2003, PASP, 115, 953
  \bibitem[Benjamin et~al.(2003)]{Benjamin2003}
Beuther, H., et al. 2012, A\&A, 538, 11
  \bibitem[Benjamin et~al.(2003)]{Benjamin2003}
Beuther, H., Meidt, S., Schinnerer, E., Paladino, R., \& Leroy, A., 2017, A\&A, 597, A85
 \bibitem[Aauthor \& Bauthor(2003a)]{key-2}
Bialy, S., Bihr, S., Beuther, H., Henning, T., \& Sternberg, A., 2017, ApJ, 835, 126
 \bibitem[Aauthor \& Bauthor(2003a)]{key-2}
Bihr, S, et al., 2015, A\&A, 580, A112
 \bibitem[Aauthor \& Bauthor(2003a)]{key-2}
Bik, A., Henning, Th., Wu, S. -W., Zhang, M., Brandner, W., Pasquali, A., \& Stolte, A., 2019, A\&A, 624, A63
 \bibitem[Aauthor \& Bauthor(2003a)]{key-2}
Binder, B. A., \& Povich, M. S., 2018, ApJ, 864, 136
 \bibitem[Aauthor \& Bauthor(2003a)]{key-2}
Blitz, L. 1993, in Protostars and Planets III, ed. E. H. Levy \& J. I. Lunine (Tucson: University of Arizona Press), 125
 \bibitem[Aauthor \& Bauthor(2003a)]{key-2}
Blitz, L., Fukui, Y., Kawamura, A., Leroy, A., Mizuno, N., \& Rosolowsky, E. 2007, in Protostars and Planets V, ed. B. Reipurth et al. (Tucson, AZ: University of Arizona Press), 81
 \bibitem[Aauthor \& Bauthor(2003a)]{key-2}
Blum, R. D., Damineli, A., \& Conti, P. S., 1999, AJ, 117, 1392
 \bibitem[Aauthor \& Bauthor(2003a)]{key-2}
Bolatto, A. D., Wolfire, M., \& Leroy, A. K. 2013, ARA\&A, 51, 207
  \bibitem[Benjamin et~al.(2003)]{Benjamin2003}
Braiding, C., et al. 2018, PASA, 35, 29
 \bibitem[Aauthor \& Bauthor(2003a)]{key-2}
Buckle, J. V., et al. 2009, MNRAS, 399, 1026
  \bibitem[Benjamin et~al.(2003)]{Benjamin2003}
Burton, M., et al. 2013, PASA, 30, 44
  \bibitem[Aauthor \& Bauthor(2003a)]{key-2} 
Carey, S. J., et al. 2009, PASP, 121, 76
\bibitem[Aauthor \& Bauthor(2003a)]{key-2}
Carlhoff, P, et al., 2013, A\&A, 560, 24
\bibitem[Aauthor \& Bauthor(2003a)]{key-2}
Carpenter, John M., \& Sanders, D. B., 1998, AJ, 116, 1856
  \bibitem[Aauthor \& Bauthor(2003a)]{key-2} 
Cesaroni, R., Churchwell, E., Hofner, P., Walmsley, C. M., \& Kurtz, S., 1994, A\&A, 288, 903
  \bibitem[Aauthor \& Bauthor(2003a)]{key-2} 
Cesaroni, R., Hofner, P., Walmsley, C. M., \& Churchwell, E., 1998, A\&A, 331, 709
  \bibitem[Aauthor \& Bauthor(2003a)]{key-2} 
Cesaroni, R., et al. 2017, A\&A, 602, 59
\bibitem[Churchwell, E., et al(2006)]{key-2}
Chibueze, J. O., et al. 2016, MNRAS, 460, 1839
\bibitem[Churchwell, E., et al(2006)]{key-2}
Chini, R., \& Hoffmeister, V. 2008, in Handbook of Star Forming Regions, Volume II: The Southern Sky, ed. B. Reipurth (San Francisco: ASP), 625
\bibitem[Churchwell, E., et al(2006)]{key-2}
Churchwell, E., et al. 2006, ApJ, 649, 759
\bibitem[Churchwell, E., et al(2006)]{key-2}
{Churchwell, E., et al. 2009, PASP, 121, 213}
\bibitem[Aauthor \& Bauthor(2003a)]{key-2}
Combes, F. 1991, ARA\&A, 29, 195
\bibitem[Aauthor \& Bauthor(2003a)]{key-2}
Condon, J. J., Cotton, W. D., Greisen, E. W., Yin, Q. F., Perley, R. A., Taylor, G. B., \& Broderick, J. J., 1998, AJ, 115, 1693
\bibitem[Aauthor \& Bauthor(2003a)]{key-2}
Cortes, P., \& Crutcher, R. M., 2006, ApJ, 639, 965
\bibitem[Aauthor \& Bauthor(2003a)]{key-2}
Cortes, P. C., Parra, R., Cortes, J. R., \& Hardy, E., 2010, A\&A, 519, A35
\bibitem[Aauthor \& Bauthor(2003a)]{key-2}
Cortes, P. C., 2011, ApJ, 743, 194
\bibitem[Aauthor \& Bauthor(2003a)]{key-2}
Cortes, P. C., et al., 2016, ApJL, 825, 15
\bibitem[Aauthor \& Bauthor(2003a)]{key-2}
Cortes, P. C., et al., 2019, ApJ, 884, 48
\bibitem[Churchwell, E., et al(2006)]{key-2}
{Dale, J. E., Ngoumou, J., Ercolano, B., \& Bonnell, I., 2013, MNRAS, 436, 3430}
\bibitem[Churchwell, E., et al(2006)]{key-2}
{Dame, T. M., Elmegreen, B. G., Cohen, R. S., \& Thaddeus, P., 1986, ApJ, 305, 892}
\bibitem[Churchwell, E., et al(2006)]{key-2}
{Dame, T. M., \& Thaddeus, P., 2008, ApJ, 683, L143}
\bibitem[Churchwell, E., et al(2006)]{key-2}
{Dame, T. M., Hartmann, D., \& Thaddeus, P., 2001, ApJ, 547, 792}
\bibitem[Churchwell, E., et al(2006)]{key-2}
De Buizer, J. M, Watson, A. M., Radomski, J. T., Pi\~na, R. K., \& Telesco, C. M., 2002, ApJ, 546, 101
\bibitem[Churchwell, E., et al(2006)]{key-2}
{Deharveng, L., et al. 2010, A\&A, 523, A6}
\bibitem[Churchwell, E., et al(2006)]{key-2}
Dempsey, J. T., Thomas, H. S., \& Currie, M. J. 2013, ApJS, 209, 8
\bibitem[Churchwell, E., et al(2006)]{key-2}
Dewangan, L. K., \& Ojha, D. K., 2017, ApJ, 849, 65
\bibitem[Churchwell, E., et al(2006)]{key-2}
Dewangan, L. K., Ojha, D. K., \& Zinchenko, I., 2017, ApJ, 851, 140
\bibitem[Churchwell, E., et al(2006)]{key-2}
Dewangan, L. K., Dhanya, J. S., Ojha, D. K., \& Zinchenko, I. 2018, ApJ, 866, 20
\bibitem[Churchwell, E., et al(2006)]{key-2}
Dewangan, L. K., Sano, H., Enokiya, R., Tachihara, K., Fukui, Y., \& Ojha, K., 2019, ApJ, 878, 26
\bibitem[Churchwell, E., et al(2006)]{key-2}
Dobashi, K., Shimoikura, T., Katakura, S., Nakamura, F., \& Shimajiri, Y., 2019, PASJ, 71, S12
 \bibitem[Aauthor \& Bauthor(2003a)]{key-2}
Draine, B. T. 2003, ARA\&A, 41, 241
 \bibitem[Aauthor \& Bauthor(2003a)]{key-2}
Draine, B. T., \& Li, A. 2007, ApJ, 657, 810
 \bibitem[Aauthor \& Bauthor(2003a)]{key-2}
Eden, D. J., Moore, T. J. T., Plume, R., \& Morgan, L. K., 2012, MNRAS, 422, 3178
 \bibitem[Aauthor \& Bauthor(2003a)]{key-2}
Elmegreen, B. G. 1998, in ASP Conf. Ser., 148, Origins, ed. C. E., Woodward et al. (San Francisco: ASP), 150
   \bibitem[Aauthor \& Bauthor(2003a)]{key-2}
Enokiya, R., et al., 2018, PASJ, 70S, 49
 \bibitem[Aauthor \& Bauthor(2003a)]{key-2}
Evans, N. J., et al., 2009, ApJS, 181, 321
 \bibitem[Aauthor \& Bauthor(2003a)]{key-2}
Fazio, G. G., et al., 2004, ApJS, 154, 10
 \bibitem[Aauthor \& Bauthor(2003a)]{key-2}
Fey, A. L., Caume, R. A., Claussen, M. J., \& Vrba, F., 1995, ApJ, 453, 308
 \bibitem[Aauthor \& Bauthor(2003a)]{key-2}
Frerking, M. A., Langer, W. D., \& Wilson, R. W. 1982, ApJ, 262, 590
 \bibitem[Aauthor \& Bauthor(2003a)]{key-2}
Fujimoto, M. 1968, in IAU Symp. 29, Nonstable Phenomena in Galaxies, ed. M. Arakeljan (Yerevan: Publishing House of the Academy of Sciences of Armenian SSR), 453.
 \bibitem[Aauthor \& Bauthor(2003a)]{key-2}
{Fujimoto, Y., Tasker., E. J., Wakayama, M., \& Habe, A. 2014a, MNRAS, 439, 936}
 \bibitem[Aauthor \& Bauthor(2003a)]{key-2}
Fujita, S., et al., 2019a, ApJ, 872, 49
 \bibitem[Aauthor \& Bauthor(2003a)]{key-2}
Fujita, S., et al., 2019b, PASJ, in press (DOI: 10.1093/pasj/psz028)
 \bibitem[Aauthor \& Bauthor(2003a)]{key-2}
Fukui, Y., \& Kawamura, A., 2010, ARA\&A, 48, 547
 \bibitem[Aauthor \& Bauthor(2003a)]{key-2}
{Fukui, Y., et al., 2012, ApJ, 746, 82}
 \bibitem[Aauthor \& Bauthor(2003a)]{key-2}
Fukui, Y., et al., 2014, ApJ, 780, 36
 \bibitem[Aauthor \& Bauthor(2003a)]{key-2}
Fukui, Y., et al., 2016, ApJ, 820, 26
 \bibitem[Aauthor \& Bauthor(2003a)]{key-2}
Fukui, Y., et al., 2018a, PASJ, 70S, 41
 \bibitem[Aauthor \& Bauthor(2003a)]{key-2}
Fukui, Y., et al., 2018b, PASJ, 70S, 44
 \bibitem[Aauthor \& Bauthor(2003a)]{key-2}
Fukui, Y., et al., 2018c, PASJ, 70S, 46
 \bibitem[Aauthor \& Bauthor(2003a)]{key-2}
Fukui, Y., et al., 2018d, ApJ, 859, 166
 \bibitem[Aauthor \& Bauthor(2003a)]{key-2}
{Fukui, Y., Inoue, T., Hayakawa, T., \& Torii, K., 2019, PASJ, submitted, (arXiv: 1909.08202)}
 \bibitem[Aauthor \& Bauthor(2003a)]{key-2}
Furukawa, N., Dawson, J. R., Ohama, A., Kawamura, A., Mizuno, N., Onishi, T., \& Fukui, Y. 2009, ApJ, 696, L115
 \bibitem[Aauthor \& Bauthor(2003a)]{key-2} 
Gao, Y., \& Solomon, P. M. 2004, ApJ, 606, 271
\bibitem[Aauthor \& Bauthor(2003a)]{key-2}
G\'enova-Santos, R., et al., 2017, MNRAS, 464, 4107
\bibitem[Aauthor \& Bauthor(2003a)]{key-2}
Ginsburg, A., Bally, J., Battersby, C., Youngblood, A., Darling, J., Rosolowsky, E., Arce, H., \& Lebron, S. M. E., 2015, A\&A 573, 106
 \bibitem[Aauthor \& Bauthor(2003a)]{key-2} 
Goldreich, P., \& Kwan, J. 1974, ApJ, 189, 441
 \bibitem[Aauthor \& Bauthor(2003a)]{key-2}
Habe, A., \& Ohta, K. 1992, PASJ, 44, 203
 \bibitem[Aauthor \& Bauthor(2003a)]{key-2}
Hanaoka, M., et al., 2019, PASJ, 71, 6
 \bibitem[Aauthor \& Bauthor(2003a)]{key-2}
Handa, T., Sofue, Y., Nakai, N., Hirabayashi, H., \& Inoue, M. 1987, PASJ, 39, 709
 \bibitem[Aauthor \& Bauthor(2003a)]{key-2}
Harayama, Y., Eisenhauer, F., \& Martins, F., 2008, ApJ, 675, 1319
 \bibitem[Aauthor \& Bauthor(2003a)]{key-2}
Hasegawa, T., Sato, F., Whiteoak, J. B., \& Miyawaki, R. 1994, ApJ, 429, L77
 \bibitem[Aauthor \& Bauthor(2003a)]{key-2}
Hattori, Y., et al., 2016, PASJ, 68, 37
 \bibitem[Aauthor \& Bauthor(2003a)]{key-2}
Haworth, T. J., et al. 2015a, MNRAS, 450, 10
 \bibitem[Aauthor \& Bauthor(2003a)]{key-2}
Haworth, T. J., Shima, K., Tasker, E. J., Fukui, Y., Torii, K., Dale, J. E., Takahira, K., \& Habe, A. 2015b, MNRAS, 454, 1634
  \bibitem[Aauthor \& Bauthor(2003a)]{key-2}
Hayashi, K., et al., 2018, PASJ, 70S, 48
 \bibitem[Aauthor \& Bauthor(2003a)]{key-2}
Herpin, F., et al., 2012, A\&A, 542, 76
 \bibitem[Aauthor \& Bauthor(2003a)]{key-2}
Heyer, M. H., \& Brunt, C. M., 2004, ApJ, 615, 45
 \bibitem[Aauthor \& Bauthor(2003a)]{key-2}
Heyer, M., \& Dame, T. M., 2015, ARA\&A, 53, 583
 \bibitem[Aauthor \& Bauthor(2003a)]{key-2}
Hoffman, I. M., Goss, W. M., Palmer, P., \& Richards, A. M. S., 2003, ApJ, 598, 1061
 \bibitem[Aauthor \& Bauthor(2003a)]{key-2}
Hoffmeister, V. H., Chini, R., Scheyda, C. M., Schulze, D., Watermann, R., Nurnberger, D., \& Vogt, N., 2008, ApJ, 686, 310 
 \bibitem[Aauthor \& Bauthor(2003a)]{key-2}
Honma, M., et al., 2012, PASJ, 64, 136
 \bibitem[Aauthor \& Bauthor(2003a)]{key-2}
Hou, L. G., \& Han, J. L., 2014, A\&A, 569, A125
 \bibitem[Aauthor \& Bauthor(2003a)]{key-2}
Hunter, J. D., 2007, Comput. Sci. Eng., 9, 90
 \bibitem[Aauthor \& Bauthor(2003a)]{key-2}
Inoue, T., \& Fukui, Y. 2013, ApJ, 774, L31
 \bibitem[Aauthor \& Bauthor(2003a)]{key-2}
Inoue, T., Hennebelle, P., Fukui, Y., Matsumoto, T., Iwasaki, K., \& Inutsuka, S. 2018, PASJ, 70S, 53
 \bibitem[Aauthor \& Bauthor(2003a)]{key-2}
Jacq, T., Braine, J., Herpin, F., van der Tak, F., \& Wyrowski, F., 2016, A\&A, 595, 66
  \bibitem[Aauthor \& Bauthor(2003a)]{key-2}
{Kauffmann, J., Bertoldi, F., Burke, T. L., Evens, N.J., \& Lee, C. W., 2008, A\&A, 487, 993}
  \bibitem[Aauthor \& Bauthor(2003a)]{key-2}
Kamazaki, T., et al. 2012, PASJ, 64, 29
\bibitem[Aauthor \& Bauthor(2003a)]{key-2}
Kang, M., Bieging, J. H., Kulesa, C. A., Lee, Y., Choi, M., \& Peters, W. L. 2010, ApJS, 190, 58
\bibitem[Aauthor \& Bauthor(2003a)]{key-2}
Kawamura, A., et al. 2009, ApJ, 184, 1
 \bibitem[Aauthor \& Bauthor(2003a)]{key-2}
Kenney, J. D. P., \& Lord, S. D., 1991, ApJ, 381, 118
\bibitem[Churchwell, E., et al(2006)]{key-2}
{Kirk, J. M., et al. 2010, A\&A, 518, L82}
 \bibitem[Aauthor \& Bauthor(2003a)]{key-2}
Kohno, M., et al., 2018a, PASJ, 70S, 50
 \bibitem[Aauthor \& Bauthor(2003a)]{key-2}
Kohno, M., et al., 2018b, PASJ, in press (doi:10.1093/pasj/psy109)
   \bibitem[Aauthor \& Bauthor(2003a)]{key-2}
Kuno, N., et al. 2011, in Proc. 2011 XXXth URSI General Assembly
and Scientific Symposium (New York: IEEE), 3670 \footnote{${\rm http://ieeexplore.ieee.org/xpl/articleDetails.jsp?arnumber=6051296}$}
   \bibitem[Aauthor \& Bauthor(2003a)]{key-2}
Kutner, M. L., \& Ulich, B. L. 1981, ApJ, 250, 341
\bibitem[Aauthor \& Bauthor(2003a)]{key-2}
Lada, C. J., \& Lada, E. A., 2003, ARA\&A, 41, 57
\bibitem[Aauthor \& Bauthor(2003a)]{key-2}
Lada, C. J., Forbrich, J., Lombardi, M., \& Alves, J. F., 2012, ApJ, 745, 190
\bibitem[Aauthor \& Bauthor(2003a)]{key-2}
Langer, W. D, Velusamy, T., Goldsmith, P. F., Pineda, J. L., Chambers, E. T., Sandell, G., Risacher, C., \& Jacobs, K., 2017, A\&A, 607, A59
\bibitem[Aauthor \& Bauthor(2003a)]{key-2}
Larson, R. B., 1981, MNRAS, 194, 809
\bibitem[Aauthor \& Bauthor(2003a)]{key-2}
Lemoine-Goumard, M., Ferrara, E., Grondin, M.-H., Martin, P., \& Renaud, M, 2011, MmSAI, 82, 739
\bibitem[Aauthor \& Bauthor(2003a)]{key-2}
Lester, D. F., Dinerstein, H. L., Werner, M. W., Harvey, P. M., Evans II, N. J., \& Brown, R. L., 1985, ApJ, 296, 565
\bibitem[Aauthor \& Bauthor(2003a)]{key-2}
Lin, Y, et al., 2016, ApJ, 828, 32
 \bibitem[Aauthor \& Bauthor(2003a)]{key-2}
Liszt, H, S., Braun, R., \& Greisen, E. W., 1993, AJ, 106, 2349L
 \bibitem[Aauthor \& Bauthor(2003a)]{key-2}
Liszt, H, S., 1995, AJ, 109, 1204L
 \bibitem[Aauthor \& Bauthor(2003a)]{key-2}
Lockman, F, J., 1989, ApJS, 71, 469
\bibitem[Aauthor \& Bauthor(2003a)]{key-2}
Longmore, S. N., et al., 2014, Protostars and Planets VI, ed. H. Beuther et al. (Tucson, AZ: University of Arizona Press), 291
 \bibitem[Aauthor \& Bauthor(2003a)]{key-2}
L\'opez-Corredoira, M., Cabrera-Lavers, A., Mahoney, T. J., Hammersley, P. L., Garz\'on, F., \& Gonz\'alex-Fern\'andez, C., 2007, AJ, 133, 154
 \bibitem[Aauthor \& Bauthor(2003a)]{key-2}
Louvet, F, et al., 2014, A\&A, 570, 15
 \bibitem[Aauthor \& Bauthor(2003a)]{key-2}
Louvet, F, et al., 2016, A\&A, 595, 122
 \bibitem[Aauthor \& Bauthor(2003a)]{key-2}
Luisi, M., Anderson, L. D., Balser, D. S., Wenger, T. V., \& Bania, T. M., 2017, ApJ, 849, 117
\bibitem[Aauthor \& Bauthor(2003a)]{key-2}
Lumsden, S. L., \& Hoare, M. G., 1996, ApJ, 464, 272
\bibitem[Aauthor \& Bauthor(2003a)]{key-2}
Lumsden, S. L., \& Hoare, M. G., 1999, MNRAS, 305, 701
\bibitem[Aauthor \& Bauthor(2003a)]{key-2}
Luque-Escamilla, P. L., Mu\~noz-Arjonilla, A. J., S\'anchez-Sutil, J. R., Marti, J., Comb\'i, J. A., \& S\'anchez-Ayaso, E., 2011, A\&A, 532, 92
 \bibitem[Aauthor \& Bauthor(2003a)]{key-2}
Mangum, J. G., \& Shirley, Y. L., 2015, PASP, 127, 266
 \bibitem[Aauthor \& Bauthor(2003a)]{key-2}
Mart\'in-Hern\'andez, N. L., Bik, A., Kaper, L., Tielens, A. G. G. M., \& Hanson, M. N., 2003, A\&A, 405, 175
 \bibitem[Aauthor \& Bauthor(2003a)]{key-2}
Martins, F, Schaerer, D., \& Hillier, D. J., 2005, A\&A, 436, 1049
   \bibitem[Aauthor \& Bauthor(2003a)]{key-2}
Matsumoto,T., Dobashi, K., \& Shimoikura, T., 2015. ApJ, 801, 77
   \bibitem[Aauthor \& Bauthor(2003a)]{key-2}
Maxia, C., Testi, L., Cesaroni, R., \& Walmsley, C. M., 2001, A\&A, 371, 287
 \bibitem[Aauthor \& Bauthor(2003a)]{key-2}
McKee, C. F., \& Ostriker, E. C. 2007, ARA\&A, 45, 565
   \bibitem[Aauthor \& Bauthor(2003a)]{key-2}
{Minamidani, T., et al.  2015, Nobeyama CO Galactic Plane Survey: New Chapter of the Nobeyama 45-m Telescope, in EAS Publications Series, Vol. 75-76, Conditions and Impact of Star Formation, ed. R. Simon, R. Schaaf, \& J. Stutzki, 193, doi:10.1051/eas/1575036}  
\bibitem[Aauthor \& Bauthor(2003a)]{key-2}
{ Minamidani, T., et al. 2016, SPIE Proc., 9914, 99141Z}
 \bibitem[Aauthor \& Bauthor(2003a)]{key-2}
Miyawaki, R., Hayashi, M., \& Hasegawa, T. 1986, ApJ, 305, 353 
 \bibitem[Aauthor \& Bauthor(2003a)]{key-2}
Miyawaki, R., Hayashi, M., \& Hasegawa, T. 2009, PASJ, 61, 39
 \bibitem[Aauthor \& Bauthor(2003a)]{key-2}
Mizuno, A., \& Fukui, Y., 2006, in ASP Conf. Ser., 317, Milky Way Surveys: The Structure and Evolution of our Galaxy, ed. D. Clemens et al. (San Francisco: ASP), 59
 \bibitem[Aauthor \& Bauthor(2003a)]{key-2}
Molet, J, et al., 2019, A\&A, 626, A132
 \bibitem[Aauthor \& Bauthor(2003a)]{key-2}
Morisset, C., Schaerer, D., Mart\'in-Hern\'andez, N. L., Peeters, E., Damour, F., Baluteau, J.-P., Cox, P., \& Roelfsema, P., 2002, A\&A, 386, 558
 \bibitem[Aauthor \& Bauthor(2003a)]{key-2}
{Mooney, T., Sievers, A., Mezger, P. G., Solomon, P. M., Kreysa, E., Haslam, C. G. T., \& Lemke, R., 1995, A\&A, 299, 869}
 \bibitem[Aauthor \& Bauthor(2003a)]{key-2}
Moore, T. J. T., et al., 2015, MNRAS, 453, 4264
 \bibitem[Aauthor \& Bauthor(2003a)]{key-2}
Motte, F, Schilke, P., \& Lis, D. C., 2003, ApJ, 582, 277
 \bibitem[Aauthor \& Bauthor(2003a)]{key-2}
Motte, F, et al., 2014, A\&A, 571, A32
 \bibitem[Aauthor \& Bauthor(2003a)]{key-2}
Motte, F, Louvet, F., \& Nguyen-Luong, Q., 2017, IAU Symp, 316, 9
 \bibitem[Aauthor \& Bauthor(2003a)]{key-2}
Motte, F, Bontemps, S., \& Louvet, F., 2018a, ARA\&A ,56 ,41
 \bibitem[Aauthor \& Bauthor(2003a)]{key-2}
Motte, F, et al., 2018b, Nature Astronomy ,2, 478
 \bibitem[Aauthor \& Bauthor(2003a)]{key-2}
Mufson, S. L., \& Liszt, H. S., 1977, ApJ, 212, 664
\bibitem[Nakajima et al.(2019)]{2019PASJ...71..S17N} 
{Nakajima, T., Inoue, H., Fujii, Y., Miyazawa, C., Iwashita, H., Sakai, T., Noguchu, T., \& Mizuno, A., \ 2019, \pasj, 71, S17}
  \bibitem[Aauthor \& Bauthor(2003a)]{key-2}
Nakanishi, H., \& Sofue, Y., 2016, PASJ, 68, 5
 \bibitem[Aauthor \& Bauthor(2003a)]{key-2}
Nguyen Luong, Q., et al., 2011, A\&A, 529, A41
 \bibitem[Aauthor \& Bauthor(2003a)]{key-2}
Nguyen Luong, Q., et al., 2013, A\&A, 775, 88
 \bibitem[Aauthor \& Bauthor(2003a)]{key-2}
Nguyen Luong, Q., et al., 2017, ApJ, 844, 25
 \bibitem[Aauthor \& Bauthor(2003a)]{key-2}
Nishitani, H., et al., 2012, PASJ, 64, 30
 \bibitem[Aauthor \& Bauthor(2003a)]{key-2}
Nishimura, A., et al., 2015, ApJS, 216, 18
 \bibitem[Aauthor \& Bauthor(2003a)]{key-2}
Nishimura, A., et al., 2018, PASJ, 70S, 42
 \bibitem[Aauthor \& Bauthor(2003a)]{key-2}
Nony, T., et al., 2018, A\&A, 618L, 5
  \bibitem[Aauthor \& Bauthor(2003a)]{key-2}
Ohama, A., et al. 2010, ApJ, 709, 975
  \bibitem[Aauthor \& Bauthor(2003a)]{key-2}
Ohama, A., et al. 2018a, PASJ, 70S, S45
  \bibitem[Aauthor \& Bauthor(2003a)]{key-2}
Ohama, A., et al. 2018b, PASJ, 70S, S47
  \bibitem[Okumura et~al.(2001)]{Okumura2001}
Okumura, S., Miyawaki, R., Sorai, K., Yamashita, T., \& Hasegawa, T., 2001, PASJ, 53, 793
 \bibitem[Aauthor \& Bauthor(2003a)]{key-2}
Olmi, L., Cesaroni, R., Hofner, P., Kurtz, S., Churchwell, E., \& Walmsley, C. M., 2003, A\&A, 407, 225
 \bibitem[Aauthor \& Bauthor(2003a)]{key-2}
Paron, S., Areal, M. B., \& Ortega, M. E., 2018, A\&A, 617, 14
\bibitem[Aauthor \& Bauthor(2003a)]{key-2}
Pe\'rez, F., \& Granger, B. 2007, Comput. Sci. Eng., 9, 21
\bibitem[Aauthor \& Bauthor(2003a)]{key-2}
Parsons, H., Thompson, M. A., Clark, J. S., \& Chrysostomou, A., 2012, MNRAS, 424, 1658
\bibitem[Aauthor \& Bauthor(2003a)]{key-2}
Pineda, J. L., Goldsmith, P. F., Chapman, N., Snell, R. L., Li, D., Cambr\'esy, L., \& Brunt, C. 2010, ApJ, 721, 686
\bibitem[Aauthor \& Bauthor(2003a)]{key-2}
Pipher, J. L., Grasdalen, G. L., \& Soifer, B. T., 1974, ApJ, 193, 283
\bibitem[Churchwell, E., et al(2006)]{key-2}
Pillai, T., Kauffmann, J., Wyrowski, F., Hatchell, J., Gibb, A. G., \& Thompson, M. A. 2011, A\&A, 530, A118
\bibitem[Aauthor \& Bauthor(2003a)]{key-2}
Povich, M. S., et al., 2007, ApJL 660, 346
\bibitem[Aauthor \& Bauthor(2003a)]{key-2}
Povich, M. S., \& Whitney, B. A., 2010, ApJL, 714, L285
\bibitem[Aauthor \& Bauthor(2003a)]{key-2}
Pratap, P., Menten, K. M., Snyder, L. E., 1994, ApJ, 430, 129
\bibitem[Aauthor \& Bauthor(2003a)]{key-2}
Pratap, P., Megeath, S. T., Bergin, E. A., 1999, ApJ, 517, 799
\bibitem[Aauthor \& Bauthor(2003a)]{key-2}
Reid, M. J., Dame, T. M., Menten, K. M., \& Brunthaler, A. 2016, ApJ, 823, 77
\bibitem[Aauthor \& Bauthor(2003a)]{key-2}
Renaud, F., et al., 2015, MNRAS, 454, 3299
\bibitem[Aauthor \& Bauthor(2003a)]{key-2}
Rieke, G. H., et al. 2004, ApJS, 154, 25 
  \bibitem[Aauthor \& Bauthor(2003a)]{key-2}
Rigby, A. J., et al. 2016, MNRAS, 456, 2885
  \bibitem[Aauthor \& Bauthor(2003a)]{key-2}
{Rigby, A. J., et al. 2019, 2019, A\&A, 632, A58)}
  \bibitem[Aauthor \& Bauthor(2003a)]{key-2}
Rizzo, J. R., Fuente, A., Rodri\'guez-Franco, A., \& Garc\'ia-Burillo, S., 2003, ApJ, 597, L153
  \bibitem[Aauthor \& Bauthor(2003a)]{key-2}
Roberts, W. W., Jr. 1969, ApJ, 158, 123
  \bibitem[Aauthor \& Bauthor(2003a)]{key-2}
Roshi, D. A., Churchwell, E., \& Anderson, L. D., 2017, ApJ, 838, 144
  \bibitem[Aauthor \& Bauthor(2003a)]{key-2}
Salpeter, E. E., 1955, ApJ, 121, 161
  \bibitem[Aauthor \& Bauthor(2003a)]{key-2}
Sano, H., et al., 2018, PASJ, 70S, 43
  \bibitem[Aauthor \& Bauthor(2003a)]{key-2}
Sato, F., Hasegawa, T., Whiteoak, J. B., \& Miyawaki, R. 2000, ApJ, 535, 857 
  \bibitem[Aauthor \& Bauthor(2003a)]{key-2}
Sato, M., Reid, M. J., Brunthaler, A., \& Menten, K., 2010, ApJ, 720, 1055
  \bibitem[Aauthor \& Bauthor(2003a)]{key-2}
Sato, M., et al. 2014, ApJ, 793,  72
  \bibitem[Aauthor \& Bauthor(2003a)]{key-2}
Sawada, T., et al. 2008, PASJ, 60, 445 
  \bibitem[Aauthor \& Bauthor(2003a)]{key-2}
Sawada, T., Hasegawa, T., Sugimoto, M., Koda, J., \& Handa, T., 2012a, ApJ, 752, 118
  \bibitem[Aauthor \& Bauthor(2003a)]{key-2}
Sawada, T., Hasegawa, T., \& Koda, J., 2012b, ApJL, 759, L26
 \bibitem[Aauthor \& Bauthor(2003a)]{key-2}
{Sawada, T., Koda, J., \& Hasegawa, T., 2018, ApJ, 867, 166}
 \bibitem[Aauthor \& Bauthor(2003a)]{key-2}
Scoville, N. Z., Sanders, D. B., \& Clemens, D. P., 1986, ApJ, 310L, 77
 \bibitem[Aauthor \& Bauthor(2003a)]{key-2}
Scoville, N. Z., Yun, Min Su., Clemens, D. P., Sanders, D. B., \& Waller, W. H., 1987, ApJS, 63, 821
 \bibitem[Aauthor \& Bauthor(2003a)]{key-2}
Shima, K., Tasker, E. J., Federrath, C., \& Habe, A. 2018, PASJ, 70, S54
 \bibitem[Aauthor \& Bauthor(2003a)]{key-2}
Smith, L. F., Biermann, P., \& Mezger, P. G., 1978, A\&A, 66, 65
  \bibitem[Aauthor \& Bauthor(2003a)]{key-2}
Sofue, Y., 1985, PASJ, 37, 507
  \bibitem[Aauthor \& Bauthor(2003a)]{key-2}
{Sofue, Y., et al., 2019, PASJ, 71, S1} 
  \bibitem[Aauthor \& Bauthor(2003a)]{key-2}
Solomon, P. M., Rivolo, A. R., Barrett, J., \& Yahil, A., 1987, ApJ, 319, 730
  \bibitem[Aauthor \& Bauthor(2003a)]{key-2}
Su, Y., et al., 2019, ApJS, 240, 9
  \bibitem[Aauthor \& Bauthor(2003a)]{key-2}
Subrahmanyan, R., \& Goss, W. M., 1996, MNRAS, 281, 239
\bibitem[Sridharan2014]{Sridharan2014}
Sridharan, T. K., Rao R., Qiu, K., Cortes, P., Li, H., Pillai, T., Patel, N. A, \& Zhang, Q., 2014, ApJ, 783, L31
 \bibitem[Takahira et~al.(2014)]{Takahira2014}
Takahira, K., Tasker, E. J., \& Habe, A., 2014, ApJ, 792, 63  
\bibitem[Takahira et~al.(2014)]{Takahira2014}
Takahira, K., Shima, K., Habe, A., \& Tasker, E. J., 2018, PASJ, 70, 58  
 \bibitem[Aauthor \& Bauthor(2003a)]{key-2}
Torii, K., et al., 2011, ApJ, 738, 46
  \bibitem[Aauthor \& Bauthor(2003a)]{key-2}
Torii, K., et al., 2015, ApJ, 806, 7
 \bibitem[Aauthor \& Bauthor(2003a)]{key-2}
Torii, K., et al., 2017, ApJ, 835, 142
  \bibitem[Aauthor \& Bauthor(2003a)]{key-2}
Torii, K., et al., 2018a, PASJ, 70S, 51
  \bibitem[Aauthor \& Bauthor(2003a)]{key-2}
Torii, K., et al., 2018b, PASJ, in press (DOI: 10.1093/pasj/psy098)
  \bibitem[Aauthor \& Bauthor(2003a)]{key-2}
{Torii, K., et al., 2019, \pasj, 71, S2}
  \bibitem[Aauthor \& Bauthor(2003a)]{key-2}
Tosa, M. 1973, PASJ, 25, 191
  \bibitem[Aauthor \& Bauthor(2003a)]{key-2}
Townsley, L. K., Broos, P. S., Garmire, G. P., Bouwman, j., Povich, M. S., Feigelson, E. D., Getman, K. V., Kuhn, M. A., 2014, ApJS, 213,1
  \bibitem[Aauthor \& Bauthor(2003a)]{key-2}
Tsuboi, M., Miyazaki, A., \& Uehara, K., 2015, PASJ, 67, 109
   \bibitem[Aauthor \& Bauthor(2003a)]{key-2}
{ Uehara, K., Tsuboi, M., Kitamura, Y., Miyawaki, R., \& Miyazaki, A., 2019, ApJ, 872, 121}
   \bibitem[Aauthor \& Bauthor(2003a)]{key-2}
{ Ulich, B. L., \& Haas, R. W., 1976, ApJS, 30, 247}
  \bibitem[Aauthor \& Bauthor(2003a)]{key-2}
{ Umemoto, T., et al. 2017, PASJ, 69, 78}
  \bibitem[Aauthor \& Bauthor(2003a)]{key-2}
Vall\'ee, J. P., 2014, ApJS, 215, 1
  \bibitem[Aauthor \& Bauthor(2003a)]{key-2}
van der Hutch, K. A. 2001, New Astron. Rev., 45, 135
  \bibitem[Aauthor \& Bauthor(2003a)]{key-2}
Van der Tak, F.F.S., Black, J.H., Schoier, F.L., Jansen, D.J., \& van Dishoeck, E.F., 2007, A\&A 468, 627
  \bibitem[Aauthor \& Bauthor(2003a)]{key-2}
van der Walt, S., Colbert, S. C., \& Varoquaux, G. 2011, Comput. Sci. Eng., 13, 22
 \bibitem[Aauthor \& Bauthor(2003a)]{key-2}
Veneziani, M., et al. 2017, A\&A, 599, A7
 \bibitem[Aauthor \& Bauthor(2003a)]{key-2}
{ Walsh, A. J., et al. 2016, MNRAS, 455, 3494}
  \bibitem[Aauthor \& Bauthor(2003a)]{key-2}
Watarai, H., Matsuhara, H., Takahashi, H \& Matsumoto, T., 1998, ApJ, 507, 263
 \bibitem[Aauthor \& Bauthor(2003a)]{key-2}
Watson, A. M., \& Hanson, M. M. 1997, ApJ, 490, L165
  \bibitem[Aauthor \& Bauthor(2003a)]{key-2}
{Werner, M. W., et al. 2004, ApJS, 154, 1}
  \bibitem[Aauthor \& Bauthor(2003a)]{key-2}
Westerhout, G. 1958, Bull. Astron. Inst. Netherlands, 14, 215
  \bibitem[Aauthor \& Bauthor(2003a)]{key-2}
Wilson, T. L., \& Rood, R. 1994, ARA\&A, 32, 191
  \bibitem[Aauthor \& Bauthor(2003a)]{key-2}
Wilson, T. L., Rohlfs, K, \& Huttemeister, S. 2009, Tools of Radio Astronomy, 5th ed. (Berlin: Springer-Verlag)
  \bibitem[Aauthor \& Bauthor(2003a)]{key-2}
{Wolfire, M. G., \& Cassinelli, J. P. 1987, ApJ, 319, 850}
  \bibitem[Aauthor \& Bauthor(2003a)]{key-2}
Wood, D. O. S., \& Churchwell, E., 1989, ApJS, 69, 831
  \bibitem[Aauthor \& Bauthor(2003a)]{key-2}
Wu, B., Van Loo, S., Tan, J. C., \& Bruderer, S. 2015, ApJ, 811, 56
  \bibitem[Aauthor \& Bauthor(2003a)]{key-2}
Wu, B., Tan, J. C., Nakamura, F., Van, Loo. S., Christie, D., \& Collins, D. 2017a, ApJ, 835, 137
  \bibitem[Aauthor \& Bauthor(2003a)]{key-2}
Wu, B., Tan, J. C., Christie, D., Nakamura, F., Van, L S., \& Collins, D. 2017b, ApJ, 841, 88
  \bibitem[Aauthor \& Bauthor(2003a)]{key-2}
Wu, B., Tan, J. C., Nakamura, F., Christie, D., Li, Q., 2018, PASJ, 70S, 57
  \bibitem[Aauthor \& Bauthor(2003a)]{key-2}
Yamagishi, M., et al., 2016, ApJ, 833, 163
  \bibitem[Aauthor \& Bauthor(2003a)]{key-2}
Zhang, B., et al., 2014, ApJ, 781, 89
  \bibitem[Aauthor \& Bauthor(2003a)]{key-2}
Zhu, Q-F., Lacy, J. H., Jaffe, D. T., Richter, M. J., \& Greathouse, T. K., 2005, ApJ, 631, 381
  \bibitem[Aauthor \& Bauthor(2003a)]{key-2}
{Zinnecker, H., \& Yorke, H. W. 2007, ARA\&A, 45, 481}
\end{thebibliography}
\end{document}